\documentclass[a4paper,fleqn,usenatbib]{mnras}

\usepackage[T1]{fontenc}
\usepackage{aecompl}

\usepackage[utf8]{inputenc}
\usepackage{bm}
\usepackage{caption}

\usepackage[dvipsnames]{xcolor}
\definecolor{mycolor}{RGB}{255,160,0}
\definecolor{mycolor_1}{RGB}{0,190,190}
\definecolor{mycolor_2}{RGB}{11,220,220}

\usepackage{graphicx}	
\usepackage{amsmath}	
\usepackage{amssymb}	
\usepackage{xcolor}
\usepackage{ulem}
\usepackage{rotating}
\usepackage{orcidlink}

\newcommand{\tavg}{\Delta t_{\rm avg}}
\newcommand{\dMS}{\delta\text{MS}}
\newcommand{\SFS}{\text{SFMS}}
\newcommand{\SFR}{\text{SFR}}
\newcommand{\sSFR}{\text{sSFR}}
\newcommand{\SFE}{\text{SFE}}
\newcommand{\Dmax}{D_{\rm max}}
\newcommand{\Qorig}{q_{\rm orig}}
\newcommand{\mol}{{\mathrm{H}_2}}
\newcommand{\fmol}{f_\mol}
\newcommand{\neutr}{{\mathrm{HI}+\mathrm{H}_2}}
\newcommand{\ximol}{\xi_\mol}
\newcommand{\SB}{{\rm SB}}
\newcommand{\tdepl}{t_{\rm depl, \mathrm{H}_2}}
\newcommand{\SF}{{\rm SF}}
\newcommand{\Myr}{~\text{Myr}}
\newcommand{\Gyr}{~\text{Gyr}}
\newcommand{\Msun}{~\text{M}_\odot}
\newcommand{\pc}{~\text{pc}}
\newcommand{\kpc}{~\text{kpc}}
\newcommand{\vir}{{\text{vir}}}
\newcommand{\percent}{~\text{per~cent}}
\newcommand{\avgSF}{\langle \ximol \, t^{-1}_{\rm ff}\rangle_\SF}
\newcommand{\willSB}{{\rm{ctrl}\Rightarrow\SB}}

\title[What drives a starburst]{Starbursts driven by central gas compaction}

\author[Cenci et al.]{
\parbox{\textwidth}{
Elia Cenci$^{1}$\orcidlink{0000-0002-0766-1704}\thanks{E-mail:\href{mailto:elia.cenci@uzh.ch}{elia.cenci@uzh.ch}},
Robert Feldmann$^{1}$\orcidlink{0000-0002-1109-1919},
Jindra Gensior$^{1}$\orcidlink{0000-0001-6119-9883},
Jorge Moreno$^{2,3}$\orcidlink{0000-0002-3430-3232},
Luigi Bassini$^{1}$\orcidlink{0000-0002-6864-7762},
Mauro Bernardini$^{1}$\orcidlink{0000-0002-2930-9509}
\vspace{6pt}
}\\
$^{1}$Institute for Computational Science, University of Zurich, Winterthurerstrasse 190, Zurich CH-8057, Switzerland \\
$^{2}$Department of Physics and Astronomy, Pomona College, Claremont, CA 91711, USA \\
$^{3}$Centre for Computational Astrophysics, Flatiron Institute, 162 Fifth Avenue, New York, NY 10010, USA
}

\date{Accepted XXX. Received YYY; in original form ZZZ}

\pubyear{2023}

\begin{document}
\label{firstpage}
\pagerange{\pageref{firstpage}--\pageref{lastpage}}
\maketitle

\begin{abstract}
Starburst (SB) galaxies are a rare population of galaxies with star formation rates (SFRs) greatly exceeding those of the majority of star-forming galaxies with similar stellar mass. It is unclear whether these bursts are the result of either especially large gas reservoirs or enhanced efficiencies in converting gas into stars. Tidal torques resulting from gas-rich galaxy mergers are known to enhance the SFR by funneling gas towards the centre. However, recent theoretical works show that mergers do not always trigger a SB and not all SB galaxies are interacting systems, raising the question of what drives a SB. We analyse a large sample of SB galaxies and a mass- and redshift-matched sample of control galaxies, drawn from the FIREbox cosmological volume at $z=0-1$. We find that SB galaxies have both larger molecular gas fractions and shorter molecular depletion times than control galaxies, but similar total gas masses. Control galaxies evolve towards the SB regime by gas compaction in their central regions, over timescales of $\sim 70\Myr$, accompanied by an increase in the fraction of ultra-dense and molecular gas. The driving mechanism behind the SB varies depending on the mass of the galaxy. Massive ($M_\star\gtrsim 10^{10}~\rm{M}_\odot$) galaxies undergoing intense, long-lasting SBs are mostly driven by galaxy interactions. Conversely, SBs in non-interacting galaxies are often triggered by a global gravitational instability, that can result in a ‘breathing’ mode in low-mass galaxies.
\end{abstract}


\section{Introduction}
The star formation rate ($\SFR$) of galaxies scales with their stellar mass ($M_\star$), both at low- \citep[][]{Brinchmann2004,Salim2007} and high-redshift \citep[][]{Daddi2007}. This relation is often referred to as the star-forming main-sequence ($\SFS$). The normalisation of the $\SFS$ increases with redshift, as gas fractions are higher at earlier epochs \citep[][]{Daddi2010a,Leslie2020}. The slope and scatter only mildly change with redshift and stellar mass \citep[][]{Noeske2007,Whitaker2012,Speagle2014,Schreiber2015}.

Starburst (SB) galaxies are the most intensely star-forming outliers above the $\SFS$ \citep[e.g.,][]{Sanders&Mirabel1996}. The fractional contribution of SB galaxies to the population of star-forming galaxies increases with redshift \citep[e.g.,][]{Rodighiero2011,Sargent2012,Caputi2017}, from about $1\percent$ at $z\lesssim 0.4$ \citep[e.g.,][]{Bergvall2016} to about $5\percent$ at $0.5<z<1$ \citep[][]{Bisigello2018,Rinaldi2022}.


The nature and origin of SBs is still debated. Two main, possibly concurrent, scenarios have been proposed to explain their enhanced star forming nature: (\textit{i}) a larger molecular gas reservoir to fuel star formation \citep[e.g.,][]{Combes1994,Casasola2004,Scoville2016,Scoville2017,Tacconi2018}; and (\textit{ii}) an increased efficiency in converting molecular gas into stars, or, equivalently, shorter (molecular gas) depletion times, i.e the time it would take to convert the available molecular gas into stars given the current $\SFR$ \citep[e.g.,][]{Sofue1993,Solomon_&_Sage1998,Sargent2014,Michiyama2016,Silverman2015,Silverman2018,Feldmann2020}. In fact, SB galaxies are often thought to form a separate sequence to normal star-forming disc galaxies, with systematically shorter depletion times, related to having a larger $\SFR$ surface density ($\Sigma_{\SFR}$) at fixed gas surface density ($\Sigma_{\rm gas}$) \citep[][]{Schmidt1959,KennicuttJr.1998,Daddi2010b,Genzel2010,Kennicutt&Evans2012,Kennicutt2021}. However, selection biases and the use of different conversion factors for SB galaxies with respect to normal star-forming discs, to retrieve the $H_2$ content from molecular tracers (e.g., CO), could lead to an apparent bimodality in depletion times. In fact, unbiased observations of the local Universe reveal continuously distributed depletion times, from values that are typical for normal star-forming disc galaxies to SB galaxies \citep[e.g.,][]{Saintonge2011a,Saintonge2011b,Feldmann2020}. Furthermore, \citet[][]{Krumholz2012} argue that local, normal discs and SB galaxies obey a universal star formation law, where the difference between the two regimes is set by how decoupled giant molecular clouds are from the average density of the interstellar medium (ISM).


High-resolution simulations provide a powerful theoretical framework to probe these scenarios and to investigate the nature of SB galaxies and their triggering mechanisms. Recent simulations demonstrated that galaxy mergers and interactions can drive SBs through tidal torques \citep[e.g.,][]{Cox2008,Capelo2015,Renaud2014,Moreno2015,Hopkins2018}, that can funnel the gas toward the centre, and enhance both the density \citep[][]{Genzel2010,Moreno2019,Renaud2019,Moreno2021,SegoviaOtero2022} and total mass of $\mathrm{H}_2$ \citep[][]{Ellison2013,Pan2018}. Local ultra-luminous infrared galaxies (ULIRGs) are a class of intensely star-forming galaxies \citep[e.g.,][]{Sanders&Mirabel1996} that is typically associated with strongly interacting systems, because of the presence of clear tidal features and disturbed morphologies, with depletion times found that are $\gtrsim 10$ times shorter than in normal star-forming galaxies \citep[e.g.,][]{Gao&Solomon2004,Garcia-Burillo2012,Hung2013,Pereira-Santaella2021}. However, non-interacting ULIRGs have depletion times comparable to those of interacting ones \citep[][]{Violino2018}. Indeed, recent theoretical and observational works suggest that mergers do not always trigger a SB, \citep[e.g.,][]{Bergvall2003c,Di_Matteo2007,Di_Matteo2008,Sparre&Springel2016,Pearson2019}, especially at high redshift \citep[e.g.,][]{Fensch2017,Shah2020}, and not all SBs are associated with merging systems \citep[e.g.,][]{Wilkinson2018,Diaz-Garcia&Knapen2020,Wilkinson2022,Li2023}.

In this paper, we analyse SB galaxies and non-starbursting (control) galaxies from the FIREbox cosmological volume \citep[][]{Feldmann2023}, over a wide range in masses and redshifts. FIREbox captures the multi-phase structure of the ISM, unlike previous works on SB galaxies in modern cosmological simulations, like, e.g., SIMBA \citep[][]{Rodriguez-Montero2019}{}{}, Illustris \citep[][]{Wilkinson2018}{}{}, and IllustrisTNG \citep[][]{Patton2020}{}{}. The high spatial and mass resolution of FIREbox and its accurate treatment of the physics of the interstellar medium (e.g., stellar feedback and gas cooling down to 10 K), make it ideally suited to explore the nature of SB events.
The paper is structured as follows. In Section ~\ref{sec:methods}, we describe the simulation and our samples. In Section~\ref{sec:results}, we report the properties of SB galaxies, also comparing to control galaxies, analyse the temporal evolution of galaxies in the period preceding the SB, and investigate the role of interactions in triggering SBs and the driving mechanism in non-interacting systems. Finally, in Section~\ref{sec:discussion}, we summarize and discuss our results.

\section{Methods}\label{sec:methods}
\subsection{FIREbox}\label{sec:methods_FIREbox}
In this work, we make use of the FIREbox cosmological volume simulation \citep[][]{Feldmann2023}. This volume has a length of $22.1~\text{cMpc}$ per side and is part of the \textit{Feedback In Realistic Environments}\footnote{\url{https://FIRE.northwestern.edu}} (FIRE) project \citep{Hopkins2014a,Hopkins2018}. Initial conditions at $z=120$ were created with MUlti Scale Initial Conditions \citep[MUSIC;][]{Hahn2011} using cosmological parameters consistent with Planck 2015 results \citep{Planck2015}: $\Omega_{\rm m}=0.3089$, $\Omega_\Lambda=1-\Omega_{\rm m}$, $\Omega_{\rm b}=0.0486$, $h=0.6774$, $\sigma_8=0.8159$, $n_{\rm s}=0.9667$ and a transfer function calculated with \textsc{camb}\footnote{\url{http://camb.info}} \citep{Lewis2000,Lewis2011}.

The simulation is run with \textsc{gizmo} \citep{Hopkins2015}. Gravitational forces between particles are calculated with a heavily modified version of the parallelisation and tree gravity solver of GADGET-3 \citep{Springel2005} allowing for adaptive force softening, while hydrodynamics is solved with the meshless-finite-mass method introduced in \citet{Hopkins2015}.

FIREbox is run with the FIRE-2 model that includes gas cooling and heating, star formation, and stellar feedback \citep{Hopkins2018}. Feedback from supermassive black holes is not included. Gas cooling down to 10 K naturally results in a multi-phase interstellar medium with a cold component. Star formation occurs in dense ($n>300$ cm$^{-3}$), self-shielding, self-gravitating, and Jeans-unstable gas, with an instantaneous, expected star formation rate:

\begin{equation}
   \SFR = \ximol \, m_{\rm gas} \; t^{-1}_{\rm ff} ~,
   \label{eqn:SFR_prescription}
\end{equation}

\noindent where $m_{\rm gas}$ is the mass of the gas particle, $\ximol$ is the fraction of molecular gas in the particle, and $t_{\rm ff}$ is the free-fall time, i.e., the time it would take for the gas particle to collapse under its self-gravity:

\begin{equation}
   t_{\rm ff} = \sqrt{\frac{3\pi}{32 G \rho}\,\,}~,
   \label{eqn:tff}
\end{equation}

where $\rho$ is the local mass density of the gas particle and $G$ is Newton's gravitational constant. The gas to star conversion takes place on a local free-fall time with a $100\percent$ efficiency. Due to stellar feedback, the integrated local star formation efficiency is lower, consistent with the Schmidt relation \citep{Schmidt1959,KennicuttJr.1998,Orr2018}. Stellar feedback includes energy, momentum, mass, and metal injections from supernovae (type II and type Ia) and stellar winds (OB and AGB stars). Radiative feedback (photo-ionisation and photo-electric heating) and radiation pressure from young stars is accounted for in the Locally Extincted Background Radiation in Optically thin Networks (LEBRON) approximation \citep{Hopkins2012a}. The FIRE-2 model has been extensively validated in a number of publications analysing properties of galaxies across a range in stellar masses and numerical resolutions \citep{Wetzel2016,Hopkins2018,Ma2018a,Ma2018}. Specifically, FIREbox reproduces key observed galaxy properties \citep[see][for further details]{Feldmann2023}{}{} and has been further validated in several recent studies \citep[e.g.,][]{Bernardini2022,Rohr2022,Gensior2023}{}{}.

At the initial redshift, FIREbox contains $N_{\rm b}=1024^3$ gas and $N_{\rm DM}=1024^3$ dark matter particles, corresponding to a mass resolution of $m_{\rm b}=6.3\times 10^4\Msun$ for baryonic (gas and star) particles and $m_{\rm DM}=3.3\times 10^5\Msun$ for dark matter particles. The minimum gas softening length is fixed to $12\pc$ (physical, up to $z=9$; comoving for $z>9$) for stars and $80\pc$ for dark matter. The force softening of gas particles is adaptive and coupled to their smoothing length down to a minimum of $1.5\pc$, which is reached only in the dense ISM. The force resolution is set such that the highest density we formally resolve is 1000 times the star formation threshold \citep[see][Section~2.2 for more details]{Hopkins2018}.

In the following, we analyse galaxies in the redshift range $z=0-1$, to have a representative, mass-complete sample of both low- and high-redshift galaxies after cosmic noon.

\subsection{Definitions}\label{sec:methods_definitions}
To identify dark matter haloes we employ the AMIGA Halo Finder (AHF)\footnote{\url{http://popia.ft.uam.es/AHF/Download.htmlTable}} \citep[][]{Gill2004,Knollmann2009}. We only consider haloes containing at least 100 particles of any type, which corresponds to a minimum halo mass of $M_\vir\sim 10^7\Msun\,h^{-1}$. The halo radius $R_\vir$ is defined based on the virial overdensity criterion, so that the halo virial mass is:

\begin{equation}
   M_\vir = \frac{4\pi}{3} \Delta\left(z\right) \rho_{\rm m}\left(z\right) R_\vir^3 ~,
   \label{eqn:Mvir}
\end{equation}

where $\rho_{\rm m}\left(z\right)$ is the critical density at a given redshift, $\Delta\left(z\right)=\left( 18\pi^2 - 82 \Omega_\Lambda\left(z\right) - 39\left[\Omega_\Lambda\left(z\right)\right]^2 \right)/\Omega_{\rm m}\left(z\right)$ is the overdensity parameter, and $\Omega_\Lambda\left(z\right)$, $\Omega_{\rm m}\left(z\right)$ are the cosmological parameters at redshift $z$ \citep[][]{Bryan&Norman1998}. Halo centres are defined as the loci with the highest total matter density (AHF's maximum density (MAX) setting). To compute galaxy sizes, we assume the total galaxy radius to be $10\percent$ of the halo virial radius \citep[][]{Price2017}. The galaxy stellar mass $M_\star$, as well as other galaxy properties such as total gas masses and $\SFR$s, are then calculated by collecting all particles within $0.1 R_\vir$. We can then compute half-mass radii for different components (e.g., stars, molecular gas, neutral gas) by linearly interpolating between log-binned radii and the cumulative mass. We compute the $\SFR$ as 

\begin{equation}
    \SFR_{\tavg} = M_\star\left(\rm{age}_\star\leq\tavg\right)/\tavg ~,
\end{equation}

\noindent where $M_\star\left(\rm{age}\leq\tavg\right)$ is the sum of the at-birth masses of stars that formed within the past time interval $\tavg$. Different values for the averaging time $\tavg$ can be associated with different observational tracers for $\SFR$ \citep[e.g.,][]{Sparre2017,Flores-Velazquez2021}. We choose $\tavg=5,20,100\Myr$, approximating the characteristic timescales probed by the most extensively employed $\SFR$ tracers: $H_\alpha$ nebular emission and IR/UV continuum from dust and young stars \citep[e.g.,][]{Calzetti2013}. We refer to galaxies with a total $\SFR_{\tavg}\leq m_{\rm b}/\tavg\sim 10^{-11}\Msun\,\rm{yr}^{-1}$ as having negligible or zero $\SFR$, where $m_{\rm b}$ is the fixed mass resolution for star particles.

A non-zero $\SFR$ computed by means of equation~\eqref{eqn:SFR_prescription} is associated with self-gravitating, Jeans-unstable, self-shielding gas particles with densities $n>300~\rm{cm}^{-3}$ (see Section~\ref{sec:methods_FIREbox}). The total instantaneous $\SFR$ of a galaxy is thus given by the sum of all $\SFR$s of its gas particles:

\begin{equation}
    \SFR = \sum_i \frac{\xi_{\mol,i}}{t_{\rm ff, i}}\, m_{\rm gas, i}\; \delta_{\SF, i} ~,
\end{equation}

\noindent where $\delta_{\SF, i}$ is a binary function being 1 in case the $i-$th particle is eligible to form stars and 0 otherwise. The molecular gas fraction $\xi_{\mol,i}$ of each particle is modelled as a function of its metallicity and dust optical depth for Lyman-Werner photons \citep[][]{Krumholz2008,McKee&Krumholz2010,Krumholz&Gnedin2011}. The model assumes photo-dissociation and two-phase equilibrium. The dust optical depth is calculated via a local Sobolev-length approximation, estimating the inter-particle separation using the kernel length of the particle \citep[][]{Hopkins2015,Hopkins2018}.
 
In our simulation, all gas particles have approximately the same mass $m_{\rm gas}$. Moreover, if we suppose to have $\mathcal{N}_\SF$ star-forming particles\footnote{Note that $\mathcal{N}_\SF\equiv\sum_i\delta_{\SF, i}$.}, then we can write:

\begin{align}
    \SFR &= m_{\rm gas}\,\frac{\mathcal{N}_\SF}{\mathcal{N}_\SF}\,\sum_i \frac{\xi_{\mol ,i}}{t_{\rm ff, i}}\, \delta_{\SF, i}\\
        &= \mathcal{N}_\SF\,m_{\rm gas}\,\avgSF ~.
\end{align}

\noindent where $\langle\,\cdot\,\rangle_\SF$ represents the average restricted to star-forming particles only. Let us introduce the fraction of star-forming gas mass $f_\SF\equiv M_\SF/M_{\rm gas}$, where $M_{\rm gas}$ is the sum of the masses of all gas particles belonging to the galaxy and $M_\SF\equiv\mathcal{N}_\SF\,m_{\rm gas}$ is the total mass of star-forming particles particles, i.e., satisfying all star formation criteria. The galaxy total (instantaneous) $\SFR$ can thus be written as follows:

\begin{equation}
    \SFR = f_\SF \, M_{\rm gas} \, \avgSF ~.
    \label{eqn:SFR_gal}
\end{equation}

From equation~\eqref{eqn:SFR_gal} follows the expression for the galaxy (molecular) depletion time $\tdepl\equiv M_\mol/\SFR$ (or, equivalently, for the star formation efficiency $\SFE\equiv \tdepl^{-1}$), where $M_\mol$ is the mass of molecular gas, that accounts for any mechanism that leads to a change in $\SFR$ without affecting the total molecular gas budget in the galaxy:

\begin{equation}
    \SFE = \tdepl^{-1} = \frac{f_\SF}{\fmol} \, \avgSF ~,
    \label{eqn:tdepl}
\end{equation}

\noindent where $\fmol\equiv M_\mol/M_{\rm gas}$ is the fraction of molecular gas in the galaxy. While there are multiple criteria that a gas particle has to satisfy in order to be eligible to form stars, the most stringent is the density threshold criterion, given its high value of $300~\rm{cm}^{-3}$ in FIREbox. Particles above $n>300~\rm{cm}^{-3}$ are typically also self-gravitating, self-shielded, and Jeans-unstable. Therefore, any change in the $f_\SF$ implies a change in the mass of gas with high-enough density to be eligible for star formation.

\subsection{Sample selection}\label{sec:methods_samples}
In this work, we only consider FIREbox galaxies with stellar masses $M_\star\geq 10^8\Msun$ in the redshift range $z=0-1$, in order to avoid potential spurious effects related to the resolution of the simulation and have a representative, mass-complete sample of low- and high-redshift galaxies after cosmic noon. 

We assume that the $\SFS$ is a linear relation\footnote{Considering either a bending or an additional constant to the $\SFS$ fit function as in, e.g., \citet[][]{Lee2015,Daddi2022} would not improve the fit results.} between $\lg\sSFR\equiv\lg\left(\SFR/M_\star\right)$ and $\lg M_\star$, with an explicit dependence on the cosmological redshift $1+z$:

\begin{equation}
    \lg \sSFR_{_{\SFS}} = A\left(1+z\right)^{\alpha}\,\lg M_\star\,+\,B\,\left(1+z\right)^{\beta}~.
    \label{eqn:SFS_fit_function}
\end{equation}

\begin{figure}
    \centering
	\includegraphics[width=\columnwidth]{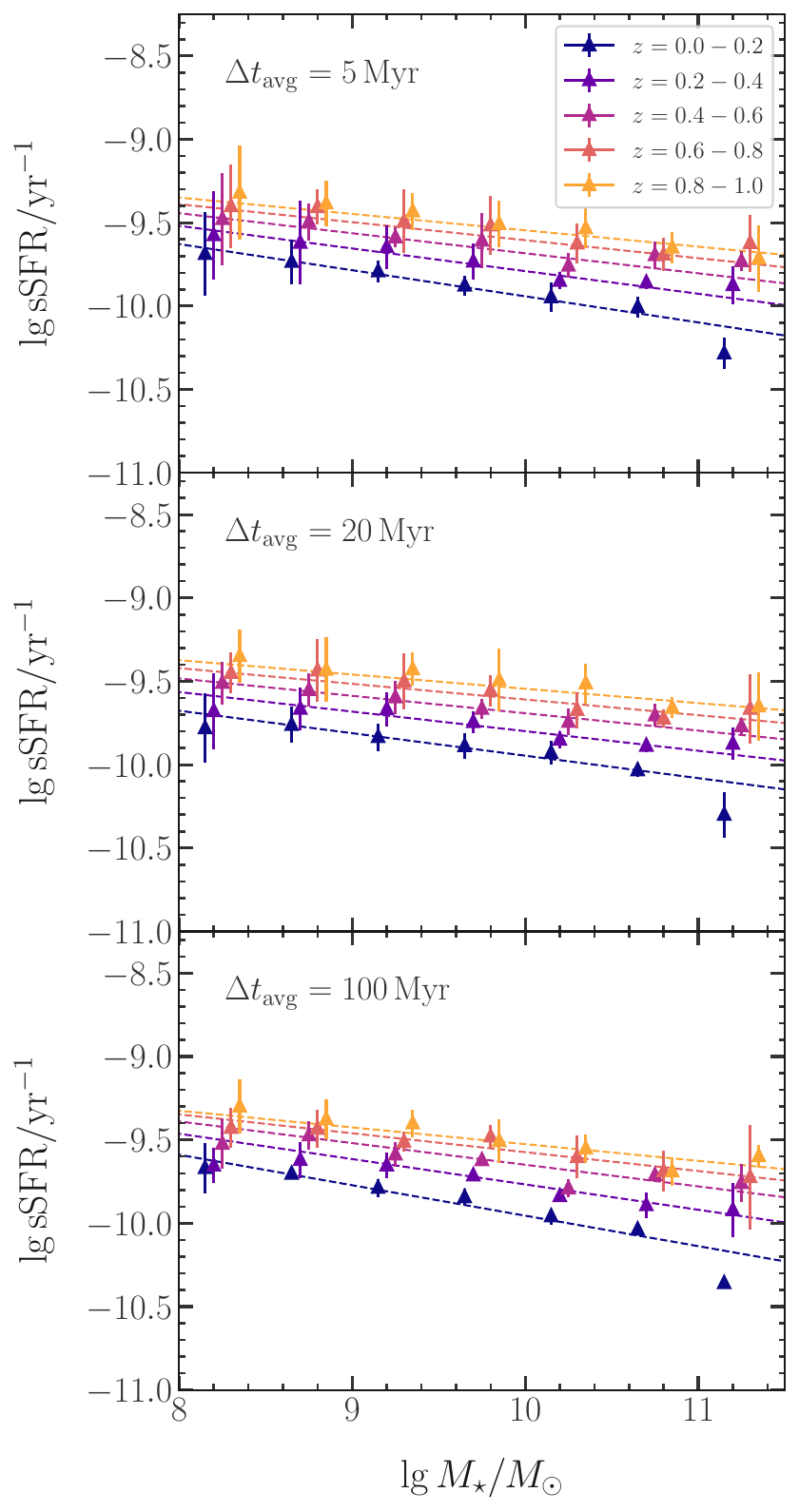}
    \caption{Specific star formation rate (s$\SFR$) as a function of stellar mass of FIREbox galaxies in the redshift range $z=0-1$, colour-coded by redshift, for different $\tavg$. Data points show the mode (with $2\sigma$ bootstrapped errors) of the s$\SFR$ in every 2D bin of stellar mass and redshift. For presentation purposes, the data points are slightly offset to the right with increasing redshift. The dashed, coloured lines show the $\SFS$ fit to the data points.}
    \label{fig:SFMS_fit}
\end{figure}

\noindent We estimate the free-parameters $\left\{A,B,\alpha,\beta\right\}$ by fitting equation~\eqref{eqn:SFS_fit_function} to the mode of the $\lg \sSFR$ distribution in each 2D bin of the $\left(1+z,\lg M_\star\right)$-space. Using the ridge-line of the distribution we are not biased by low-$\SFR$ galaxies and SBs, that would instead affect the median and the moments of the distribution. In Table~\ref{tab:SFS_fit_results}, we report the $\SFS$ fit results for different choices of $\tavg$, with bootstrapped $1\sigma$-errors. The fit parameters weakly depend on the choice of $\tavg$, with a steeper relation for $\tavg=100\Myr$. Figure~\ref{fig:SFMS_fit} shows the resulting $\SFS$ as a function of stellar mass and colour-coded according to the redshift. The data points refer to the mode $\sSFR$ to which we fit equation~\eqref{eqn:SFS_fit_function} in the corresponding stellar mass and redshift bin, for all choices of $\tavg$.

\begin{table}
    \centering
    \caption{Best-fit estimate for the parameters in equation~\eqref{eqn:SFS_fit_function}, with bootstrapped $1\sigma$ errors, for different choices of $\tavg$.}
    \begin{tabular}{ccc}
        \hline
        $\tavg$ [Myr] & $A$ & $\alpha$ \\
        \hline
        $5$ & $-0.17\pm0.03$ & $-0.91\pm0.36$\\
        $20$ & $-0.15\pm0.02$ & $-0.77\pm0.45$\\
        $100$ & $-0.22\pm0.04$ & $-1.17\pm0.32$\\
        \hline
    \end{tabular}
    
    \begin{tabular}{ccc}
        \hline
        $\tavg$ [Myr] & $B$  & $\beta$ \\
        \hline
        $5$ & $-8.36\pm0.24$ & $\quad 0.04\pm0.05$ \\
        $20$ & $-8.58\pm0.22$ & $\quad 0.02\pm0.06$ \\
        $100$ & $-7.92\pm0.34$ & $\quad 0.11\pm0.06$ \\
        \hline
    \end{tabular}
    \label{tab:SFS_fit_results}
\end{table}

We define SB galaxies as having a main-sequence offset of $\dMS\equiv \lg\left(\SFR/\SFR_{_{\SFS}}\right)\geq 0.6~\mathrm{dex}$, where, $\sSFR_{_{\SFS}}$ is the expected value on the $\SFS$, given their redshift and $M_\star$. This definition for SB galaxies is designed to select galaxies lying well above (at least of a factor of 2) the typical $\SFS$ scatter of about $0.2-0.3$ dex \citep[][]{Noeske2007,Daddi2007,Whitaker2012,Speagle2014,Wilkinson2018}{}{}. We also create a redshift- and mass-matched control sample of non-starbursting galaxies: for each SB galaxy, we select its control counterpart by randomly picking a galaxy with $\dMS<0.6~\mathrm{dex}$, at the same redshift, that matches the stellar mass of the SB galaxy within 0.1 dex.

\subsection{Interactions}\label{sec:methods_interactions}
We classify the galaxies in our samples as interacting, if another galaxy with a similar stellar mass approached them within the past $100\Myr$. Specifically, for each galaxy in the SB and control samples (referred to as the \textit{central} galaxy, in this section), we collected all other galaxies (\textit{satellites}, in this section) within a distance $\Dmax=200\kpc$ from its centre, as the impact of galaxy interactions can manifest even at these large separations \citep[e.g.,][]{Moreno2012,Moreno2013,Patton2013,Patton2016}{}{}. The coordinates of galaxies' centres are those given by the halo finder. We followed each satellite back in time until it first approached a distance $\leq\Dmax$ from the central galaxy. The ratio between the stellar masses of each satellite and its corresponding central galaxy, when the satellite first approached $\Dmax$ defines its original mass-ratio $\Qorig$. Furthermore, we also collected and traced all satellites that are found at $\leq\Dmax$ within the last 100 Myr. Therefore, we additionally account for galaxies that have already merged with the central galaxy. We also account for those satellites that entered and then escaped the neighbourhood of radius $\Dmax$ in the past 100 Myr \citep[i.e., `\textit{backsplash galaxies}' and `\textit{fly-bys}' ,][]{Benavides2021,Haggar2021}{}{}, potentially influencing the present observed properties of the central galaxy. Galaxies are followed back in time through their most similar progenitor, based on their position and stellar mass. A central galaxy is classified as interacting if it has at least one satellite with $\Qorig\geq 1:10$, as $\SFR$ can be enhanced as a result of these low-mass ratio galaxy interactions \citep[e.g.,][]{Cox2008,Jackson2019}{}{}. Otherwise, the galaxy is classified as non-interacting. Changing either the $\Dmax$ or the threshold on $\Qorig$, would affect the fraction of interacting galaxies the same way in both the SB and the control sample, but does not strongly affect our conclusions when comparing the SB and control sample.

\subsection{Starburst duration}\label{sec:methods_durations}
\begin{figure}
	\includegraphics[width=\columnwidth]{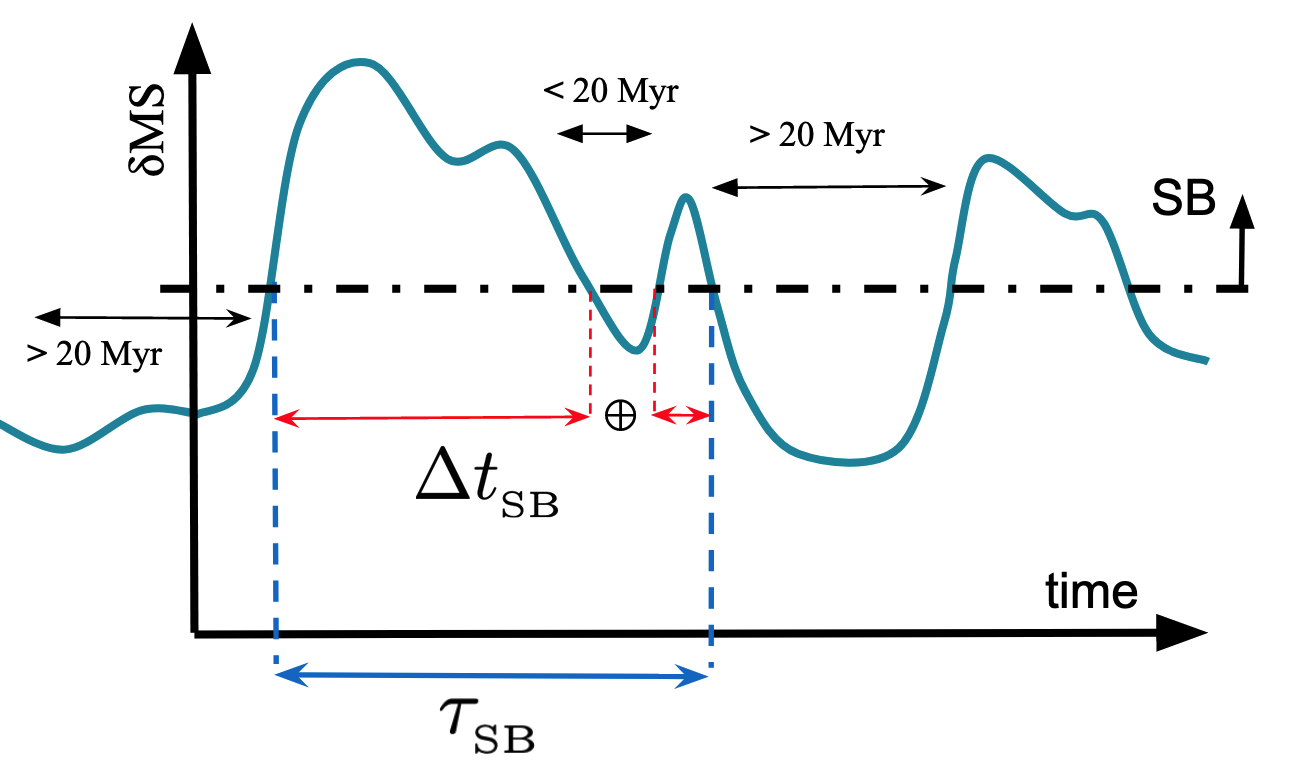}
    \caption{Schematic illustration for the two definitions employed to measure the duration of a SB event, $\tau_\SB$ and $\Delta t_\SB$. A mock star formation history of a SB galaxy is shown in terms of $\dMS$ as a function of cosmic time. The archaeological $\SFR$s are interpolated as a function of time in between snapshots. The effective beginning and end of SB events is computed assuming that distinct SB events taking place in the same galaxy are separated by $\geq20\Myr$ where $\dMS<0.6~\mathrm{dex}$. The duration $\tau_\SF$ is defined as the time interval between the effective beginning and end of the SB event, i.e., the SB-phase; $\Delta t_\SB$ is the time that the galaxy spends having $\dMS\geq 0.6~\mathrm{dex}$ during the SB-phase.}
    \label{fig:SB_duration_cartoon}
\end{figure}

In order to calculate the duration of SB events, we track our galaxies in time through their most similar progenitors and successors. By interpolating the archaeological $\SFR$s between snapshots as a function of cosmic time, we can recover the star formation history of any galaxy, which is then smoothed by taking the rolling average with an averaging time window equal to the chosen $\tavg$. Since the typical stars with a measurable ionising photon flux are O/B stars with main-sequence lifetimes $\lesssim 20\Myr$, we define a SB to end whenever $\dMS<0.6~\mathrm{dex}$ for a period of time $\geq20\Myr$. In other words, a galaxy undergoes multiple SB events if the $\dMS\geq 0.6~\mathrm{dex}$ periods are separated by $>20\Myr$ with $\dMS<0.6~\mathrm{dex}$. Conversely, if the $\SFS$ offset of the galaxy drops below 0.6 dex for less than $20\Myr$, before rising to $\dMS\geq 0.6~\mathrm{dex}$ again, this event is classified as a single SB. The beginning of the SB event is defined as the time at which the interpolated star formation history satisfies $\dMS\geq 0.6~\mathrm{dex}$, prior to the snapshot at which the galaxy is first identified as a SB. The end of the SB is defined as the time at which $\dMS<0.6~\mathrm{dex}$ for $\geq20\Myr$, after the time at which the galaxy is identified as a SB. We refer to the period of time between the effective beginning and end of the SB event as the SB-phase, that lasts for a time $\tau_\SB$. As an alternative measure of the SB duration, we consider the time $\Delta t_\SB$ that the galaxy has $\dMS\geq 0.6~\mathrm{dex}$ during the SB-phase. Figure~\ref{fig:SB_duration_cartoon} illustrates the employed definitions for the duration of a SB, by showing a fictitious star formation history (as the time-series for $\dMS$) including a SB event.

\section{Results}\label{sec:results}
\subsection{Starburst sample}\label{sec:results_SB_sample}

\begin{figure}
    \centering
	\includegraphics[width=\columnwidth]{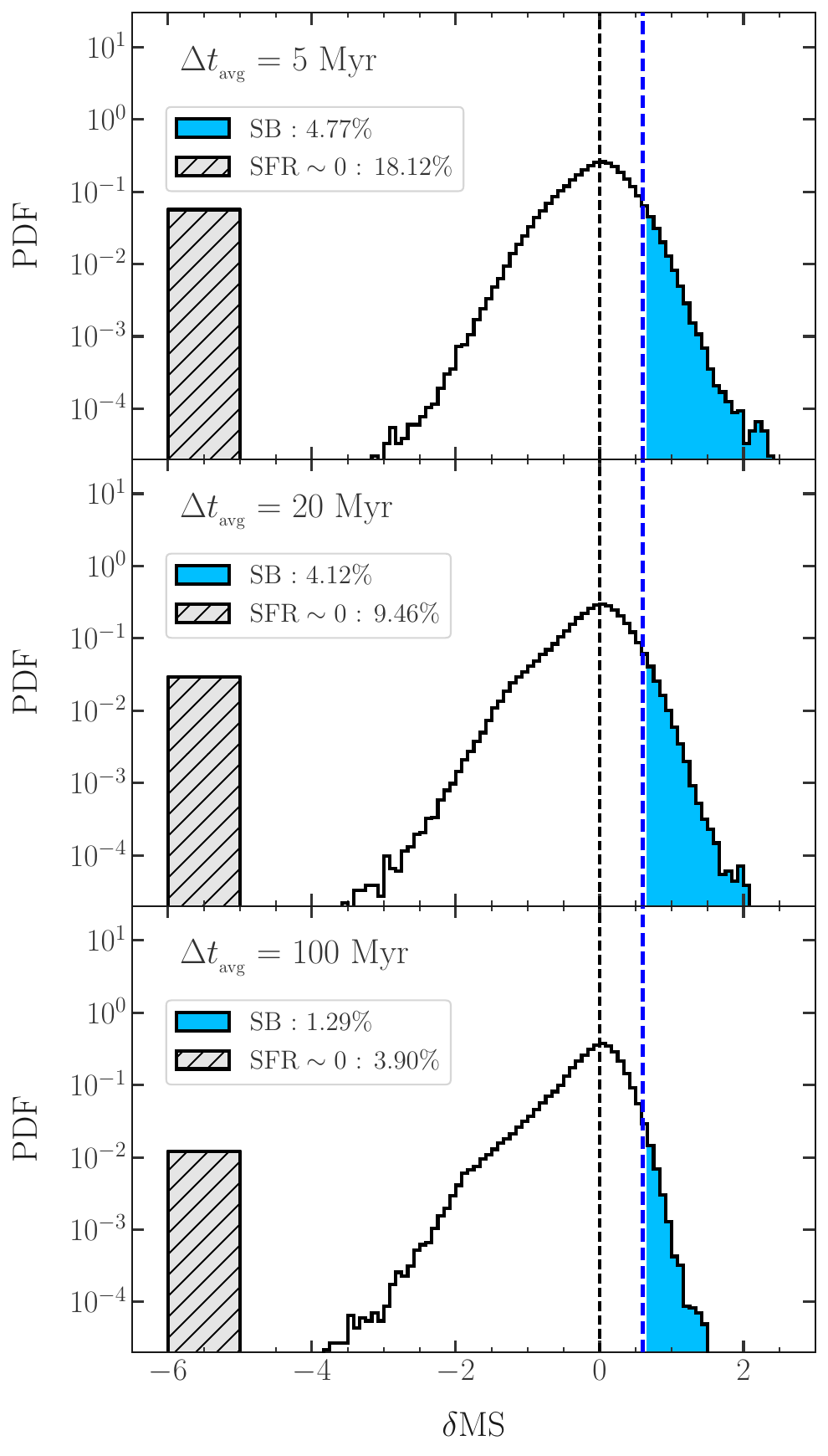}
    \caption{Probability distribution functions (PDFs) for the offsets $\dMS$ from the $\SFS$, for different choices of $\tavg$, for all galaxies in FIREbox with $M_\star\geq10^8\Msun$ and at $z=0-1$. The blue, shaded areas represents the SB contribution (i.e., galaxies with $\dMS\geq 0.6~\mathrm{dex}$), that decreases as $\tavg$ becomes longer. The grey area represents the fraction of non star-forming galaxies with $\dMS\leq-5$.}
    \label{fig:dMS}
\end{figure}

Figure~\ref{fig:dMS} shows the the probability distribution function (PDF) for $\dMS$, for different choices of $\tavg$. The area of the grey bars equals the fraction of non-star-forming galaxies, defined as having $\dMS\leq-5$. The blue-shaded area corresponds to SBs, i.e., those galaxies with $\dMS\geq 0.6~\mathrm{dex}$ (blue, dashed vertical line). The SB population contributes $\lesssim 5\percent$ to the total galaxy population with $M_\star\geq 10^8\Msun$ between $z=0-1$ for $\tavg=5\Myr$. The fraction of SB galaxies decreases with increasing $\tavg$ to $\sim 4$ and $\sim 1.3\percent$ for $\tavg =20\Myr$ and $100\Myr$, respectively. Longer $\tavg$ are less sensitive to bursts of star formation on short timescales, thus leading to a smaller fraction of galaxies identified as SB. The fraction of SBs over the total number of galaxies with $M_\star\geq 10^8\Msun$ and $z=0-1$ is $\sim5, 4, 1\percent$, for $\tavg=5, 20, 100\Myr$, respectively.

\begin{figure}
    \centering
	\includegraphics[width=\columnwidth]{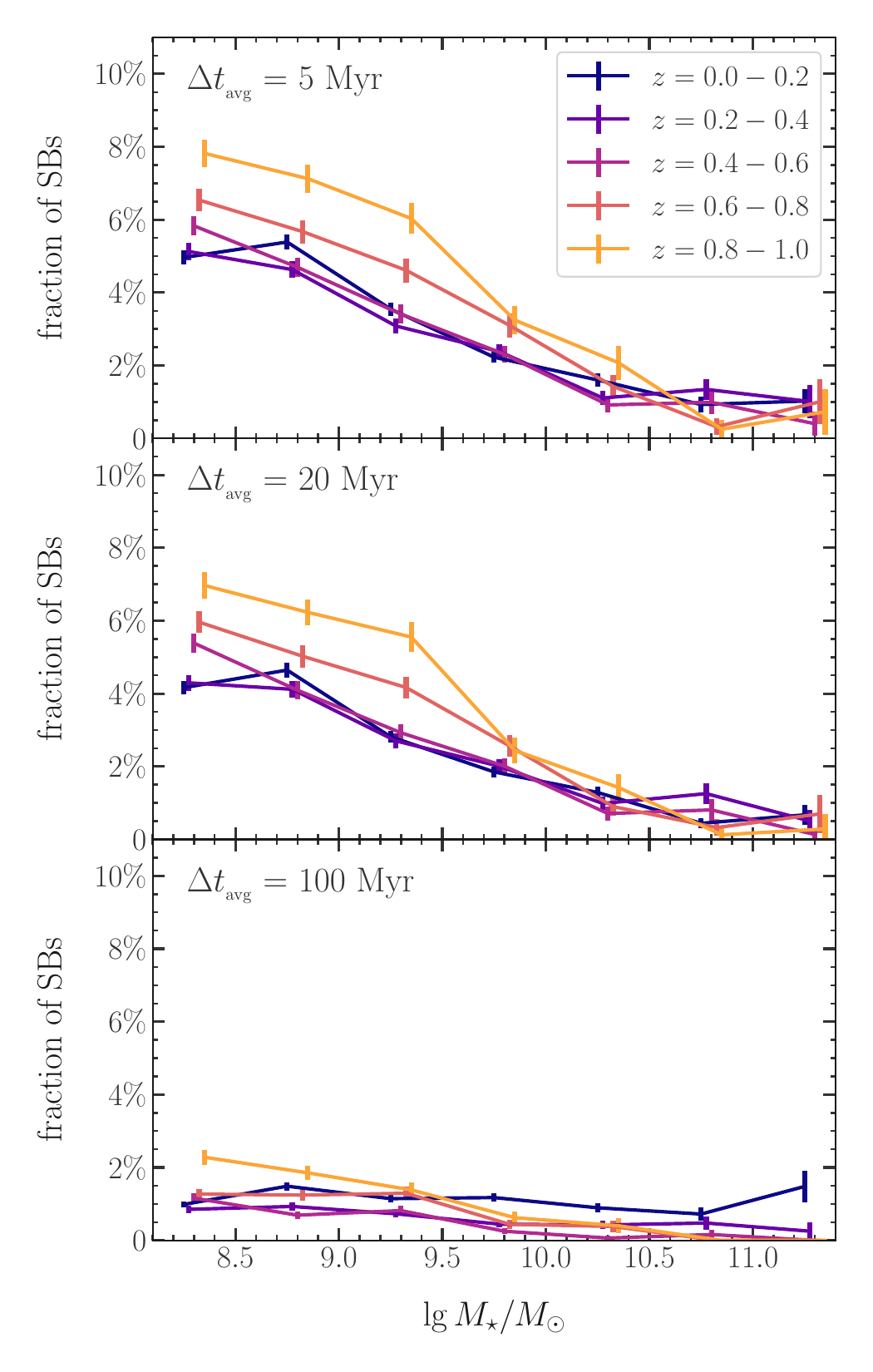}
    \caption{Fraction of SB galaxies as a function of stellar mass, colour-coded by redshift, with error bars representing the $2\sigma$ bootstrapped errors, for different $\tavg$. The SB fraction increases with decreasing stellar mass for $M_\star\lesssim 10^{10}\Msun$, and displays a moderate redshift evolution with more SB galaxies at high redshift. For $\tavg=100\Myr$, the SB fraction is generally below $2\percent$ in our sample of FIREbox galaxies.}
    \label{fig:SB_fraction}
\end{figure}

Figure~\ref{fig:SB_fraction} shows the fractional contribution of SBs to the total number of galaxies per stellar mass bin, colour-coded by redshift bin. Different panels correspond to different choices of $\tavg$, from $5\Myr$ (top panel) to $100\Myr$ (bottom panel). The SB fraction generally decreases with increasing $M_\star$, for $\tavg=5,20\Myr$, whereas it remains at a constant value of $\lesssim 2\percent$ independent of stellar mass for $\tavg=100\Myr$. The SB fraction shows a moderate redshift dependence, especially in the low-mass regime, with more SBs at higher redshift (note that our SB definition does depend on redshift, accounting for the increased normalisation of the $\SFS$ at higher redshifts). This is in agreement with the observed higher gas content and $\SFR$s of higher redshift galaxies \citep[see e.g.,][and references therein]{Tacconi2020}. For $M_\star\gtrsim 10^{10}\Msun$, the SB fraction remains constant at $\lesssim 2\percent$, independent of redshift.

\begin{figure}
	\includegraphics[width=\columnwidth]{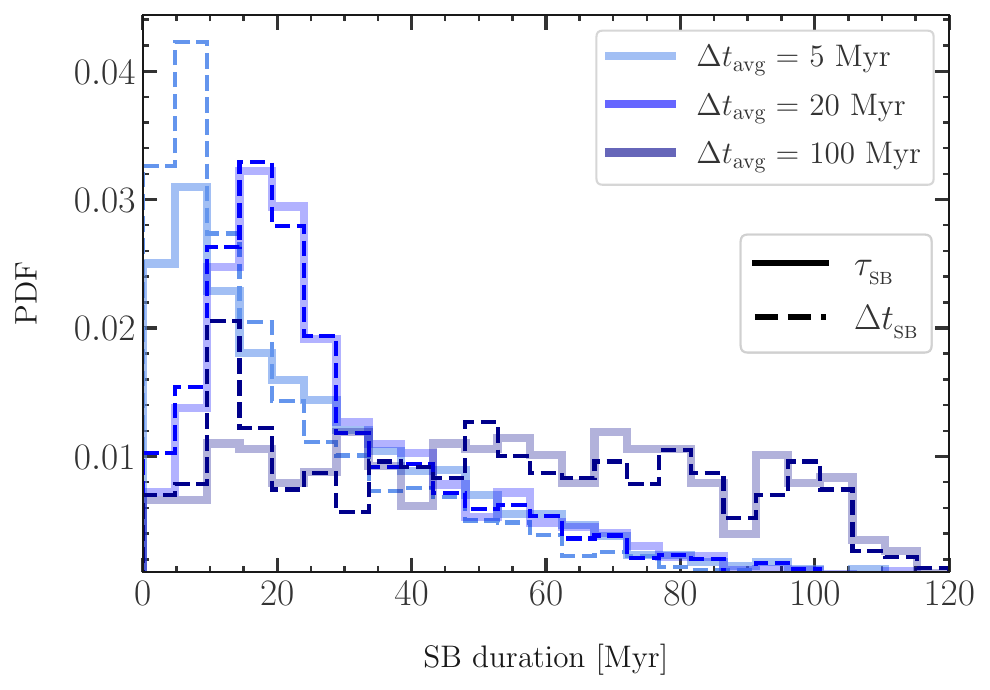}
    \caption{Probability distribution functions (PDFs) for the SB duration, as measured by either $\tau_\SB$, i.e., the duration of the SB-phase (solid lines), or $\Delta t_\SB$, i.e., the time the galaxy spends with $\dMS\geq 0.6~\mathrm{dex}$ during the SB-phase (dashed lines; see Section~\ref{sec:methods_durations} for more details on these definitions), colour-coded by $\tavg$. The two different SB duration definitions result in very similar PDFs for the SB duration, for all $\tavg$. The PDFs for $\tavg = 5\Myr$ and $20\Myr$ are strongly peaked at durations of approximately the averaging time scale, before declining sharply, while the PDF for $\tavg = 100\Myr$ is approximately flat.}
    \label{fig:SB_duration_pdf}
\end{figure}

Figure~\ref{fig:SB_duration_pdf} shows the PDFs for the SB duration, colour-coded by $\tavg$. We show the distributions for both the measures for the SB duration, $\tau_\SB$ and $\Delta t_\SB$, defined in Section~\ref{sec:methods_durations} (see Figure~\ref{fig:SB_duration_cartoon}). The PDFs for $\tavg =5,20\Myr$ peak at SB durations that are approximately the employed averaging time scale, and sharply decline towards longer durations. The PDF for $\tavg =100\Myr$ is essentially flat,  with a similar contribution from $\sim 80\Myr$ long SBs and those lasting for only $\sim 5\Myr$. The distributions for $\tau_\SB$ and $\Delta t_\SB$ are qualitatively similar. By definition, $\Delta t_\SB\leq\tau_\SB$, implying that the distribution for $\Delta t_\SB$ have a larger contribution from short SB durations, especially for $\tavg=5\Myr$. As we move to a longer $\tavg$, some short-duration SBs might be `washed-out' when averaging the $\SFR$ over long timescales. In general, a longer $\tavg$ is biased toward longer SBs. We expect that the duration of a SB is positively correlated with the star-forming gas availability and the duration of the driving-mechanism. For instance, tidal features arising from torques during galaxy interactions could persist up to about $1\Gyr$ \citep[see, e.g.,][]{Lotz2008}, implying that interactions might affect the galaxy dynamics over relatively long time-scales. We investigate this aspect of interactions in Section~\ref{sec:results_interactions} (see Figure~\ref{fig:INT_fraction_2}).

\subsection{Molecular gas masses and depletion times}\label{sec:results_tdepl}
In this section, we address the question whether it is the star formation efficiency (or, equivalently, the gas depletion time) or the total gas content of a galaxy that characterise SB galaxies when compared to control galaxies. Here, we only consider the sample with $\tavg =20\Myr$, but we find that choosing $\tavg=5, 20, 100\Myr$ does not qualitatively change our results. We use $\tavg =20\Myr$ as fiducial averaging time because 
(\textit{i}) we want to exclude galaxies that briefly enter the SB regime due to stochastic noise in their bursty star formation histories; (\textit{ii}) it is generally of the order of the lifetime of typical massive stars with significant ionising flux.
 
\begin{figure}
	\includegraphics[width=\columnwidth]{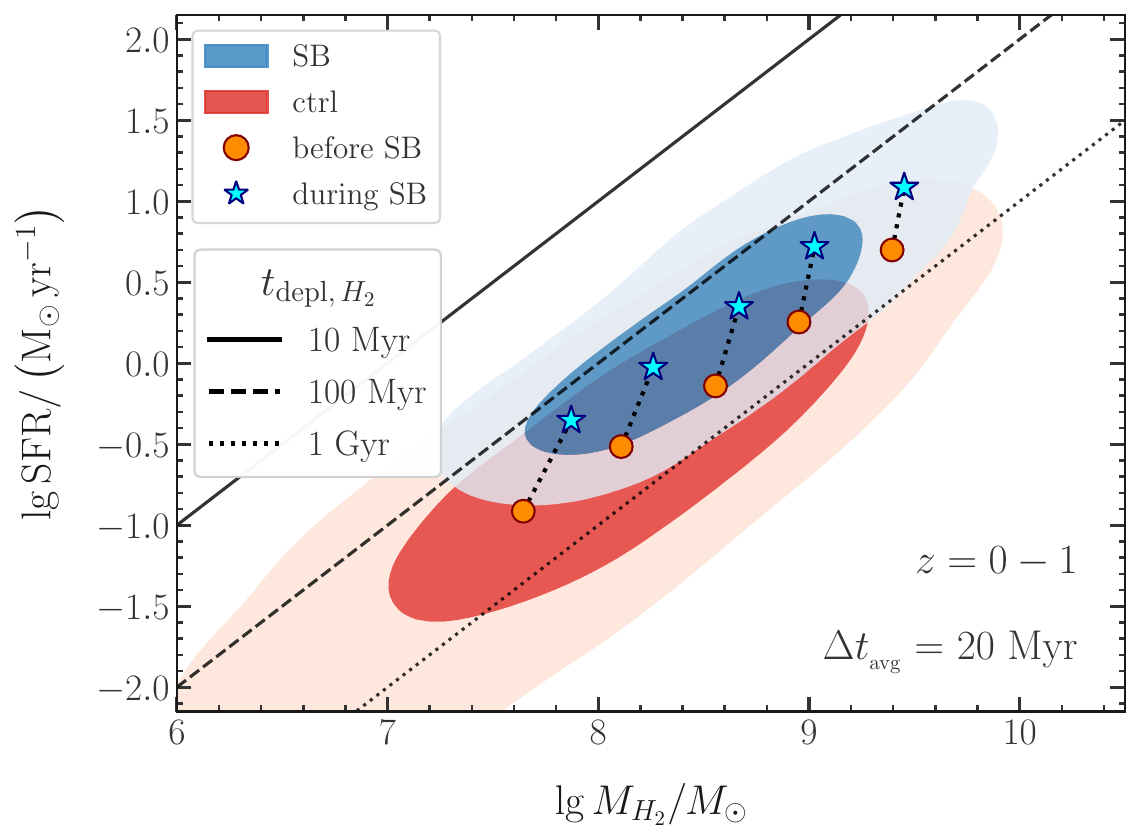}
    \caption{Distributions (1$\sigma$ and 2$\sigma$ contours) of the SB (blue contours) and control (red contours) samples in the $\left(\lg M_\mol, \lg \SFR\right)$-plane, for $\tavg=20\Myr$, i.e., our fiducial sample. The black lines show loci of constant molecular depletion time ($\tdepl =10\Myr, 100\Myr, 1\Gyr$). Orange circles and cyan stars show the median $M_\mol$ and $\SFR$, respectively, $50\Myr$ prior to the SB and during the SB event, for stellar mass bins of width 0.5 dex in the range $\lg M_\star/\Msun=8-10.5$ (left to right; the relevant stellar mass is the one at the beginning of the SB). The black, dotted lines connect to the cyan stars representing the median $M_\mol$ and $\SFR$ during the SB event. SB galaxies have systematically shorter depletion times and both SB and control galaxies extend to similarly large molecular gas masses, although SB galaxies have a significantly larger lower limit. The transition to the SB regime is mainly determined by a shortening in $\tdepl$, but it is also accompanied by an increase in $M_\mol$, especially in low-mass galaxies.}
    \label{fig:SFR_vs_MH2}
\end{figure}

Figure~\ref{fig:SFR_vs_MH2} shows the 1$\sigma$ and 2$\sigma$ contours of the 2D-distributions of all SB and control galaxies in the $\left(\lg M_\mol, \lg \SFR\right)$-plane, for $\tavg =20\Myr$. Both samples show a close to linear correlation between $\SFR$ and $M_\mol$, implying approximately constant molecular depletion times $\tdepl\equiv M_\mol/\SFR$. SB galaxies have $\tdepl\sim 100\Myr$, whereas control galaxies exhibit longer $\tdepl\sim 1\Gyr$, implying that SB galaxies are more efficient at transforming their gas into stars, at fixed $M_\mol$, in agreement with observations \citep[e.g.,][]{Feldmann2020}. SB galaxies occupy a narrower range of $\mathrm{H}_2$ masses ($\sim 10^{7.5}-10^{9.5}\Msun$) compared to the control sample ($\sim 10^{6.5}-10^{9.5}\Msun$), although both samples extend to similarly large molecular gas masses $M_\mol\gtrsim 2-3\times 10^9\Msun$ (considering the 1$\sigma$ contours). This implies that a large $M_\mol$ is a necessary but not sufficient condition for having a SB. We show the median value of $M_\mol$ and $\SFR$ at $50\Myr$ prior to the beginning of the SB (orange circles), connected (dotted lines) to the median values during the SB event (cyan stars). From left to right, the different points correspond to increasing 0.5 dex stellar mass (at the beginning of the SB) bins in the range $\lg M_\star/\Msun=8-10.5$. While star-forming galaxies transition to the SB regime by increasing both their molecular star formation efficiency (i.e., shortening their molecular depletion time) and total molecular gas content, the decrease in $\tdepl$ appears overall to be the dominant driver for the enhanced $\SFR$ in SB galaxies.

\subsection{Fraction of high-density gas}\label{sec:results_fSF}
Equation~\eqref{eqn:SFR_gal} and \eqref{eqn:tdepl} show that the main quantities governing the $\SFR$ and depletion time of a galaxies are: the total gas mass, $M_{\rm gas}$; the fraction of gas mass $f_\SF$ that is eligible for star formation in our model; the average fraction of molecular-in-neutral gas divided by the local free-fall time for particles that are eligible for star formation, $\avgSF$; the fraction of molecular gas in the galaxy, $\fmol\equiv M_\mol/M_{\rm gas}$. In the following, we explore how these quantities differ in our SB galaxies with respect to control galaxies, and which ones are a good predictor for whether a galaxy is a SB.

\begin{figure*}
    \centering
	\includegraphics[width=\textwidth]{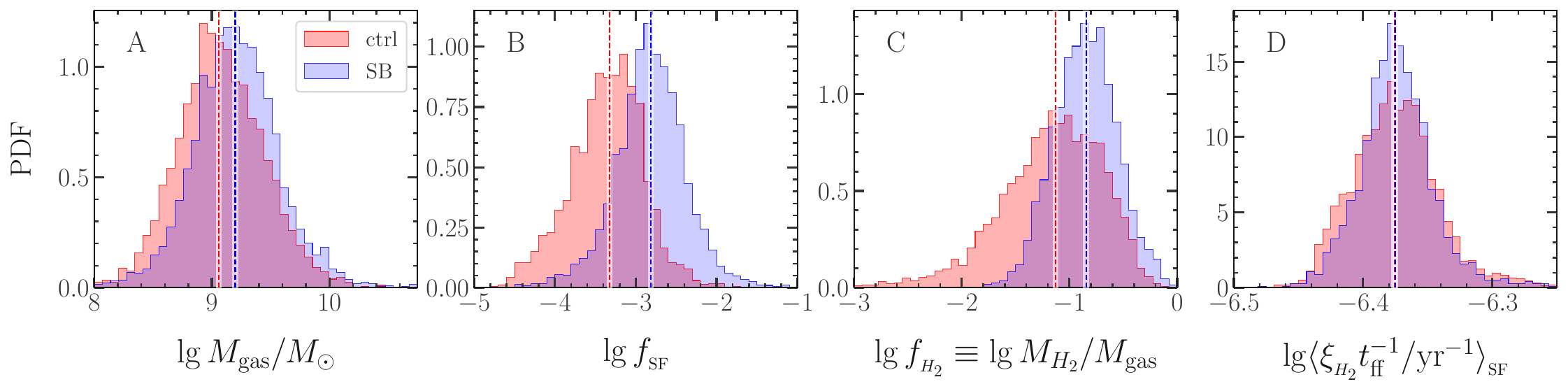}
    \caption{Comparing the probability distribution functions (PDFs) of the SB (blue) and control (red) galaxies with $\tavg=20\Myr$ for the quantities contributing to the $\SFR$ according to equation~\eqref{eqn:SFR_gal}. Specifically, we show the PDFs for the total gas mass ($M_{\rm gas}$; panel A), the fraction of (dense and self-gravitating) gas eligible for star formation ($f_\SF$; panel B), the molecular gas fraction ($\fmol$; panel C), and the quantity $\avgSF$ (panel D) that reflects the star formation prescription at a particle level (see Section~\ref{sec:methods_definitions} for further details). The median of each distribution is indicated by the vertical dashed line of the same colour. $M_{\rm gas}$, $\fmol$, and $f_\SF$ are all larger in SB galaxies than in control galaxies, with the difference in $f_\SF$ ($\sim 0.5-0.6$ dex) being the largest. $\avgSF$ spans a narrow range of values of $\sim 0.2$ dex, with little difference between SB and control galaxies.}
    \label{fig:SB_vs_ctrl}
\end{figure*}

In Figure~\ref{fig:SB_vs_ctrl}, we compare the SB and control samples for $\tavg=20\Myr$, by displaying the PDFs of the main quantities related to a difference in $\SFR$, i.e., those highlighted by equations~\eqref{eqn:SFR_gal} and ~\eqref{eqn:tdepl}. Panel A shows the PDF for the total gas mass, $M_{\rm gas}$. SB galaxies have slightly larger gas masses, with a difference of $\sim 0.15$ dex in their median with respect to the one of control galaxies. Panel B shows the fraction gas that is eligible for star formation according to our model, $f_\SF$. This fraction is significantly larger in SB galaxies by $\sim 0.5$ dex. Panel C shows the molecular gas fraction, $\fmol$. Both the median and the shape of the PDF for $\fmol$ differ between the SB and control sample. The $\fmol$ distribution for control galaxies is much broader, ranging from $\sim 10^{-3}-1$, with a median value of $\sim 10^{-1.2}-10^{-1.1}$. By contrast, SB galaxies display a narrower $\fmol$ distribution, ranging from $\sim 10^{-1.8}-1$, and with a median at $\sim 10^{-0.9}-10^{-0.8}$. Thus, the median fraction of molecular gas is larger by $\sim 0.3$ dex in SB galaxies with respect to that in control galaxies. Again, a high fraction of molecular gas appears to be a necessary, but not sufficient SB criterion.  Finally, panel D shows the distributions of $\avgSF$, that span a narrow range of values of $\sim 0.2$ dex, with little difference between SB and control galaxies. The choice of $\tavg$ does not qualitatively affect any of the aforementioned results. Furthermore, binning galaxies based on either stellar mass or redshift would not affect our conclusions.

The  ratio of total gas masses, $M_{\rm gas}$, of SB and control galaxies is a factor of 2-3 smaller than their ratio of $\fmol$ and $f_\SF$, making $M_{\rm gas}$ a poor predictor to distinguish SB galaxies from control galaxies. Star-forming particles with densities above the density threshold for star formation have similar free-fall times. Furthermore, we expect $\fmol\sim 1$ for all star-forming particles. Therefore, it is expected that $\avgSF$ is not significantly different in SB galaxies, as shown in the Figure. We conclude that the main quantities characterising SB galaxies are $f_\SF$ and $\fmol$. These results highlight that the molecular gas content alone is not sufficient for driving a SB. We therefore seek for a mechanism that can increase the fraction of both molecular and high-density gas in the galaxy, at approximately fixed total gas mass.


\begin{figure}
    \centering
	\includegraphics[width=\columnwidth]{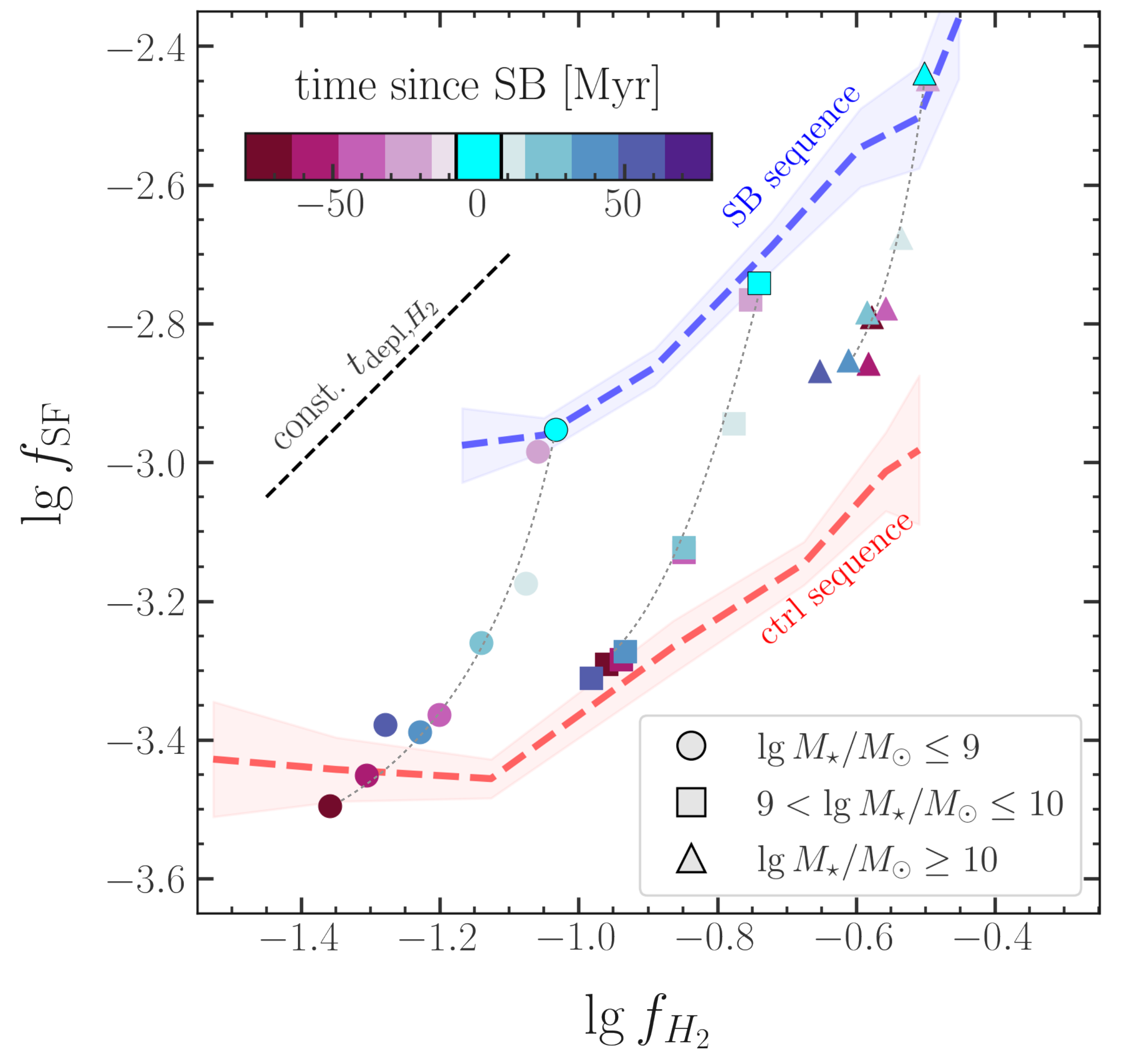}
    \caption{Median evolution of the mass fraction of molecular gas ($\fmol$) and the fraction of gas mass eligible for star formation in our model ($f_\SF$) in SB galaxies, for a $\sim 150\Myr$ time window centred around the time when the galaxies are identified as a SB, colour-coded by the time difference with respect to the beginning of the SB event. The blue and red dashed curves represent the median trends for SB and control galaxies, respectively, parameterised by their $M_\star$ (i.e., connecting the median values in this plane for SB and control galaxies in different stellar mass bins), with $2\sigma$ bootstrapped error (shaded area). The different marker styles refer to different stellar mass bins (see legend). Light grey, dashed tracks show the approximate trends. Galaxies move towards the SB sequence by increasing both $\fmol$ and, disproportionately, $f_\SF$, i.e., by reducing their $\tdepl$. Following the SB event, galaxies move towards the control sequence along the same tracks, reducing both $\fmol$ and $f_\SF$.}
    \label{fig:fSF_vs_fH2}
\end{figure}

Figure~\ref{fig:fSF_vs_fH2}, shows the median path of SB galaxies for a time period of $\sim 150\Myr$ centred around the time when the galaxies are identified as a SB, in the $\left(f_\SF, \fmol\right)$-plane. The blue and red dashed curves represent the median trends for SB and control galaxies, respectively, parameterised by their $M_\star$, i.e., connecting the median values for SB and control galaxies in this plane in different mass bins. In general, galaxies evolve from the control sequence to the SB sequence by increasing both their $f_\SF$ and $\fmol$. The net increase in $f_\SF$ ($\sim 0.4-0.6$ dex) and $\fmol$ ($\sim 0.2-0.4$ dex) over the time period prior to the SB depends on the mass bin considered. However, the increase in $f_\SF$ is larger than the one in $\fmol$ by $\sim 0.3$ dex, independent of the mass bin. From equation~\eqref{eqn:tdepl}, we know that the molecular depletion time
scales with the ratio of $f_\SF$ to $\fmol$, assuming $\avgSF\sim 3-5\times 10^{-6}$ for all galaxies. Therefore, a larger change in $f_\SF$ compared to $\fmol$ implies that SB galaxies have on average shorter depletion times with respect to control galaxies. At first ($\sim 60-40\Myr$ prior to the SB), low and intermediate-mass galaxies ($M_\star\leq 10^{10}\Msun$) proceed by increasing $f_\SF$ and $\fmol$ at a similar rate, retaining a nearly constant $\tdepl$, while increasing their $M_\mol$. Approaching the SB event, the increase in $f_\SF$ becomes steeper than the increase in $\fmol$. Hence, galaxies with $M_\star\leq 10^{10}\Msun$ evolve towards significantly shorter $\tdepl$ and large $M_\mol$. More massive galaxies exhibit a much steeper increase in $f_\SF$ with respect to $\fmol$, indicating that the decrease in $\tdepl$ dominates over the increase in $M_\mol$ in driving the enhanced $\SFR$. After the SB event, galaxies reduce both their $f_\SF$ and $\fmol$, following the same track (but reversed) that brought them to the SB sequence.

In the time period of $\gtrsim 70\Myr$ prior to the SB, massive ($M_\star\gtrsim 10^{10}\Msun$) SB galaxies are offset ($\sim 0.3$ dex) from the corresponding control sequence. This offset suggests that the most massive SB galaxies could be predominantly driven by a different mechanism compared to the lower mass galaxies. As we will demonstrate in Section~\ref{sec:results_interactions}, the main driver for SBs in the high-mass ($M_\star\gtrsim 10^{10}\Msun$) regime are interactions. Indeed, restricting the analysis to non-interacting SB galaxies (see Appendix~\ref{app:INT_split}) results in a similar trend as the ones observed for lower mass galaxies, i.e., departure from and then return to the control sequence over a period of $\sim 150\Myr$ encompassing the SB event, starting with only a $\lesssim 0.1$ dex offset from the control sequence. We therefore conclude that interactions in $M_\star\gtrsim 10^{10}\Msun$ galaxies can enhance $f_\SF$ prior to the $\sim 70\Myr$ before the SB, such that at $\sim 70\Myr$ prior to the SB $f_\SF$ is already, on average, $\sim 0.2-0.3$ dex higher than the median value for control galaxies.

Our model for star-formation requires gas to be self-gravitating, Jeans-unstable, self-shielding, and above a certain density threshold ($n=300~\rm{cm}^{-3}$ in FIREbox). As discussed earlier, in our model, dense-enough gas is also typically eligible for star formation. This implies that $f_\SF$ is closely related to the fraction of gas with densities above $n=300~\rm{cm}^{-3}$. Therefore, a shift in the gas density PDF towards higher densities would accordingly increase $f_\SF$. Furthermore, gas that is dense-enough to form stars in our model is essentially fully molecular. The emerging picture is that a shift in the gas density PDF towards higher densities can explain the increase in both $f_\SF$ and $\fmol$. In order to produce a SB, the molecular depletion time has to reduce significantly. Hence, since $\tdepl\propto \fmol/f_\SF$, the fraction of high-density gas must increase more than the fraction of molecular gas. The result is that \textit{SB galaxies have shorter molecular depletion times and larger molecular gas masses than control galaxies}.

\subsection{Gas compaction in the central kpc}\label{sec:results_1kpc}
\begin{figure*}
    \centering
	\includegraphics[width=\textwidth]{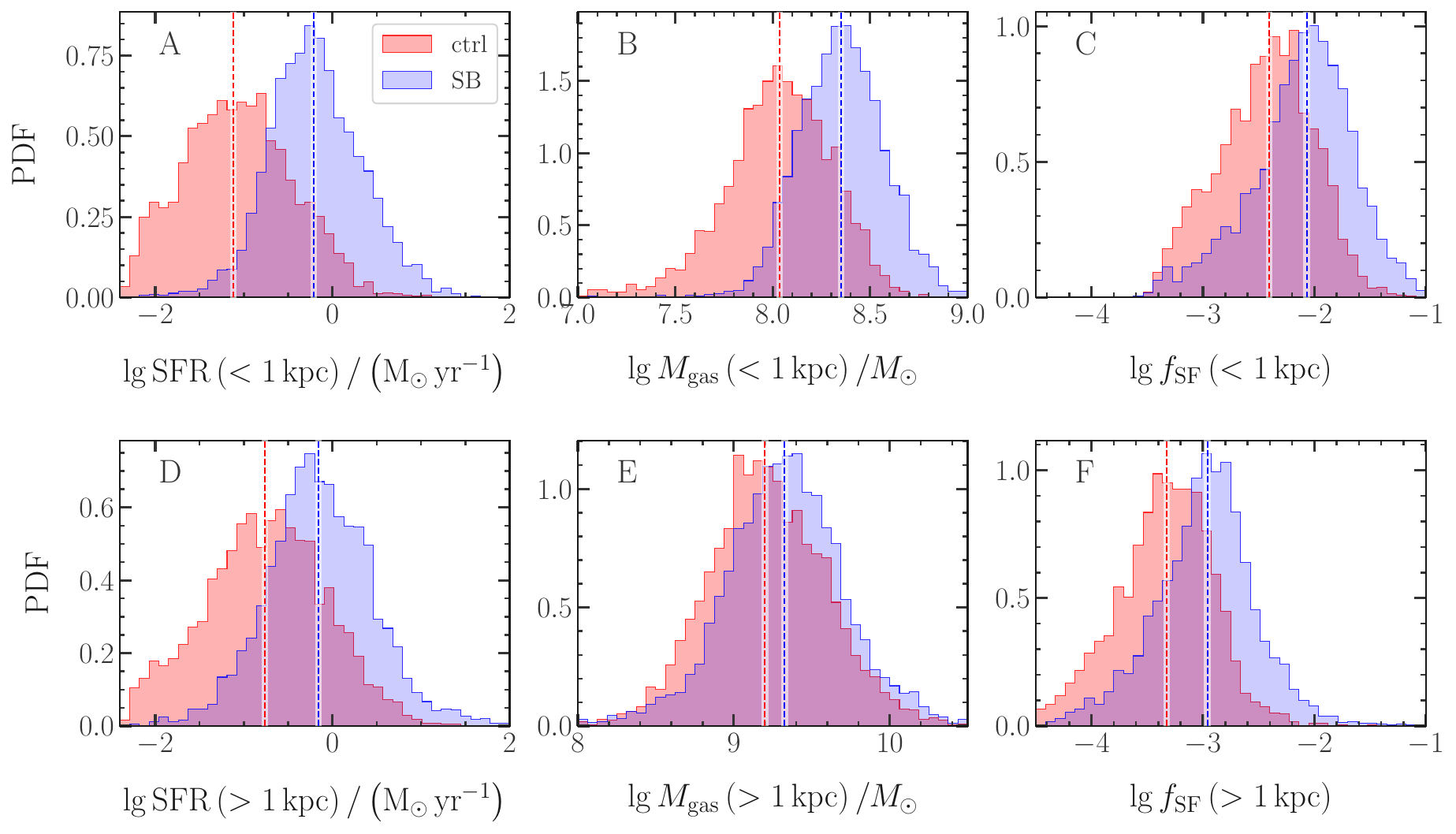}
    \caption{Comparison between the probability distribution functions (PDFs) of SB (blue) and control (red) galaxies for, from left to right, their $\SFR$, total gas mass ($M_{\rm gas}$), and fraction of total gas mass that is eligible for star formation in our model (i.e., high-density gas fraction, $f_\SF$), within and outside the central kpc (top and bottom panels, respectively). We only show the results of the fiducial sample with $\tavg=20\Myr$. The vertical, dashed lines refer to the median value of the distributions (colour-coded accordingly). Overall, SB galaxies have higher $\SFR$s at all radii, with a larger median difference with respect to control galaxies when considering the central kpc ($\sim 0.8$ dex, compared to the $\sim 0.6$ dex difference outside the central kpc). The total gas mass within the central kpc is $\sim 0.3$ dex higher in SB galaxies, compared to control galaxies, compared to the $\sim 0.1$ dex difference outside the central kpc. The fraction of high-density gas in SB galaxies is larger than in control galaxies by $\sim 0.4$ dex, both within and outside the central kpc.}
    \label{fig:SB_vs_ctrl_1kpc}
\end{figure*}

To identify the best predictor for classifying galaxies as SB/control we performed a Random Forest classification \citep[][]{Ho1995}{}{}. The features we consider for each galaxy are the gas masses (different phases), stellar masses, angular momenta, and velocity dispersion computed within the galaxy (assumed to be within $0.1\,R_{\rm vir}$) and few fixed apertures of 1, 3, 5, and 10 kpc. We find that the gas (all phases) within the central kpc is the best predictor of whether a galaxy is starbursting. Gas compaction events increase the gas density in the centres of galaxies by increasing the mass in the central regions without significantly changing the total gas mass in the galaxy, leading to a reservoir of dense, molecular gas. We therefore explore the properties of the central region of our SB and control galaxies. In the following, we only show the results relative to the fiducial sample with $\tavg=20\Myr$ since our conclusions do not qualitatively change for $\tavg=5,20,100\Myr$. Figure~\ref{fig:SB_vs_ctrl_1kpc} shows a comparison between the distributions of $\SFR$ and gas masses within the central kpc of SB and control galaxies. We show the $\SFR$ averaged over $\tavg=20\Myr$, the total gas mass ($M_{\rm gas}$), and the fraction of total gas mass that is eligible for star formation in our model (i.e., high-density gas, $f_\SF$), computed within and outside the central kpc. In general, SB galaxies exhibit an enhanced $\SFR$ both within and outside their central regions, with a median difference of $\sim 0.8$ dex and $\sim 0.6$ dex, respectively, compared to control galaxies. In SB galaxies, the $\SFR$ within the central kpc is comparable to the $\SFR$ in their outer regions, whereas control galaxies display lower values of $\SFR$ in their centres compared to their outer parts. Binning galaxies based on either stellar mass or redshifts does not affect our conclusions.

The median $M_{\rm gas}$ of SB galaxies in their central regions is higher by $\sim 0.3$ dex with respect to control galaxies. However, $M_{\rm gas}$ outside the central kpc is comparable in SB and control galaxies, within $\sim 0.1$ dex.\\
The median fraction of high-density gas is $\sim 0.4$ dex higher in SB galaxies than in control galaxies, both within and outside their central kpc. Therefore, an increase in $M_{\rm gas}$ in the central kpc is required, together with the increased fraction of high-density gas, to explain the higher median $\SFR$ in the central regions of SB galaxies compared to control galaxies. On the other hand, the larger $\SFR$ outside the central kpc is mostly explained by the enhanced fraction of high-density gas. We conclude that SB galaxies in FIREbox have more concentrated gas reservoirs than control galaxies, supporting the hypothesis of nuclear SB events \citep[e.g.,][]{Barnes&Hernquist1991,Wilkinson2018,Ellison2018}.

\begin{figure}
    \centering
	\includegraphics[width=\columnwidth]{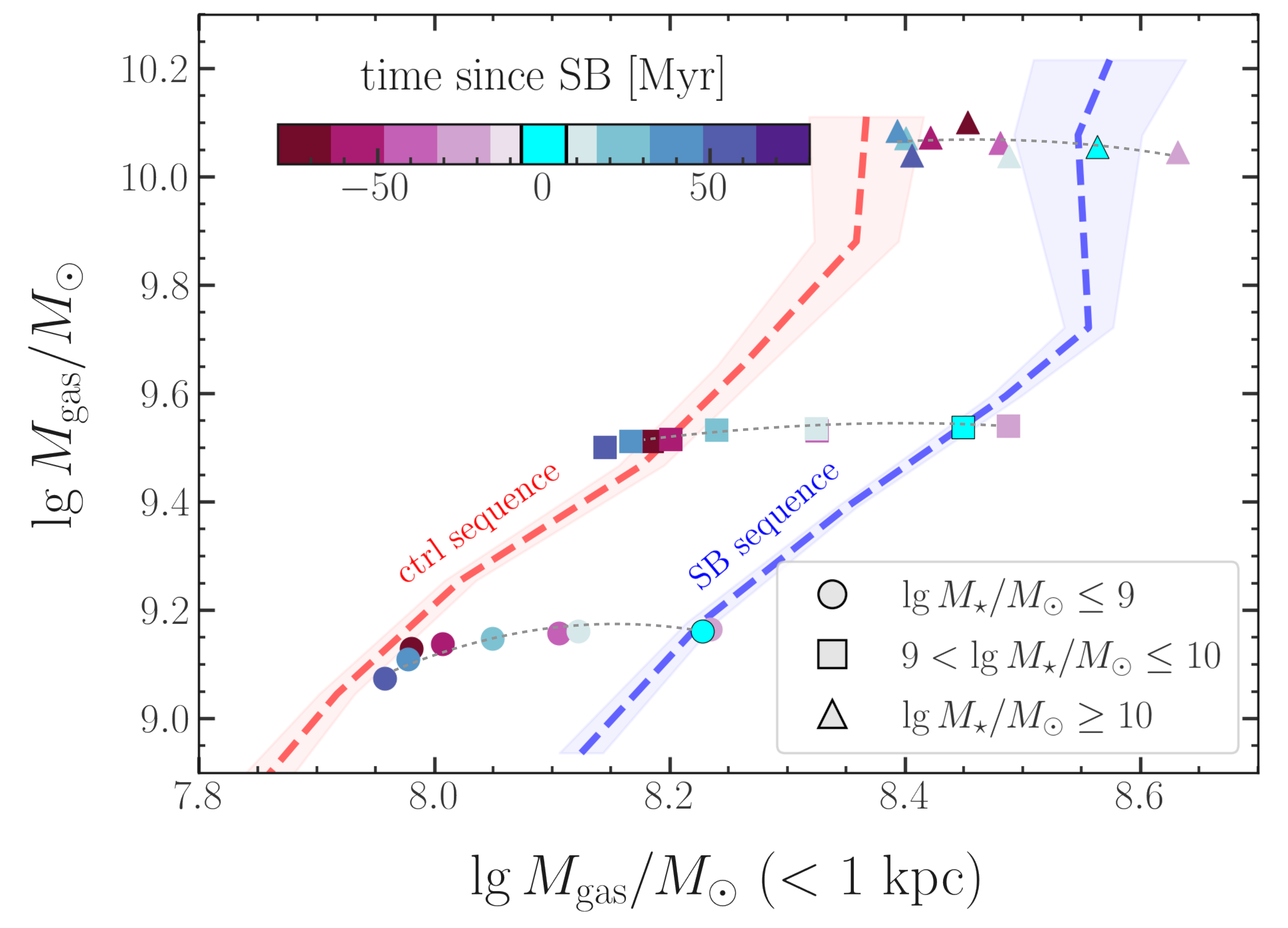}
    \caption{Median evolution of the total gas mass ($M_{\rm gas}$) and gas mass within the central kpc ($M_{\rm gas}\left(<1\kpc\right)$) of SB galaxies, in a period of time $\sim 150\Myr$ centred around the time when the galaxies are identified as a SB, colour-coded by the time difference with respect to the beginning of the SB event. The blue and red dashed curves represent the median trends for SB and control galaxies, respectively, parameterised by their $M_\star$ (i.e., connecting the median values in this plane for SB and control galaxies in different stellar mass bins), with $2\sigma$ bootstrapped error (shaded area). Light grey, dashed tracks show the approximate trends. Galaxies evolve from the sequence of control galaxies to the SB sequence by increasing their gas mass in the central kpc at a fixed total gas mass over $\sim 70\Myr$, suggesting that the SB event is driven by gas compaction in the central regions. After the SB event, they reduce $M_{\rm gas}\left(<1\kpc\right)$, returning to the control sequence in $\sim 70\Myr$.}
    \label{fig:Mgas_vs_Mgas_1kpc}
\end{figure}

Figure~\ref{fig:Mgas_vs_Mgas_1kpc} shows the median evolution of the total gas mass $M_{\rm gas}$ and the gas mass within the central kpc for SB galaxies in different stellar mass bins and identified with $\tavg=20\Myr$, in a period of time $\sim 150\Myr$ centred around the time when the galaxies are identified as a SB. All SB galaxies exhibit a similar trend, regardless of their $M_\star$: the total gas mass remains essentially constant, whereas the mass within the central kpc increases significantly ($\sim 0.3$ dex), moving galaxies from the control to the SB median sequence in this plane. We conclude that SB galaxies experienced a compaction event in the last $\sim 70\Myr$ (comparable to the galaxy free-fall timescale), enhancing the gas mass in their central regions, at fixed total $M_{\rm gas}$. After the SB event, galaxies move back to the former control sequence over $\sim 70\Myr$. Massive galaxies ($M_\star\gtrsim 10^{10}\Msun$) have slightly larger gas masses within their central kpc with respect to control galaxies in the period of time of $\pm 70\Myr$ from the SB event. Again, as in the previous Section~\ref{sec:results_fSF}, this effect can be explained by making a distinction between interacting and non-interacting SB galaxies (see the next Section~\ref{sec:results_interactions} and Appendix~\ref{app:INT_split}). In the $\sim 70\Myr$ prior to the SB, non-interacting SB galaxies have gas masses within their central kpc that are close to the median value for control galaxies.

\begin{figure}
    \centering
	\includegraphics[width=\columnwidth]{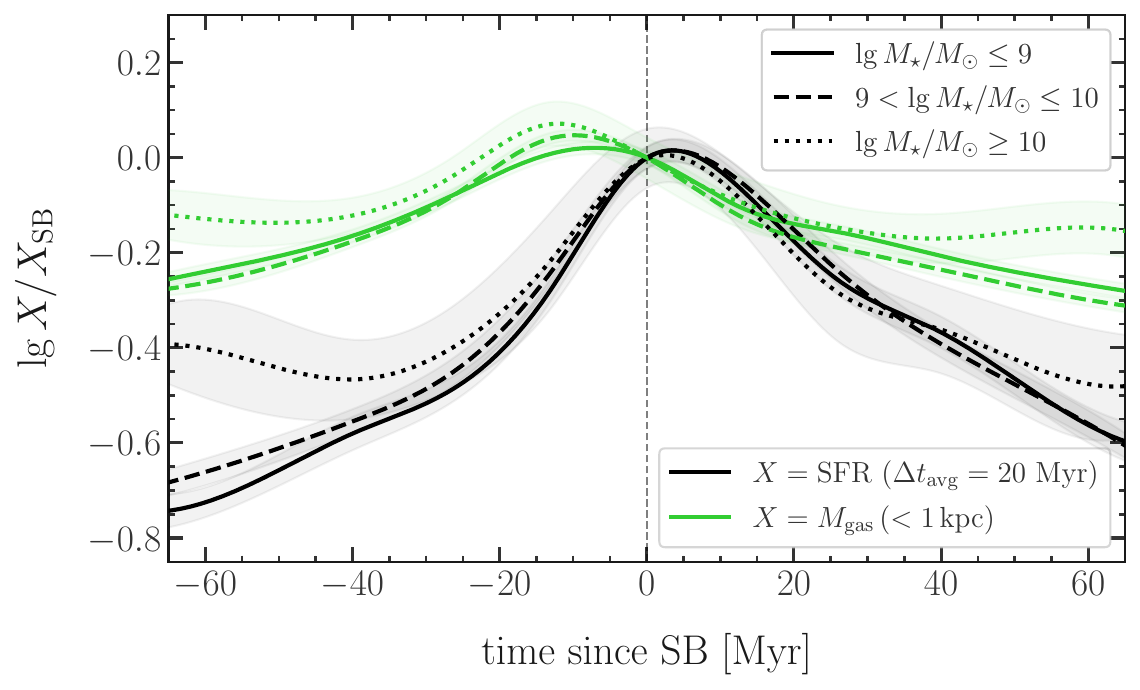}
    \caption{Median evolution of the $\SFR$ (black lines; for the fiducial averaging time of $20\Myr$) and gas mass within the central kpc (green lines; $M_{\rm gas}\left(<1\kpc\right)$) in all galaxies in the SB sample, normalised to their value at the beginning of the SB event, in a period of time of about $70\Myr$ prior to the beginning of the SB to about $70\Myr$ after the beginning of the SB. Different line styles refer to different stellar mass bins. The shaded bands represent the $2\sigma$ bootstrapped errors. Both the $\SFR$ and gas mass within the central kpc increase in the $\sim 70\Myr$ prior to the SB (see also Figure~\ref{fig:Mgas_vs_Mgas_1kpc}). The gas mass within the central kpc reaches its maximum approximately $20\Myr$ before the $\SFR$ does, suggesting that the gas compaction is indeed driving the SB across the entire mass range we considered.}
    \label{fig:median_evolution_SFR_and_Mgas_1kpc}
\end{figure}

Figure~\ref{fig:median_evolution_SFR_and_Mgas_1kpc} shows the median evolution of the $\SFR$ and the gas mass within the central kpc in SB galaxies, with respect to their values at the beginning of the SB event, from about $70\Myr$ prior to until about $70\Myr$ after the beginning of the SB. The the gas mass within the central kpc increases together with the $\SFR$, reaching its maximum about $20\Myr$ before the $\SFR$ does and before the beginning of the SB event, for all of the considered stellar masses. This suggests that the gas compaction event is driving the SB across the entire mass range we considered, rather than just correlate with the increase in the $\SFR$. Massive galaxies display a milder median enhancement in their $\SFR$ and gas mass within their central kpc, implying that they have pre-enhanced gas reservoirs and star formation activity.

\subsection{The role of galaxy interactions}\label{sec:results_interactions}
In this section, we study the connection between galaxy interactions and SBs, to understand in which regimes interactions can be considered the main SB-driving mechanism. By collecting the neighbouring galaxies of all our SB and control galaxies, we classified them as interacting/non-interacting (see Section~\ref{sec:methods_interactions} for details).\\

\begin{figure*}
    \centering
	\includegraphics[width=0.8\textwidth]{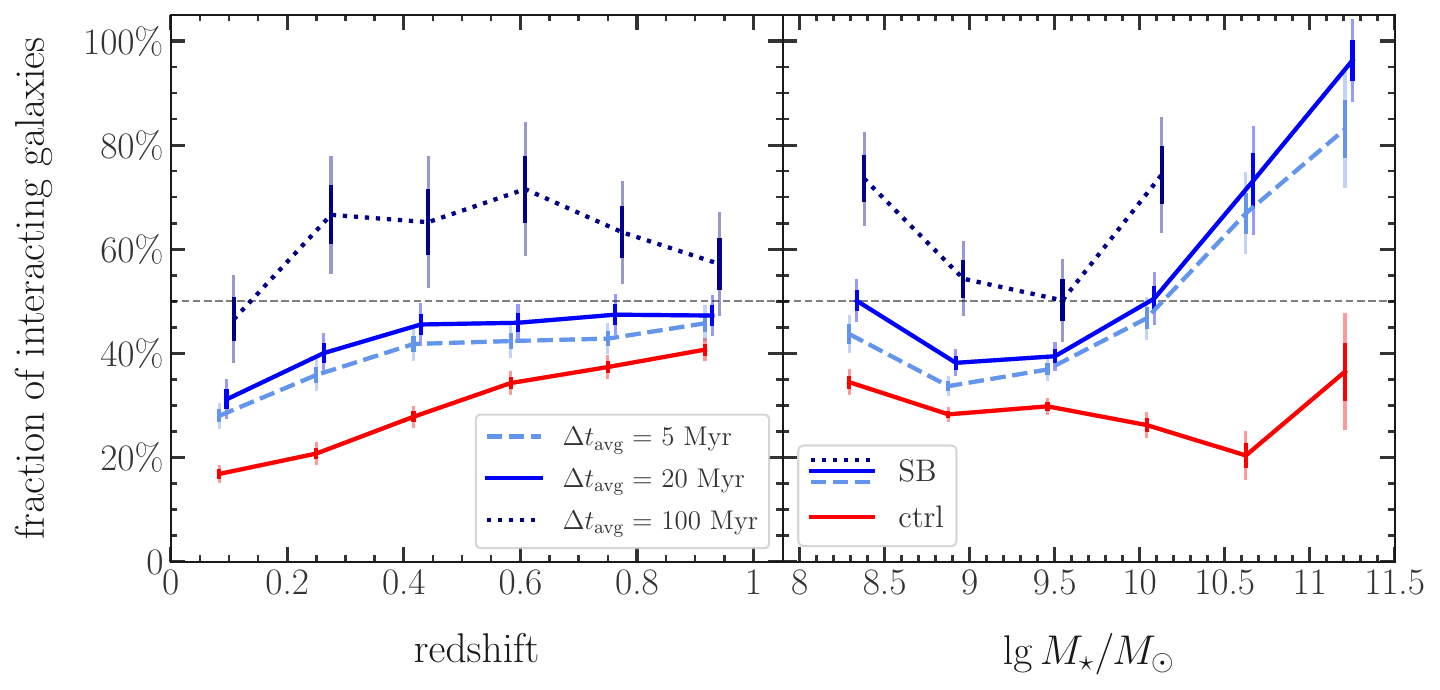}
    \caption{Fraction of interacting SB (blue curves) and control (red curve) galaxies as a function of redshift (left) and stellar mass (right), with $1\sigma$ and $2\sigma$ (increased transparency) bootstrapped error bars. Different curves for the SB sample refer to different choices for $\tavg$. For control galaxies, we show a single curve including galaxies for all different $\tavg$ since they do not depend on $\tavg$. High-mass and low-redshift SB galaxies are significantly more likely to be interacting than control galaxies, although, for $\tavg=5,20\Myr$, at low-masses and high-redshifts they exhibit a similar fraction of interacting systems as controls. For $\tavg=100\Myr$, the fraction of interacting SB galaxies is systematically higher than in control galaxies (and SB galaxies selected with $\tavg=5,20\Myr$).}
    \label{fig:INT_fraction_1}
\end{figure*}

Figure~\ref{fig:INT_fraction_1} shows the fraction of interacting galaxies as a function of redshift and stellar mass, among SB and control galaxies, for different $\tavg$. Control galaxies are shown as a single, cumulative curve including all different $\tavg$ since the interacting fraction is effectively independent of the choice of $\tavg$. In general, at any redshift and stellar mass, SB galaxies are more likely to be interacting than control galaxies. The differences between SB and control galaxies are the largest at low-redshift and for massive ($M_\star\gtrsim 10^{10}\Msun$) galaxies.\\
The behaviour of the fraction of interacting SB galaxies for $\tavg=5,20\Myr$ is consistent with that for control galaxies. For these averaging times, the fraction of interacting SB galaxies is similar to the one for control galaxies at high-redshift ($\sim 40-50\percent$) and slowly decreases to $\sim 30\percent$ moving to low-redshift, with a slower pace with respect to control galaxies ($\lesssim 20\percent$ at $z=0$). For $\tavg=100\Myr$, the fraction of interacting SB galaxies is systematically higher by $\sim20\percent$ than in the samples with a shorter $\tavg$, independent of redshift and stellar mass. Therefore, we conclude that observational tracers that are sensitive to $\SFR$ on time-scales of the order of $\sim 100\Myr$ will preferentially select SBs driven by interactions and thus overpredict the fraction of interacting SB galaxies.\\

The interacting fraction for control galaxies is almost constant ($\sim 30-40\percent$) over the whole mass range, whereas it strongly depends on stellar mass for SB galaxies, increasing from $\sim 40-50\percent$ to $\sim 80-90\percent$ for $M_\star\sim 10^8-10^{11.5}\Msun$. Interactions can enhance the $\SFR$ and drive SB events through tidal torques that funnel gas toward the centre of the galaxy \citep[e.g.,][]{Renaud2014,Hopkins2018,Moreno2019}. Hence, we conclude that galaxy interactions in FIREbox are the dominant cause of SBs at high galaxy stellar masses ($M_\star\gtrsim 10^{10}\Msun$), especially at low-redshift. Furthermore, for $\tavg=100\Myr$, the interacting fraction of SB galaxies displays an increase from $\sim 50-80\percent$ in the low-mass regime, increasing with decreasing $M_\star\sim10^{9}\Msun-10^{8}\Msun$. This behaviour is related to the systematically larger fraction of satellite (i.e., non-central) galaxies in the $\tavg=100\Myr$ SB sample, compared to the samples of SB galaxies with shorter $\tavg$ and control galaxies. Therefore, the preferred channel to drive long SBs (i.e., selected with $\tavg=100\Myr$) in low-mass galaxies is galaxy interactions. Satellite galaxies are generally less massive and more likely interacting than centrals. Moreover, during galaxy interactions, satellites are more likely to experience a significant enhancement in their $\SFR$ \citep[e.g.,][]{Moreno2021}{}{}. The low-mass SB galaxies selected with $\tavg=100\Myr$ in FIREbox are indeed dominated by satellite galaxies. 

\begin{figure*}
    \centering
	\includegraphics[width=\textwidth]{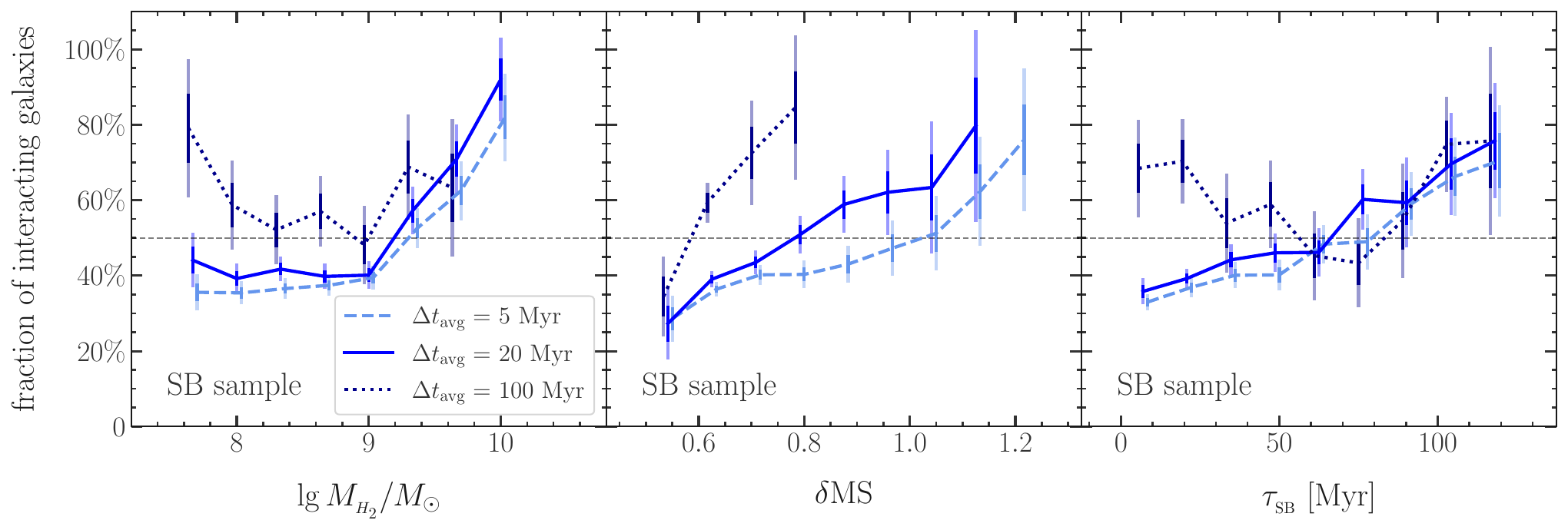}
    \caption{Fraction of interacting SB galaxies, as a function of molecular gas mass ($M_{\mathrm{H}_2}$; left panel), main-sequence offset ($\dMS$; middle panel), and SB duration ($\tau_\SB$; right panel), colour-coded by $\tavg$. The error bars represent the $1\sigma$ and $2\sigma$ (increased transparency) bootstrapped errors. The fraction of interacting SB galaxies is higher, the higher $M_\mol$, the larger $\dMS$ and the more prolonged the burst. A longer $\tavg$ systematically results in a larger fraction of interacting objects since only the most intense and prolonged SBs are selected when averaging the star formation history over long timescales.}
    \label{fig:INT_fraction_2}
\end{figure*}

In Figure~\ref{fig:INT_fraction_2}, we further investigate the role of galaxy encounters in driving SBs by showing the fraction of interacting SB galaxies as a function of their molecular gas masses, offset from the $\SFS$, and duration (for different $\tavg$). In general, the fraction of interacting SB galaxies is the highest in $\mathrm{H}_2$-rich SB galaxies with prolonged and intense SBs (i.e., with larger positive main-sequence offset $\dMS$). The fraction of interacting SB galaxies increases strongly with $M_\mol\gtrsim 10^9\Msun$ (left panel), independent of $\tavg$, from $\sim 50-80\percent$, likely because more massive SB galaxies are both more likely to be interacting (see the previous Figure~\ref{fig:INT_fraction_1}) and to have higher $M_\mol$. For $\tavg =100\Myr$, the fraction of interacting SB galaxies also increases from $\sim 50-80\percent$ for decreasing molecular gas mass from $M_\mol =10^{9}-10^{7.5}\Msun$, whereas for $\tavg =5,20\Myr$ it remains constant at $\sim 40\percent$ in the same mass regime. The fraction of interacting SB galaxies increase with increasing $\dMS$ (middle panel), from $\sim 20-80\percent$, with a strong dependence on the employed averaging time. Specifically, longer averaging times result in a steeper dependency of the fraction of interacting SB galaxies on $\dMS$. The longer SB duration ($\tau_\SB$; right panel) present in interacting systems is associated with a longer period over which the mechanism driving gas to high densities is effective, as expected for mergers and fly-bys. For $\tau_\SB\sim 70-120\Myr$, the fraction of interacting SB galaxies increases from $\sim50-80\percent$, independent of $\tavg$. For $\tavg=100\Myr$, the fraction of interacting SB galaxies increases from $\sim 50-70\percent$ with decreasing SB duration for $\tau_\SB\lesssim 70\Myr$, whereas, for $\tavg=5,20\Myr$, it decreases from $\sim 50-30\percent$ for decreasing $\tau_\SB$ in the same range. Short SBs are generally not selected in the $\tavg=100\Myr$ sample, unless they are intense (i.e., with a large $\dMS$) enough to be identified as SBs when smoothing the star formation history over 100 Myr. Since strong SBs are more likely interacting (see middle panel) than less intense ones, a large fraction of short ($\tau_\SB\lesssim 70\Myr$) SBs in the $\tavg=100\Myr$ sample is interacting. 

We find that a significant fraction ($\sim 40-50\percent$) of SB galaxies is non-interacting at all redshifts $z=0-1$, especially in the low-mass regime ($M_\star\lesssim 10^{10}\Msun$). In the next sections, we investigate the behaviour of non-interacting SBs and possible SB driving mechanisms in these galaxies.

\subsection{Breathing mode in non-interacting SB galaxies}\label{sec:results_breathing}
In this section, we explore the mechanisms that could drive gas compaction and SBs in isolated (non-interacting) systems. Low-mass galaxies ($M_\star\lesssim 10^{9.5}\Msun$) have relatively shallow potential wells and are thus likely to experience periods of strong fluctuations with periodic gas displacement. This process is associated with the bursty and stochastic star formation and the efficient stellar feedback in low-mass galaxies in simulations run with FIRE physics: for $M_\star\lesssim 10^{9.5}\Msun$ galaxies, the $\SFR$ derived with $\tavg=10\Myr$ can vary by an order of magnitude in a time interval of $\sim 200\Myr$ \citep[][]{Sparre2017}. The in-falling gas gets denser and eventually form stars, likely resulting in a SB event. Stellar feedback heats the gas, that expands to larger radii at velocities that are typically below the escape velocity, temporarily halting $\SFR$ by gas depletion in the central regions. The out-flowing gas cools over short time-scales and eventually turns-around and re-accretes onto the galaxy, restarting the cycle. This `\textit{breathing}' process has already been studied in previous theoretical works \citep[e.g.,][]{Stinson2007,Christensen2016,El-Badry2016}. Stellar feedback is capable of displacing large amounts of gas, especially in low-mass galaxies (with mass-loading factors increasing by a factor $\sim 10-100$ for decreasing galaxy stellar mass in the range $M_\star\sim 10^{8}-10^{11}\Msun$; see e.g., \citealt[][]{Muratov2015,Angles-Alcazar2017,Pandya2021}), and triggers a breathing mode that involves a large fraction of the gas in the galaxy \citep[e.g.,][]{El-Badry2016}. We investigate the properties of FIREbox galaxies in order to identify breathing galaxies and understand whether they form a separate class of systems with a specific SB triggering mechanism.

We randomly select 125 non-interacting SB galaxies (random NI-SB sample) in the sample with $\tavg=20\Myr$, uniformly distributed in stellar mass and redshift (5 randomly selected galaxies in each stellar mass and redshift bins: 5 equally spaced log stellar mass bins from $M_\star=10^{8}-10^{11.5}\Msun$ with 0.5 dex width and 5 equally spaced redshift bins from $z=0-1$). We then classify them as breathing/non-breathing by visually inspecting the evolution of their (total) gas maps in a period of time of $\sim 150\Myr$ around the SB event. In the gas maps of breathing galaxies, we expect a clear signature of a cycle of gas compaction and subsequent outflow, involving a large fraction of the gas reservoir. Moreover, we expect these global breathing modes to eventually destroy any disc structure in the gas component.

\begin{figure*}
    \centering
    \includegraphics[width=.9\textwidth]{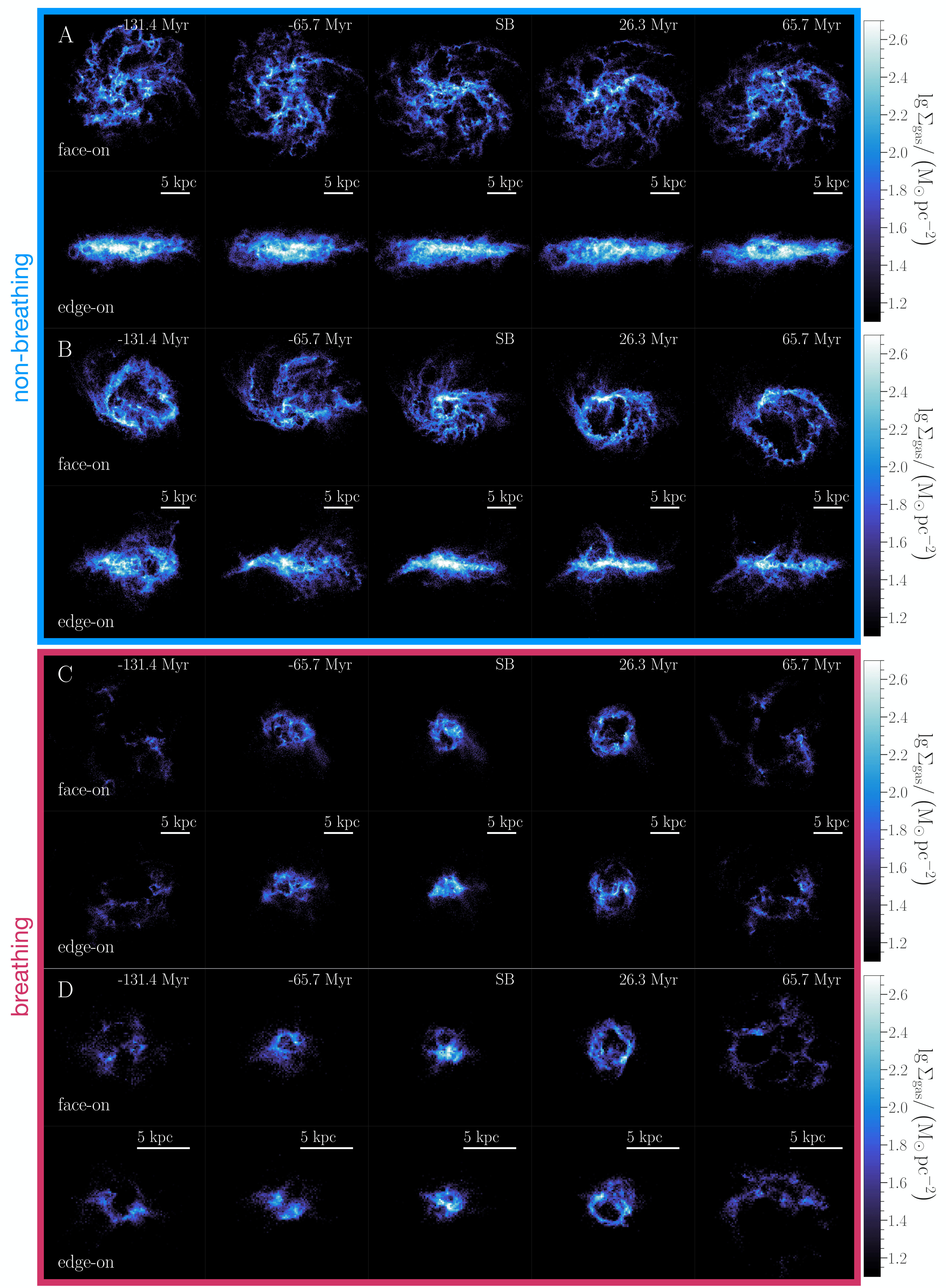}
    \caption{Total gas surface-density maps (face-on and edge-on projections) of four galaxies in our sample of 125 non-interacting SB galaxies (NI-SB sample) that were visually classified as non-breathing (panel A and B) or breathing (panel C and D). From left to right, we show the evolution of the galaxy close to the SB event. In the upper right corner of each sub-panel, we show the time relative to the beginning of the SB (sub-panels labelled with `SB'), in the sample with $\tavg=20\Myr$. See Table~\ref{tab:breathing_example_info} for information on the redshift, stellar mass, total gas mass, fraction of stellar mass within the central kpc, and main-sequence offset of these four galaxies. 
    }
    \label{fig:breathing_example}
\end{figure*}

To clarify this point, Figure~\ref{fig:breathing_example} shows the evolution of total gas surface-density maps of four non-interacting galaxies from this sample. For sake of simplicity, we only show the total gas maps at specific snapshots that helped the classification, during a time interval from $\sim 130\Myr$ before to $\sim 70\Myr$ after the beginning of the SB event. In Table~\ref{tab:breathing_example_info} we report the redshift, stellar mass, total gas mass, fraction of stellar mass within the central kpc, and main-sequence offset (for $\tavg =20\Myr$) of these four example galaxies, at the time when they are selected as SB. The galaxies in panel A and B are classified as non-breathing, whilst those in panel C and D are classified as breathing. The breathing galaxies in panel C and D undergo an episode of compaction leading to the SB and subsequently experiencing quasi-isotropic expansion of a gas shell (clearly visible in both the face-on and edge-on projections) that removes the gas supply for star formation in the centre of the galaxy. The galaxy in the panel B exhibits a clear, although disturbed, gaseous disc in the edge-on projection and is thus classified as non-breathing. Its face-on projection reveals that the central gas compaction, leading to the SB and subsequently an expanding shell/ring of gas in the central $\sim 1-2\kpc$, akin to the behaviour of breathing galaxies. This could be related to its poorly centrally concentrated potential that would potentially allow for global breathing modes to develop \citep[see e.g.,][]{Hopkins2023}, as shown by the fraction of stellar mass within its central kpc similar to that of the breathing galaxies in panel C and D (see Table~\ref{tab:breathing_example_info}). The massive, non-breathing galaxy in panel A preserves its disc structure during the SB, due to its deeper potential and well defined disc structure (see the discussion below on the role of angular momentum).

\begin{table*}
    \centering
    \caption{Redshift, stellar mass, total gas mass, fraction of stellar mass within the central kpc, and main-sequence offset (for $\tavg=20\Myr$) of the four galaxies shown in Figure~\ref{fig:breathing_example}, at the time when they are selected as SB.}
    \begin{tabular}{ccccccc}
        \hline
        Figure~\ref{fig:breathing_example} & Breathing? & redshift & $\lg M_\star/M_\odot$ & $\lg M_{\rm gas}/M_\odot$ & $\lg M_\star\left(<1\kpc\right)/M_\odot$ & $\dMS$ [dex]\\
        \hline
        panel A & No & 0.12 & 10.13 & 10.14 & -1.71 & 0.61 \\
        panel B & No &  0.61 & 10.27 & 10.03 & -0.80 & 0.62 \\
        panel C & Yes & 0.94 & 8.87 & 9.47 & -0.73 & 1.22 \\
        panel D & Yes & 0.86 & 8.43 & 9.02 & -0.60 & 0.89 \\
        \hline
    \end{tabular}
    \label{tab:breathing_example_info}
\end{table*}

We expect the breathing mode to preferentially take place in low-mass galaxies and with low rotational support for their gas component, related to a low specific gas angular momentum and therefore a less well-defined disc structure. Moreover, we expect breathing galaxies to have a relatively small gas reservoir with respect to non-breathing galaxies, in order for stellar feedback to effectively alter the dynamics of a large fraction of the total gas reservoir, i.e., allowing the SB in the central regions to easily displace most of the gas.

\begin{figure}
    \centering
	\includegraphics[width=\columnwidth]{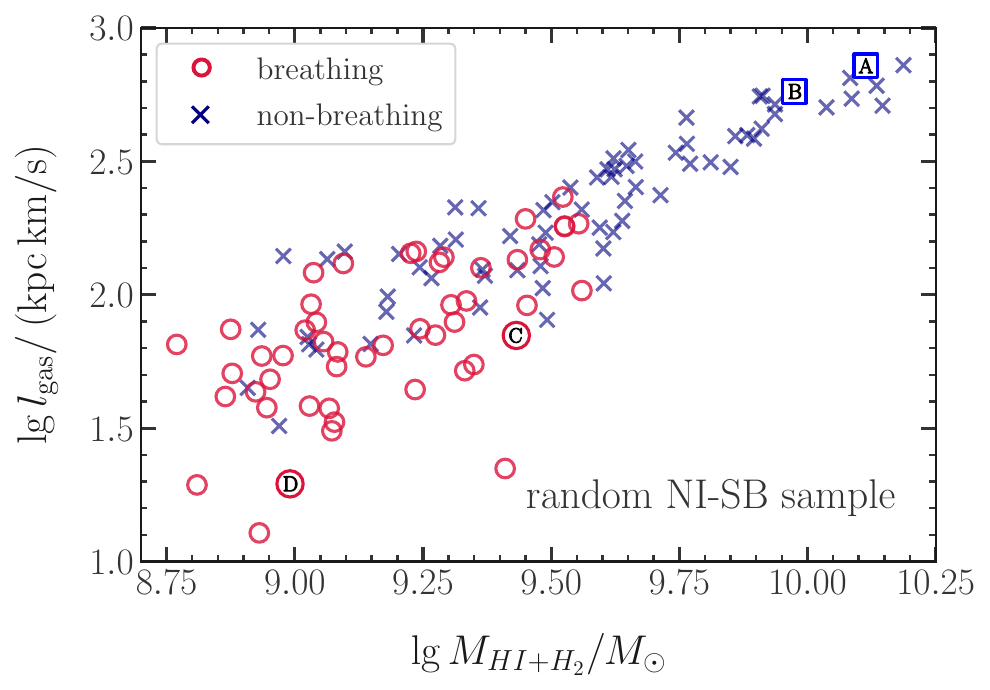}
    \caption{Total atomic+molecular (i.e., neutral) hydrogen mass ($M_\neutr$) and gas specific angular momentum ($l_{\rm gas}$) of the 125 randomly selected, non-interacting SB galaxies (random NI-SB sample) that we visually classified as breathing (red circles) or non-breathing (blue crosses), by inspecting the evolution of their gas surface-density maps (see examples in Figure~\ref{fig:breathing_example}, here indicated with A, B, C, and D). Breathing galaxies have the lowest total $M_\neutr$ and $l_{\rm gas}$, reflecting the fact that stellar feedback (associated with the SB event) is more efficient in affecting the gas dynamics in low-mass galaxies with little rotational support. The overlap between the two classes in this plane suggests that the breathing mode is not a distinct channel associated with SBs. The text labels refer to the galaxies shown in Figure~\ref{fig:breathing_example}, surrounded by a red circle (blue square) when the galaxy is classified as breathing (non-breathing).}
    \label{fig:Lgas_vs_MHI_breathing}
\end{figure}

Figure~\ref{fig:Lgas_vs_MHI_breathing} shows the sample of 125 randomly selected non-interacting SB (random NI-SB sample) galaxies in the plane of total gas specific angular momentum and total neutral gas mass. Galaxies that are visually-classified as breathing (red circles) have on average smaller total neutral gas masses and specific gas angular momenta, compared to non-breathing galaxies (blue crosses).

To further test our hypothesis that gas content and gas specific angular momentum are the best predictors to classify breathing galaxies, we made use of a Random Forest classifier. The employed features include the mass, angular momentum, and velocity dispersion of different components (stars, dark matter and gas at different phases), computed within the galaxy size (assumed to be $0.1\,R_{\rm vir}$) and fixed apertures of 1, 3, 5, and 10 kpc. From the feature importance, the total neutral gas mass, $M_\neutr$ and the total gas specific angular momentum, $l_{\rm gas}\equiv J_{\rm gas}/M_{\rm gas}$, where $J_{\rm gas}$ is the total angular momentum of the gas in the galaxy, are indeed the best predictors to determine whether a non-interacting SB galaxy is experiencing a breathing mode.

We also run a linear discriminant analysis \citep[LDA,][]{Fisher1936}{}{} to find the combination of $l_{\rm gas}$ and $M_{\mathrm{HI}+\mathrm{H}_2}$ that best separates the two classes of breathing/non-breathing galaxies. The result is the axis defined by a quantity $\sim\lg \left[ l_{\rm gas}/\left(\mathrm{kpc}~\mathrm{km}/\mathrm{s}\right)\right] - 0.01\,\lg \left[ M_{\mathrm{HI}+\mathrm{H}_2}/M_\odot\right]$. The overlap between the two classes (breathing/non-breathing) in the $\left(\lg l_{\rm gas}-\lg M_\neutr\right)$-plane, or, equivalently, the absence of a clear bimodality, suggests that the breathing mode is not a distinct channel to trigger (as well as triggered by) the SB event through central gas compaction, but rather an extreme case where the stellar feedback is effectively displacing a large fraction of the total gas reservoir of the galaxy. As a galaxy increases in mass and angular momentum, it becomes consistently harder to trigger an evident breathing mode. Therefore, a similar mechanism could drive SBs in all non-interacting galaxies.

\subsection{You too can become a starburst: occurrence of SBs in former control galaxies}
We showed that neutral gas mass and specific gas angular momentum are the best predictors for whether galaxies experience breathing, based on the properties of their gas and stellar content. The lack of a strong rotational support (low $l_{\rm gas}$) and a relatively small gas reservoir seems to favour strong fluctuations in the $\SFR$ that lead to one or multiple SBs in non-interacting galaxy. We will now investigate where non-interacting, control galaxies in the sample with $\tavg=20\Myr$ (NI-ctrl sample) fall within this parameter space and how it relates to the probability of these galaxies undergoing a SB in the future.

\begin{figure}
    \centering
	\includegraphics[width=\columnwidth]{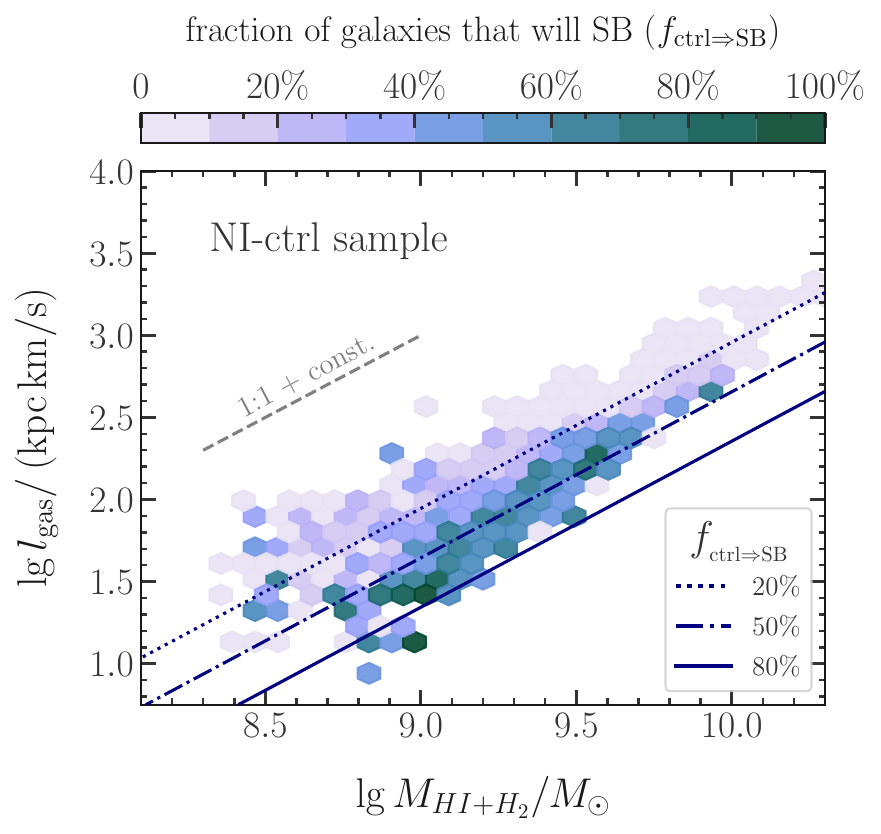}
    \caption{Total neutral hydrogen mass ($M_\neutr$) and gas specific angular momentum ($l_{\rm gas}$) of non-interacting control galaxies, colour-coded by the probability that these galaxies will undergo a SB in the next Gyr ($f_\willSB$). At any given $M_\neutr$, that is enough to sustain a high $\SFR$, only galaxies with low-enough $l_{\rm gas}$ are likely becoming a SB in the next Gyr. The fraction $f_\willSB$ increases almost linearly with $\sim M_\neutr/l_{\rm gas}$, as shown by the dark blue contours resulting from fitting $f_\willSB$ as a function of $l_{\rm gas}$ and $M_\neutr$.}
    \label{fig:Lgas_vs_MHI_fraction_willSB}
\end{figure}

Figure~\ref{fig:Lgas_vs_MHI_fraction_willSB} shows the specific angular momentum of the total gas ($l_{\rm gas}$) as a function of total neutral gas mass ($M_\neutr$) for the non-interacting control galaxies, colour-coded by the fraction of galaxies that will undergo a SB within the next Gyr. Galaxies with the lowest angular momentum at a given $M_\neutr$ have the highest likelihood of becoming a SB. This can be understood in the context of gas compaction in the central kpc triggering starbursts. To funnel gas to the centre, effective rotational support is crucial, thus galaxies with lower specific angular momentum are more likely to host a SB in the next Gyr. Only galaxies with $M_\neutr\gtrsim 7\times 10^8\Msun$ are likely to become a SB (where galaxies with higher $M_\neutr$ can become a SB with larger $l_{\rm gas}$), because large amounts of gas are required to sustain $\dMS\gtrsim 0.6~\mathrm{dex}$. The fraction of non-interacting, control galaxies that will become a SB, $f_\willSB$, increases in a direction that is almost orthogonal to the 1:1 relation. To show this more clearly, we report the resulting contours for $f_\willSB$ (dark blue lines)\footnote{Specifically, we performed a multi-linear regression fit of $f_\willSB$ as a function of both $l_{\rm gas}$ and $M_\neutr$: $f_\willSB = \frac{1}{2}\left\{\,1+ \tanh\left[\,\mathcal{C}_1\,\lg l_{\rm gas} + \mathcal{C}_2\,\lg M_\neutr + \mathcal{C}_3\,\right]\,\right\}$}. Hence, NI-ctrl galaxies that will become a SB are characterized by a lower gas specific angular momentum at fixed neutral gas mass, such that the likelihood of becoming a SB in the next Gyr scale as $f_\willSB\sim M_\neutr/l_{\rm gas}$.

\begin{figure}
    \centering
	\includegraphics[width=\columnwidth]{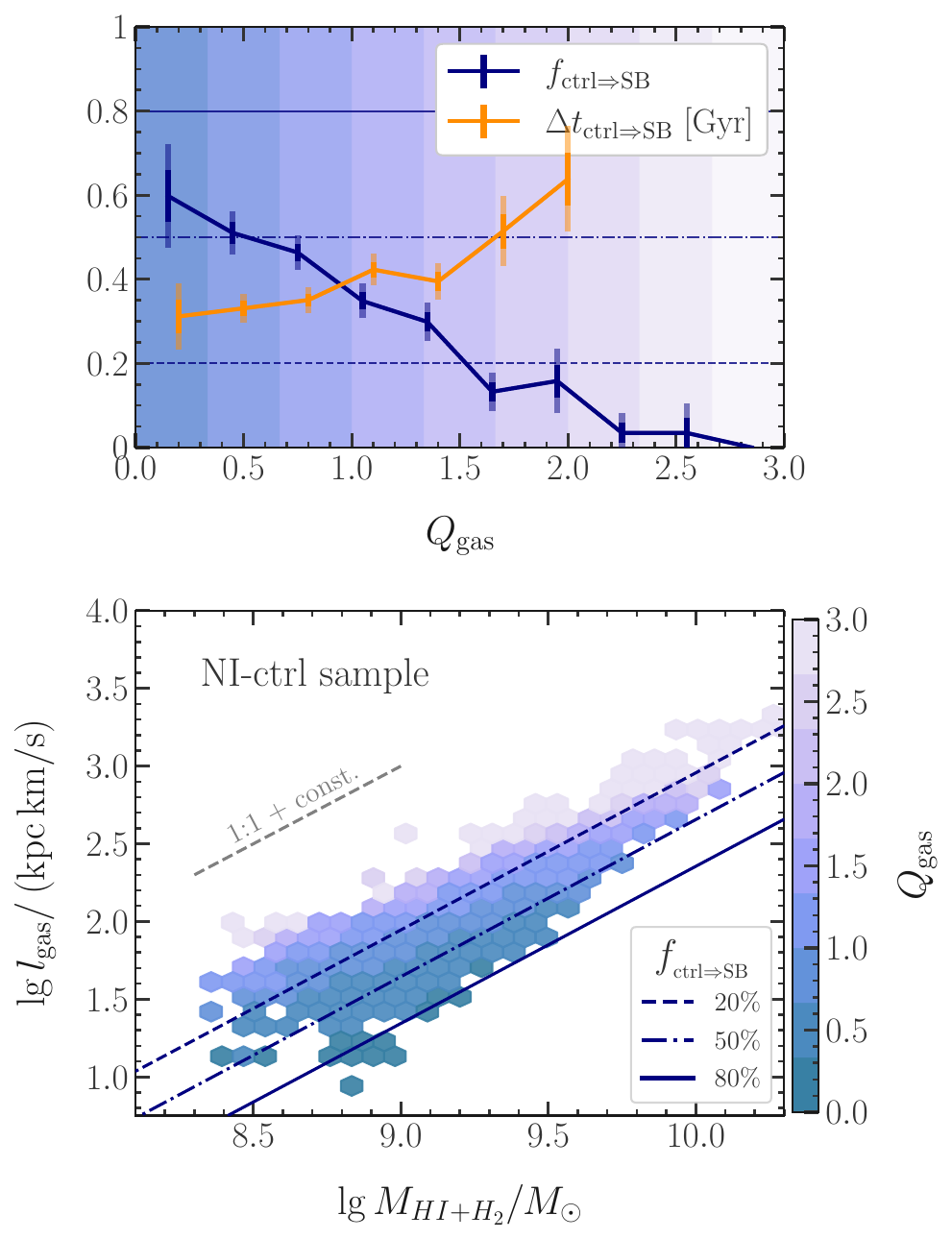}
    \caption{Upper panel: fraction of non-interacting control galaxies (NI-ctrl sample) that will undergo a SB in the next Gyr ($f_\willSB$) and the average time until the next SB ($\Delta t_\willSB$; among galaxies that will undergo a SB in the next Gyr), as a function of the Toomre parameter for gas ($Q_{\rm gas}$). Lower panel: total neutral hydrogen mass ($M_\neutr$) and gas specific angular momentum ($l_{\rm gas}$) of NI-ctrl galaxies, colour-coded by their average $Q_{\rm gas}$. The fraction $f_\willSB$ increases with decreasing $Q_{\rm gas}$. Galaxies with $Q\lesssim 1$ are globally unstable against gravitational collapse and will likely evolve towards more compact configuration, eventually triggering a SB. To have a low $Q_{\rm gas}$ appears to be a necessary, although not sufficient condition for NI-ctrl galaxies to undergo a SB in the next Gyr. On average, the lower the $Q_{\rm gas}$, the sooner the galaxy will undergo a SB.}
    \label{fig:Lgas_vs_MHI_Toomre}
\end{figure}

The \citet[][]{Toomre1964} $Q$ parameter measures the susceptibility of gas discs to axisymmetric gravitational instability. Galaxies with $Q\lesssim 1$ are globally unstable against gravitational collapse \citep[][]{Safronov1960,Toomre1964}, thus we expect gas to move to the galactic centre and their central gas reservoir to compactify over a dynamical timescale, rendering them likely to become a SB. Let us consider $Q$ for the gas component only, namely:

\begin{equation}
    Q_{\rm gas} = \frac{\sigma_{\rm gas}\,l_{\rm gas}}{G\,M_{\rm gas}} ~,
    \label{eqn:gas_Toomre}
\end{equation}

\noindent where $\sigma_{\rm gas}$ is the gas velocity dispersion in the galaxy and $G$ is Newton's gravitational constant. For a fixed velocity dispersion, $Q_{\rm gas}$ will scale as $Q_{\rm gas}\sim l_{\rm gas}/ M_{\rm gas} \sim l_{\rm gas}/M_\neutr$. The upper panel of Figure~\ref{fig:Lgas_vs_MHI_Toomre} shows the fraction of NI-ctrl galaxies that will undergo a SB in the next Gyr ($f_\willSB$) and the average time until the next SB ($\Delta t_\willSB$; among galaxies that will undergo a SB in the next Gyr), as a function of $Q_{\rm gas}$. In the lower panel, we show the total neutral hydrogen mass ($M_\neutr$) and gas specific angular momentum ($l_{\rm gas}$) for NI-ctrl galaxies, colour-coded by their average $Q_{\rm gas}$. The fraction $f_\willSB$ increases with decreasing $Q_{\rm gas}$, going from essentially zero for $Q_{\rm gas}\gtrsim 2.5$ up to $\sim 60\percent$ for $Q_{\rm gas}\lesssim 0.2$. Moreover since $f_\willSB\sim 0$ for $Q_{\rm gas}\gtrsim 2.5$, we conclude that \textit{only galaxies with a gravitationally unstable gas reservoir will be able to undergo a SB}, meaning that $Q_{\rm gas}\lesssim 2.5$ is a necessary, although not sufficient condition to have a SB. For NI-ctrl galaxies that will undergo a SB in the next Gyr, the average $\Delta t_\willSB$ decreases for decreasing $Q_{\rm gas}$, going from $\Delta t_\willSB\sim 300\Myr$ for $Q_{\rm gas}\lesssim 0.5$ to $\Delta t_\willSB\sim 600\Myr$ for $Q_{\rm gas}\sim 2$. Therefore, on average, the more gravitationally unstable is the gas reservoir, the sooner the galaxy will undergo a SB.

\section{Summary and discussion}\label{sec:discussion}
We analysed galaxies with $M_\star\geq 10^8 \Msun$, at redshift $z=0-1$, from the FIREbox cosmological volume \citep[][]{Feldmann2023}, to understand the mechanisms driving starbursts (SBs). We defined a starburst galaxy as having an $\SFR$ offset from the star-forming main-sequence exceeding 0.6 dex and we compiled a redshift- and stellar mass-matched control sample. We employed different $\SFR$ averaging times $\tavg =5,20,100\Myr$ to mimic different star formation tracers. For all galaxies in our SB and control samples, we computed the properties of their stellar and gas content and, specifically for SB galaxies, the duration of the SB event. Furthermore, we classified all SB and control galaxies as interacting/non-interacting, based on the mass-ratio of other galaxies that came within 200 kpc of the galaxy centre in the past 100 Myr. In the following, we summarize our findings (see also Figure~\ref{fig:summary_cartoon} for a schematic overview):

\begin{figure}
    \centering
    \includegraphics[width=\columnwidth]{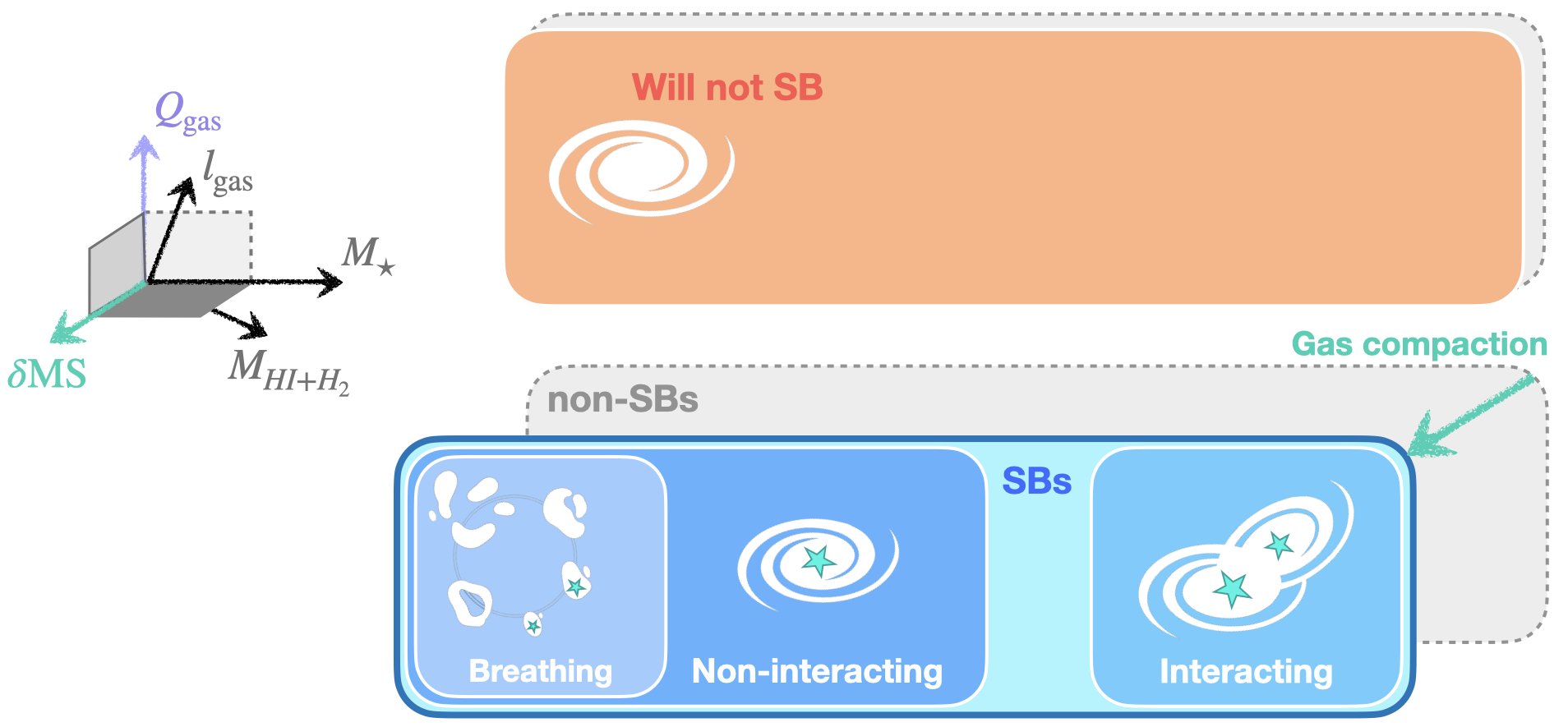}
    \caption{Schematic summary of the main findings of this work. Globally unstable gas reservoirs or galaxy interactions lead to central gas compaction events that increase the amount of ultra-dense gas and shorten the galaxy depletion time (at fixed total gas mass) and hence trigger a SB. The level of gravitational instability of the gas reservoir can be measured by the average Toomre parameter for the gas in the galaxy, $Q_{\rm gas}$: the lower the $Q_{\rm gas}$, the sooner the galaxy will have a SB. Interactions are especially important in driving SBs in massive galaxies. Non-interacting SB galaxies with the lowest gas specific angular momentum likely exhibit global breathing modes.}
    \label{fig:summary_cartoon}
\end{figure}

\begin{itemize}
    \item
        SB galaxies make up $\sim 5,4,1\percent$ of the entire galaxy population analysed in this work, for $\SFR$ averaging timescale of $\tavg=5,20,100\Myr$, respectively (see Figure~\ref{fig:SB_fraction}).
    \item
        At $z=0$, the SB fraction increases from about $2\percent$ to about $5\percent$ for decreasing stellar mass from $M_\star\sim 10^{10}\Msun$ to $10^{8}\Msun$. The SB fraction increases with redshift, increases from about $4-5\percent$ at $z=0$, to about $8\percent$ at $z=1$, for $M_\star\sim 10^{8}\Msun$.
    \item
        SB galaxies have molecular gas depletion times that are on average shorter ($\sim 50-200\Myr$) than in control galaxies ($\sim 1\Gyr$) across the whole range of masses and redshifts considered, implying larger global star-forming efficiencies (see Figure~\ref{fig:SFR_vs_MH2}). The average molecular gas reservoir hosted by SB galaxies is larger than in control galaxies.
    \item
        The median total gas mass in SB galaxies is only slightly larger (by $\sim 0.1$ dex) than in control galaxies, whereas their median fractions of molecular and high-density gas, are $\sim 0.4$ and $\sim0.6$ dex larger, respectively (see Figure~\ref{fig:SB_vs_ctrl}). The distribution of the molecular gas fraction for control galaxies exhibits a tail towards low molecular gas fractions, extending to fractions of $\sim 10^{-3}$, whereas SB galaxies have a larger lower limit of $\sim 10^{-2}$. The SFR enhancement in SB galaxies is accompanied by a shortening of their gas depletion times and an increase in their molecular gas masses, as a consequence of an increase in both the fraction of molecular and high-density star-forming gas (as per our model definition) at fixed total gas mass (see equations~\ref{eqn:SFR_gal}-\ref{eqn:tdepl})
    \item
        Galaxies become SB by shortening their molecular gas depletion time and, less significantly, by increasing their molecular gas mass over the $\sim 70\Myr$ prior to the SB (see Figure~\ref{fig:SFR_vs_MH2}). In the $\sim 70\Myr$ prior to the SB, the fractions of both molecular and high-density, star-forming gas increase on average by $\sim 0.4-0.6$ dex and $\sim 0.2-0.4$ dex, respectively, depending on stellar mass (see Figure~\ref{fig:fSF_vs_fH2}), from values that are typical of control galaxies to those for SB galaxies. On average, the fraction of high-density, star-forming gas increases more than the fraction of molecular gas, resulting in the shortening of the molecular depletion time (see equation~\eqref{eqn:tdepl}), especially in the $\sim 20\Myr$ prior to the SB, and an increase of the molecular gas mass at fixed total gas mass. After the SB, the fractions of molecular and high-density, star-forming gas reduce to typical values for control galaxies in $\sim 70\Myr$, reversing the path they followed prior to the SB.
    \item
        SB events are accompanied by a significant increase ($\sim 0.3$ dex) in the gas mass in the central kpc, at fixed total gas mass (see Figure~\ref{fig:Mgas_vs_Mgas_1kpc}). The gas mass in the central regions of SB galaxies reaches its maximum prior to the beginning of the SB event, suggesting the gas compaction events causally result in SBs (see Figure~\ref{fig:median_evolution_SFR_and_Mgas_1kpc}).
    \item
        SB galaxies display more concentrated $\SFR$s, with the median $\SFR$ within the central kpc of SB galaxies being comparable to their median global $\SFR$ and significantly larger than in control galaxies (see Figure~\ref{fig:SB_vs_ctrl_1kpc}). The higher median global $\SFR$ of SB galaxies is explained by the larger median fraction of high-density, star-forming gas, whereas, within their central kpc, also an increase in their median total gas mass has to be accounted for to explain the higher median $\SFR$.
    \item
        The fraction of SB galaxies that are interacting increases from $\sim 50$ to $\sim 90\percent$ with increasing galaxy stellar masses $\lg M_\star\sim 8-11.5\Msun$, i.e., SBs in massive ($M_\star\gtrsim 10^{10}\Msun$) galaxies are predominantly driven by galaxy interactions including mergers and flybys (see Figure~\ref{fig:INT_fraction_1}). Long and intense SBs are more likely to be interacting. The fraction of interacting SB galaxies increases with the SB duration ($\tau_\SB$), increasing from $\sim 40\percent$ to $\sim 80\percent$ for $\tau_\SB\sim 1-120\Myr$. Interacting SB galaxies also display the most extreme $\SFR$ offset from the $\SFS$ $\SFR$, $\dMS$, increasing from $\sim 30\percent$ to $\sim 80\percent$ for $\dMS\sim 0.6-1.2~\rm{dex}$  (see Figure~\ref{fig:INT_fraction_2}). The observational tracers for star formation that are sensitive to $\SFR$ on relatively long time-scales of $\sim 100\Myr$ systematically overpredict the number of interacting SB galaxies in FIREbox, and consequently reduce the observed contribution of non-interacting systems to the SB population. A large fraction ($\gtrsim 50\percent$) of low-mass ($M_\star\lesssim 10^{10}\Msun$) SB galaxies at $z=0-1$ is non-interacting, implying that SBs are not always triggered by interactions.
     \item
        Non-interacting SB galaxies with a poor rotational support (i.e., with low gas specific angular momentum) and low neutral gas masses can exhibit a `\textit{breathing}' mode \citep[see, e.g.,][]{Christensen2016,El-Badry2016}{}{}, since in this regime stellar feedback by the nuclear SB is able to displace a large fraction of the gas reservoir in a quasi-isotropic, spherical fashion, evacuate the central region of the galaxy of gas, and destroy the gas disc (see Figure~\ref{fig:Lgas_vs_MHI_breathing}). The smooth transition between breathing and non-breathing galaxies in the properties of their gas and stellar content, and star formation histories suggests that they share a common SB-driving mechanism, instead of pointing to a separate physical process triggering the breathing mode.
    \item
        The fraction of non-interacting, control galaxies that will undergo a SB in the next Gyr increases up to $\sim 40-60\percent$ for a global Toomre parameter $Q_{\rm gas}\lesssim 1-0.1$ (see Figure~\ref{fig:Lgas_vs_MHI_Toomre}). Galaxies with $Q_{\rm gas}\gtrsim 2.5$ have a low chance of undergoing a SB in the next Gyr. Moreover, the lower the $Q_{\rm gas}$, the shorter is the average time until the next SB, ranging from $\sim 300-600\Myr$ for $Q_{\rm gas}\sim 0.1-2$. Massive galaxies have on average larger $Q_{\rm gas}$ and therefore likely require galaxy interactions to funnel gas towards the central regions and initiate a nuclear SB.
\end{itemize}

The fractional contribution of SB galaxies in FIREbox is in agreement with recent observations, increasing from about $1\percent$ at $z\sim 0$ to about $5\percent$ at $z\sim 1$ \citep[e.g.,][]{Bergvall2016,Rinaldi2022}. However, the reported fractions depend on the exact definition of SB galaxies. \citet[][]{Bisigello2018} report the SB fraction in a similar redshift and mass range as presented here. They find that SB galaxies constitute $\sim 5\percent$ of all $z=0.5-1$ galaxies with $\lg M_\star/M_\odot =8.25-11.25$, with $\SFR$s estimated using either mid/far-IR or rest-UV photometry (to be compared to our estimates of $1-5\percent$ for $\tavg=20-100\Myr$). They also show that the SB fraction increases with redshift, especially at $M_\star\lesssim10^{9}\Msun$, in agreement with our results.

Gas compaction drives SBs in FIREbox, as evidenced by the increase in the fraction of total gas mass in the central regions of SB galaxies over $\sim 70\Myr$ prior to the beginning of the SB event. The fractions of molecular and high-density gas increase in the process of funnelling gas towards the centre, resulting in a global increase of both the molecular gas mass and $\SFR$. The disproportional increase in the fraction of high-density gas compared to the fraction of molecular gas results in a more important shortening of the molecular depletion time with respect to the increase in molecular gas mass. This behaviour in FIREbox owes to a shift of the gas density PDF towards high densities \citep[as it is the case in SBs triggered by interactions, e.g., in ][]{Renaud2014}{}{}. In the case of the SB galaxies studied here, the increase in the amount of gas dense enough to be eligible for star-formation in our model does not imply, on average, a proportional increase in the amount of molecular gas \citep[see, e.g., ][]{Moreno2021}{}{}. On the other hand, the turbulent nature of the ISM can also result in the opposite scenario, where only a small fraction of dense gas being star-forming, resulting in low star-forming efficiencies as in the Milky Way's central molecular zone \citep[e.g.,][]{Moreno2019,Orr2021}{}{}. To summarize, our results favour a scenario where the SBs are driven mainly by enhanced star-forming efficiencies rather that more massive gas reservoirs \citep[e.g.,][]{Ellison2020,Feldmann2020}, and support the hypothesis of a mainly nuclear, rather than a uniform, global SB event \citep[e.g.,][]{Barnes&Hernquist1991,Wilkinson2018,Ellison2018}.

Galaxy interactions can trigger gas compaction by tidal torques \citep[see ,e.g.,][]{Pan2018,Moreno2021,Garay-Solis2023,He2023}{}{}. Results from observations of local ultra-luminous infrared galaxies and recent high-resolution simulations demonstrated that SBs can be driven by interactions \citep[e.g.,][]{Hopkins2018,Renaud2019}. We found that interactions play a major role in driving SBs in $M_\star\gtrsim 10^{10}\Msun$ galaxies, aligned with the findings by \citet[][]{Hung2013}{}{} for Herschel-selected galaxies at $0.2<z<1.5$. However, a large fraction ($\gtrsim 50\percent$) of SB galaxies in FIREbox are non-interacting. This finding is consistent with an analysis of the Illustris simulation \citep[][]{Vogelsberger2014} by \citet[][]{Wilkinson2018}, who report that $z\lesssim 0.15$ SB galaxies are not triggered by mergers in $55\percent$ of the cases. Our results are also in agreement with recent observational work by \citet[][]{Wilkinson2022}, who report a (lower limit) merger fraction of about 10 to 40 per cent (depending on the exact methods employed for the selection and classification) for the observed post-starburst galaxies at $z\lesssim 0.25$, with stellar masses mostly at $M_\star\gtrsim 10^{10}\Msun$ \citep[see also, e.g.,][]{Zabludoff1996,Blake2004,Goto2005,Pawlik2018}.

Non-interacting galaxies are more likely to undergo a SB in the future if their gas reservoir is globally gravitationally unstable (as measured by their global Toomre $Q_{\rm gas}$), allowing for massive gas inflows towards the central regions \citep[e.g.,][]{Dekel&Burkert2014,Danovich2015,Zolotov2015}. Our work extends previous findings on the role of gas compaction events in the $\SFS$ scatter \citep[e.g.,][]{Tacchella2016a}{}{}. This is also consistent with theoretical work on high-redshift galaxies, on the existence of a `\textit{blue nugget}' phase caused by wet gas compaction \citep[e.g.,][]{Tacchella2016a,Lapiner2023}. This scenario is especially akin to what we found for our non-interacting, breathing, SB galaxies.

We found that the lower the $Q_{\rm gas}$ the sooner a galaxy will undergo a SB. This aligns with recent observations of a small number of local SB galaxies with $Q_{\rm gas}\lesssim 1$ from the DYNAMO (DYnamics of Newly-Assembled Massive Object) survey \citep[][]{Fisher2014,Fisher2017a,Fisher2017b,Fisher2022}. Local, non-starbursting spirals are commonly found to have $Q_{\rm gas}\gtrsim 2$ \citep[e.g.,][]{Leroy2008}. Furthermore, observations of a nearby star-forming galaxy \citep[LARS 8;][]{Ostlin2014,Hayes2014} with a stellar mass of $M_\star\sim 10^{11}\Msun$ reveal a highly globally unstable disk of gas that triggers the formation of massive, dense molecular clumps \citep[][]{Puschnig2023}, thus corroborating our findings.

We showed that SB galaxies move back to the main-sequence of star-forming galaxies after the SB event, whilst hosting a significant amount of molecular gas. Recent studies revealed that some post-SB galaxies might lie on the star-forming main-sequence and have relatively large molecular gas fractions, despite their low rate of star formation inferred from UV indicators \citep[e.g.,][]{French2021,Baron2022,Smercina2022,Baron2023}. Hence, our predictions on the fate of SB galaxies might have important implications for future research on the nature an origin of post-SB galaxies \citep[e.g.,][and references therein]{Kriek2010,Suess2019}.

In this work, we showed that global gravitational instabilities driving central gas compaction are the main cause for SBs in FIREbox. To validate this theoretical prediction, high-resolution studies of the gas properties and kinematics in star-forming galaxies are crucial, e.g., with the PHANGS (Physics at High Angular resolution in Nearby GalaxieS) survey \citep[e.g.,][]{Leroy2021,Emsellem2022,He2023}. Furthermore, upcoming data from the James Webb Space Telescope (JWST) \citep[][]{Gardner2006} will help unveiling the connection between the stability and properties of the dense interstellar medium in local and high-redshift galaxies in unprecedented detail.


\section*{Acknowledgements}
EC thanks Papa IV for His countless blessings. EC thanks Guochao Sun for inspiring discussions and helpful
comments. RF, LB acknowledge financial support from the Swiss National Science Foundation (grant no PP00P2$\_$194814). EC, RF, MB acknowledge financial support from the Swiss National Science Foundation (grant no 200021$\_$188552). JG gratefully acknowledges financial support from the Swiss National Science Foundation (grant no CRSII5$\_$193826). JM is supported by the Hirsch Foundation. We acknowledge PRACE for awarding us access to MareNostrum at the Barcelona Supercomputing Center (BSC), Spain. This research was partly carried out via the Frontera computing project at the Texas Advanced Computing Center. Frontera is made possible by National Science Foundation award OAC-1818253. This work was supported in part by a grant from the Swiss National Supercomputing Centre (CSCS) under project IDs s697 and s698. We acknowledge access to Piz Daint at the Swiss National Supercomputing Centre, Switzerland, under the University of Zurich’s share with the project ID uzh18. This work made use of infrastructure services provided by S3IT (\url{www.s3it.uzh.ch}), the Service and Support for Science IT team at the University of Zurich. All plots were created with the \textsc{matplotlib} library for visualization with Python \citep{Hunter2007}. This project is part of the FIRE simulation collaboration.

\section*{Data Availability}
The data supporting the plots within this article are available on reasonable request to the corresponding author. A public version of the \textsc{gizmo} code is available at \url{http://www.tapir.caltech.edu/~phopkins/Site/GIZMO.html}.




\bibliographystyle{mnras}
\bibliography{main}

\begin{thebibliography}{}
\makeatletter
\relax
\def\mn@urlcharsother{\let\do\@makeother \do\$\do\&\do\#\do\^\do\_\do\%\do\~}
\def\mn@doi{\begingroup\mn@urlcharsother \@ifnextchar [ {\mn@doi@}
  {\mn@doi@[]}}
\def\mn@doi@[#1]#2{\def\@tempa{#1}\ifx\@tempa\@empty \href
  {http://dx.doi.org/#2} {doi:#2}\else \href {http://dx.doi.org/#2} {#1}\fi
  \endgroup}
\def\mn@eprint#1#2{\mn@eprint@#1:#2::\@nil}
\def\mn@eprint@arXiv#1{\href {http://arxiv.org/abs/#1} {{\tt arXiv:#1}}}
\def\mn@eprint@dblp#1{\href {http://dblp.uni-trier.de/rec/bibtex/#1.xml}
  {dblp:#1}}
\def\mn@eprint@#1:#2:#3:#4\@nil{\def\@tempa {#1}\def\@tempb {#2}\def\@tempc
  {#3}\ifx \@tempc \@empty \let \@tempc \@tempb \let \@tempb \@tempa \fi \ifx
  \@tempb \@empty \def\@tempb {arXiv}\fi \@ifundefined
  {mn@eprint@\@tempb}{\@tempb:\@tempc}{\expandafter \expandafter \csname
  mn@eprint@\@tempb\endcsname \expandafter{\@tempc}}}

\bibitem[\protect\citeauthoryear{Alves, Combes, Ferrara, Forveille  \&
  Shore}{Alves et~al.}{2016}]{Planck2015}
Alves J.,  Combes F.,  Ferrara A.,  Forveille T.,   Shore S.,  2016, \mn@doi
  [Astronomy and Astrophysics] {10.1051/0004-6361/201629543}, 594

\bibitem[\protect\citeauthoryear{{Angl{\'e}s-Alc{\'a}zar},
  {Faucher-Gigu{\`e}re}, {Kere{\v{s}}}, {Hopkins}, {Quataert}  \&
  {Murray}}{{Angl{\'e}s-Alc{\'a}zar} et~al.}{2017}]{Angles-Alcazar2017}
{Angl{\'e}s-Alc{\'a}zar} D.,  {Faucher-Gigu{\`e}re} C.-A.,  {Kere{\v{s}}} D.,
  {Hopkins} P.~F.,  {Quataert} E.,   {Murray} N.,  2017, \mn@doi [\mnras]
  {10.1093/mnras/stx1517}, \href
  {https://ui.adsabs.harvard.edu/abs/2017MNRAS.470.4698A} {470, 4698}

\bibitem[\protect\citeauthoryear{{Barnes} \& {Hernquist}}{{Barnes} \&
  {Hernquist}}{1991}]{Barnes&Hernquist1991}
{Barnes} J.~E.,  {Hernquist} L.~E.,  1991, \mn@doi [\apjl] {10.1086/185978},
  \href {https://ui.adsabs.harvard.edu/abs/1991ApJ...370L..65B} {370, L65}

\bibitem[\protect\citeauthoryear{{Baron}, {Netzer}, {Lutz}, {Prochaska}  \&
  {Davies}}{{Baron} et~al.}{2022}]{Baron2022}
{Baron} D.,  {Netzer} H.,  {Lutz} D.,  {Prochaska} J.~X.,   {Davies} R.~I.,
  2022, \mn@doi [\mnras] {10.1093/mnras/stab3232}, \href
  {https://ui.adsabs.harvard.edu/abs/2022MNRAS.509.4457B} {509, 4457}

\bibitem[\protect\citeauthoryear{{Baron}, {Netzer}, {French}, {Lutz}, {Davies}
  \& {Prochaska}}{{Baron} et~al.}{2023}]{Baron2023}
{Baron} D.,  {Netzer} H.,  {French} K.~D.,  {Lutz} D.,  {Davies} R.~I.,
  {Prochaska} J.~X.,  2023, \mn@doi [\mnras] {10.1093/mnras/stad1792}, \href
  {https://ui.adsabs.harvard.edu/abs/2023MNRAS.524.2741B} {524, 2741}

\bibitem[\protect\citeauthoryear{{Benavides} et~al.,}{{Benavides}
  et~al.}{2021}]{Benavides2021}
{Benavides} J.~A.,  et~al., 2021, \mn@doi [Nature Astronomy]
  {10.1038/s41550-021-01458-1}, \href
  {https://ui.adsabs.harvard.edu/abs/2021NatAs...5.1255B} {5, 1255}

\bibitem[\protect\citeauthoryear{{Bergvall}, {Laurikainen}  \&
  {Aalto}}{{Bergvall} et~al.}{2003}]{Bergvall2003c}
{Bergvall} N.,  {Laurikainen} E.,   {Aalto} S.,  2003, \mn@doi [\aap]
  {10.1051/0004-6361:20030542}, \href
  {https://ui.adsabs.harvard.edu/abs/2003A&A...405...31B} {405, 31}

\bibitem[\protect\citeauthoryear{{Bergvall}, {Marquart}, {Way}, {Blomqvist},
  {Holst}, {{\"O}stlin}  \& {Zackrisson}}{{Bergvall}
  et~al.}{2016}]{Bergvall2016}
{Bergvall} N.,  {Marquart} T.,  {Way} M.~J.,  {Blomqvist} A.,  {Holst} E.,
  {{\"O}stlin} G.,   {Zackrisson} E.,  2016, \mn@doi [\aap]
  {10.1051/0004-6361/201525692}, \href
  {https://ui.adsabs.harvard.edu/abs/2016A&A...587A..72B} {587, A72}

\bibitem[\protect\citeauthoryear{{Bernardini}, {Feldmann},
  {Angl{\'e}s-Alc{\'a}zar}, {Boylan-Kolchin}, {Bullock}, {Mayer}  \&
  {Stadel}}{{Bernardini} et~al.}{2022}]{Bernardini2022}
{Bernardini} M.,  {Feldmann} R.,  {Angl{\'e}s-Alc{\'a}zar} D.,
  {Boylan-Kolchin} M.,  {Bullock} J.,  {Mayer} L.,   {Stadel} J.,  2022,
  \mn@doi [\mnras] {10.1093/mnras/stab3088}, \href
  {https://ui.adsabs.harvard.edu/abs/2022MNRAS.509.1323B} {509, 1323}

\bibitem[\protect\citeauthoryear{{Bisigello}, {Caputi}, {Grogin}  \&
  {Koekemoer}}{{Bisigello} et~al.}{2018}]{Bisigello2018}
{Bisigello} L.,  {Caputi} K.~I.,  {Grogin} N.,   {Koekemoer} A.,  2018, \mn@doi
  [\aap] {10.1051/0004-6361/201731399}, \href
  {https://ui.adsabs.harvard.edu/abs/2018A&A...609A..82B} {609, A82}

\bibitem[\protect\citeauthoryear{{Blake} et~al.,}{{Blake}
  et~al.}{2004}]{Blake2004}
{Blake} C.,  et~al., 2004, \mn@doi [\mnras] {10.1111/j.1365-2966.2004.08351.x},
  \href {https://ui.adsabs.harvard.edu/abs/2004MNRAS.355..713B} {355, 713}

\bibitem[\protect\citeauthoryear{{Brinchmann}, {Charlot}, {White}, {Tremonti},
  {Kauffmann}, {Heckman}  \& {Brinkmann}}{{Brinchmann}
  et~al.}{2004}]{Brinchmann2004}
{Brinchmann} J.,  {Charlot} S.,  {White} S.~D.~M.,  {Tremonti} C.,  {Kauffmann}
  G.,  {Heckman} T.,   {Brinkmann} J.,  2004, \mn@doi [\mnras]
  {10.1111/j.1365-2966.2004.07881.x}, \href
  {https://ui.adsabs.harvard.edu/abs/2004MNRAS.351.1151B} {351, 1151}

\bibitem[\protect\citeauthoryear{Bryan \& Norman}{Bryan \&
  Norman}{1998}]{Bryan&Norman1998}
Bryan G.~L.,  Norman M.~L.,  1998, \mn@doi [The Astrophysical Journal]
  {10.1086/305262}, 495, 80

\bibitem[\protect\citeauthoryear{{Calzetti}}{{Calzetti}}{2013}]{Calzetti2013}
{Calzetti} D.,  2013, in {Falc{\'o}n-Barroso} J.,  {Knapen} J.~H.,  eds, ,
  Secular Evolution of Galaxies.
p.~419

\bibitem[\protect\citeauthoryear{{Capelo}, {Volonteri}, {Dotti}, {Bellovary},
  {Mayer}  \& {Governato}}{{Capelo} et~al.}{2015}]{Capelo2015}
{Capelo} P.~R.,  {Volonteri} M.,  {Dotti} M.,  {Bellovary} J.~M.,  {Mayer} L.,
   {Governato} F.,  2015, \mn@doi [\mnras] {10.1093/mnras/stu2500}, \href
  {https://ui.adsabs.harvard.edu/abs/2015MNRAS.447.2123C} {447, 2123}

\bibitem[\protect\citeauthoryear{{Caputi} et~al.,}{{Caputi}
  et~al.}{2017}]{Caputi2017}
{Caputi} K.~I.,  et~al., 2017, \mn@doi [\apj] {10.3847/1538-4357/aa901e}, \href
  {https://ui.adsabs.harvard.edu/abs/2017ApJ...849...45C} {849, 45}

\bibitem[\protect\citeauthoryear{{Casasola}, {Bettoni}  \&
  {Galletta}}{{Casasola} et~al.}{2004}]{Casasola2004}
{Casasola} V.,  {Bettoni} D.,   {Galletta} G.,  2004, \mn@doi [\aap]
  {10.1051/0004-6361:20040283}, \href
  {https://ui.adsabs.harvard.edu/abs/2004A&A...422..941C} {422, 941}

\bibitem[\protect\citeauthoryear{{Christensen}, {Dav{\'e}}, {Governato},
  {Pontzen}, {Brooks}, {Munshi}, {Quinn}  \& {Wadsley}}{{Christensen}
  et~al.}{2016}]{Christensen2016}
{Christensen} C.~R.,  {Dav{\'e}} R.,  {Governato} F.,  {Pontzen} A.,  {Brooks}
  A.,  {Munshi} F.,  {Quinn} T.,   {Wadsley} J.,  2016, \mn@doi [\apj]
  {10.3847/0004-637X/824/1/5710.48550/arXiv.1508.00007}, \href
  {https://ui.adsabs.harvard.edu/abs/2016ApJ...824...57C} {824, 57}

\bibitem[\protect\citeauthoryear{{Combes}, {Prugniel}, {Rampazzo}  \&
  {Sulentic}}{{Combes} et~al.}{1994}]{Combes1994}
{Combes} F.,  {Prugniel} P.,  {Rampazzo} R.,   {Sulentic} J.~W.,  1994, \aap,
  \href {https://ui.adsabs.harvard.edu/abs/1994A&A...281..725C} {281, 725}

\bibitem[\protect\citeauthoryear{{Cox}, {Jonsson}, {Somerville}, {Primack}  \&
  {Dekel}}{{Cox} et~al.}{2008}]{Cox2008}
{Cox} T.~J.,  {Jonsson} P.,  {Somerville} R.~S.,  {Primack} J.~R.,   {Dekel}
  A.,  2008, \mn@doi [\mnras] {10.1111/j.1365-2966.2007.12730.x}, \href
  {https://ui.adsabs.harvard.edu/abs/2008MNRAS.384..386C} {384, 386}

\bibitem[\protect\citeauthoryear{{Daddi} et~al.,}{{Daddi}
  et~al.}{2007}]{Daddi2007}
{Daddi} E.,  et~al., 2007, \mn@doi [\apj] {10.1086/521818}, \href
  {https://ui.adsabs.harvard.edu/abs/2007ApJ...670..156D} {670, 156}

\bibitem[\protect\citeauthoryear{{Daddi} et~al.,}{{Daddi}
  et~al.}{2010a}]{Daddi2010a}
{Daddi} E.,  et~al., 2010a, \mn@doi [\apj] {10.1088/0004-637X/713/1/686}, \href
  {https://ui.adsabs.harvard.edu/abs/2010ApJ...713..686D} {713, 686}

\bibitem[\protect\citeauthoryear{{Daddi} et~al.,}{{Daddi}
  et~al.}{2010b}]{Daddi2010b}
{Daddi} E.,  et~al., 2010b, \mn@doi [\apjl] {10.1088/2041-8205/714/1/L118},
  \href {https://ui.adsabs.harvard.edu/abs/2010ApJ...714L.118D} {714, L118}

\bibitem[\protect\citeauthoryear{{Daddi} et~al.,}{{Daddi}
  et~al.}{2022}]{Daddi2022}
{Daddi} E.,  et~al., 2022, \mn@doi [\aap] {10.1051/0004-6361/202243574}, \href
  {https://ui.adsabs.harvard.edu/abs/2022A&A...661L...7D} {661, L7}

\bibitem[\protect\citeauthoryear{{Danovich}, {Dekel}, {Hahn}, {Ceverino}  \&
  {Primack}}{{Danovich} et~al.}{2015}]{Danovich2015}
{Danovich} M.,  {Dekel} A.,  {Hahn} O.,  {Ceverino} D.,   {Primack} J.,  2015,
  \mn@doi [\mnras] {10.1093/mnras/stv270}, \href
  {https://ui.adsabs.harvard.edu/abs/2015MNRAS.449.2087D} {449, 2087}

\bibitem[\protect\citeauthoryear{{Dekel} \& {Burkert}}{{Dekel} \&
  {Burkert}}{2014}]{Dekel&Burkert2014}
{Dekel} A.,  {Burkert} A.,  2014, \mn@doi [\mnras] {10.1093/mnras/stt2331},
  \href {https://ui.adsabs.harvard.edu/abs/2014MNRAS.438.1870D} {438, 1870}

\bibitem[\protect\citeauthoryear{{Di Matteo}, {Combes}, {Melchior}  \&
  {Semelin}}{{Di Matteo} et~al.}{2007}]{Di_Matteo2007}
{Di Matteo} P.,  {Combes} F.,  {Melchior} A.~L.,   {Semelin} B.,  2007, \mn@doi
  [\aap] {10.1051/0004-6361:20066959}, \href
  {https://ui.adsabs.harvard.edu/abs/2007A&A...468...61D} {468, 61}

\bibitem[\protect\citeauthoryear{{Di Matteo}, {Bournaud}, {Martig}, {Combes},
  {Melchior}  \& {Semelin}}{{Di Matteo} et~al.}{2008}]{Di_Matteo2008}
{Di Matteo} P.,  {Bournaud} F.,  {Martig} M.,  {Combes} F.,  {Melchior} A.~L.,
   {Semelin} B.,  2008, \mn@doi [\aap] {10.1051/0004-6361:200809480}, \href
  {https://ui.adsabs.harvard.edu/abs/2008A&A...492...31D} {492, 31}

\bibitem[\protect\citeauthoryear{{D{\'\i}az-Garc{\'\i}a} \&
  {Knapen}}{{D{\'\i}az-Garc{\'\i}a} \& {Knapen}}{2020}]{Diaz-Garcia&Knapen2020}
{D{\'\i}az-Garc{\'\i}a} S.,  {Knapen} J.~H.,  2020, \mn@doi [\aap]
  {10.1051/0004-6361/201937384}, \href
  {https://ui.adsabs.harvard.edu/abs/2020A&A...635A.197D} {635, A197}

\bibitem[\protect\citeauthoryear{{El-Badry}, {Wetzel}, {Geha}, {Hopkins},
  {Kere{\v{s}}}, {Chan}  \& {Faucher-Gigu{\`e}re}}{{El-Badry}
  et~al.}{2016}]{El-Badry2016}
{El-Badry} K.,  {Wetzel} A.,  {Geha} M.,  {Hopkins} P.~F.,  {Kere{\v{s}}} D.,
  {Chan} T.~K.,   {Faucher-Gigu{\`e}re} C.-A.,  2016, \mn@doi [\apj]
  {10.3847/0004-637X/820/2/131}, \href
  {https://ui.adsabs.harvard.edu/abs/2016ApJ...820..131E} {820, 131}

\bibitem[\protect\citeauthoryear{{Ellison}, {Mendel}, {Patton}  \&
  {Scudder}}{{Ellison} et~al.}{2013}]{Ellison2013}
{Ellison} S.~L.,  {Mendel} J.~T.,  {Patton} D.~R.,   {Scudder} J.~M.,  2013,
  \mn@doi [\mnras] {10.1093/mnras/stt1562}, \href
  {https://ui.adsabs.harvard.edu/abs/2013MNRAS.435.3627E} {435, 3627}

\bibitem[\protect\citeauthoryear{{Ellison}, {S{\'a}nchez}, {Ibarra-Medel},
  {Antonio}, {Mendel}  \& {Barrera-Ballesteros}}{{Ellison}
  et~al.}{2018}]{Ellison2018}
{Ellison} S.~L.,  {S{\'a}nchez} S.~F.,  {Ibarra-Medel} H.,  {Antonio} B.,
  {Mendel} J.~T.,   {Barrera-Ballesteros} J.,  2018, \mn@doi [\mnras]
  {10.1093/mnras/stx2882}, \href
  {https://ui.adsabs.harvard.edu/abs/2018MNRAS.474.2039E} {474, 2039}

\bibitem[\protect\citeauthoryear{{Ellison}, {Thorp}, {Pan}, {Lin}, {Scudder},
  {Bluck}, {S{\'a}nchez}  \& {Sargent}}{{Ellison} et~al.}{2020}]{Ellison2020}
{Ellison} S.~L.,  {Thorp} M.~D.,  {Pan} H.-A.,  {Lin} L.,  {Scudder} J.~M.,
  {Bluck} A. F.~L.,  {S{\'a}nchez} S.~F.,   {Sargent} M.,  2020, \mn@doi
  [\mnras] {10.1093/mnras/staa001}, \href
  {https://ui.adsabs.harvard.edu/abs/2020MNRAS.492.6027E} {492, 6027}

\bibitem[\protect\citeauthoryear{{Emsellem} et~al.,}{{Emsellem}
  et~al.}{2022}]{Emsellem2022}
{Emsellem} E.,  et~al., 2022, \mn@doi [\aap] {10.1051/0004-6361/202141727},
  \href {https://ui.adsabs.harvard.edu/abs/2022A&A...659A.191E} {659, A191}

\bibitem[\protect\citeauthoryear{{Feldmann}}{{Feldmann}}{2020}]{Feldmann2020}
{Feldmann} R.,  2020, \mn@doi [Communications Physics]
  {10.1038/s42005-020-00493-0}, \href
  {https://ui.adsabs.harvard.edu/abs/2020CmPhy...3..226F} {3, 226}

\bibitem[\protect\citeauthoryear{{Feldmann} et~al.,}{{Feldmann}
  et~al.}{2023}]{Feldmann2023}
{Feldmann} R.,  et~al., 2023, \mn@doi [\mnras] {10.1093/mnras/stad1205}, \href
  {https://ui.adsabs.harvard.edu/abs/2023MNRAS.522.3831F} {522, 3831}

\bibitem[\protect\citeauthoryear{{Fensch} et~al.,}{{Fensch}
  et~al.}{2017}]{Fensch2017}
{Fensch} J.,  et~al., 2017, \mn@doi [\mnras] {10.1093/mnras/stw2920}, \href
  {https://ui.adsabs.harvard.edu/abs/2017MNRAS.465.1934F} {465, 1934}

\bibitem[\protect\citeauthoryear{Fisher}{Fisher}{1936}]{Fisher1936}
Fisher R.~A.,  1936, \mn@doi [Annals of Eugenics]
  {https://doi.org/10.1111/j.1469-1809.1936.tb02137.x}, 7, 179

\bibitem[\protect\citeauthoryear{{Fisher} et~al.,}{{Fisher}
  et~al.}{2014}]{Fisher2014}
{Fisher} D.~B.,  et~al., 2014, \mn@doi [\apjl] {10.1088/2041-8205/790/2/L30},
  \href {https://ui.adsabs.harvard.edu/abs/2014ApJ...790L..30F} {790, L30}

\bibitem[\protect\citeauthoryear{{Fisher} et~al.,}{{Fisher}
  et~al.}{2017a}]{Fisher2017a}
{Fisher} D.~B.,  et~al., 2017a, \mn@doi [\mnras] {10.1093/mnras/stw2281}, \href
  {https://ui.adsabs.harvard.edu/abs/2017MNRAS.464..491F} {464, 491}

\bibitem[\protect\citeauthoryear{{Fisher} et~al.,}{{Fisher}
  et~al.}{2017b}]{Fisher2017b}
{Fisher} D.~B.,  et~al., 2017b, \mn@doi [\apjl] {10.3847/2041-8213/aa6478},
  \href {https://ui.adsabs.harvard.edu/abs/2017ApJ...839L...5F} {839, L5}

\bibitem[\protect\citeauthoryear{{Fisher}, {Bolatto}, {Glazebrook},
  {Obreschkow}, {Abraham}, {Kacprzak}  \& {Nielsen}}{{Fisher}
  et~al.}{2022}]{Fisher2022}
{Fisher} D.~B.,  {Bolatto} A.~D.,  {Glazebrook} K.,  {Obreschkow} D.,
  {Abraham} R.~G.,  {Kacprzak} G.~G.,   {Nielsen} N.~M.,  2022, \mn@doi [\apj]
  {10.3847/1538-4357/ac51c8}, \href
  {https://ui.adsabs.harvard.edu/abs/2022ApJ...928..169F} {928, 169}

\bibitem[\protect\citeauthoryear{{Flores Vel{\'a}zquez} et~al.,}{{Flores
  Vel{\'a}zquez} et~al.}{2021}]{Flores-Velazquez2021}
{Flores Vel{\'a}zquez} J.~A.,  et~al., 2021, \mn@doi [\mnras]
  {10.1093/mnras/staa3893}, \href
  {https://ui.adsabs.harvard.edu/abs/2021MNRAS.501.4812F} {501, 4812}

\bibitem[\protect\citeauthoryear{{French}}{{French}}{2021}]{French2021}
{French} K.~D.,  2021, \mn@doi [\pasp] {10.1088/1538-3873/ac0a59}, \href
  {https://ui.adsabs.harvard.edu/abs/2021PASP..133g2001F} {133, 072001}

\bibitem[\protect\citeauthoryear{{Gao} \& {Solomon}}{{Gao} \&
  {Solomon}}{2004}]{Gao&Solomon2004}
{Gao} Y.,  {Solomon} P.~M.,  2004, \mn@doi [\apj] {10.1086/382999}, \href
  {https://ui.adsabs.harvard.edu/abs/2004ApJ...606..271G} {606, 271}

\bibitem[\protect\citeauthoryear{{Garay-Solis}, {Barrera-Ballesteros},
  {Colombo}, {S{\'a}nchez}, {Lugo-Aranda}, {Villanueva}, {Wong}  \&
  {Bolatto}}{{Garay-Solis} et~al.}{2023}]{Garay-Solis2023}
{Garay-Solis} Y.,  {Barrera-Ballesteros} J.~K.,  {Colombo} D.,  {S{\'a}nchez}
  S.~F.,  {Lugo-Aranda} A.~Z.,  {Villanueva} V.,  {Wong} T.,   {Bolatto} A.~D.,
   2023, \mn@doi [\apj] {10.3847/1538-4357/acd781}, \href
  {https://ui.adsabs.harvard.edu/abs/2023ApJ...952..122G} {952, 122}

\bibitem[\protect\citeauthoryear{{Garc{\'\i}a-Burillo}, {Usero},
  {Alonso-Herrero}, {Graci{\'a}-Carpio}, {Pereira-Santaella}, {Colina},
  {Planesas}  \& {Arribas}}{{Garc{\'\i}a-Burillo}
  et~al.}{2012}]{Garcia-Burillo2012}
{Garc{\'\i}a-Burillo} S.,  {Usero} A.,  {Alonso-Herrero} A.,
  {Graci{\'a}-Carpio} J.,  {Pereira-Santaella} M.,  {Colina} L.,  {Planesas}
  P.,   {Arribas} S.,  2012, \mn@doi [\aap] {10.1051/0004-6361/201117838},
  \href {https://ui.adsabs.harvard.edu/abs/2012A&A...539A...8G} {539, A8}

\bibitem[\protect\citeauthoryear{{Gardner} et~al.,}{{Gardner}
  et~al.}{2006}]{Gardner2006}
{Gardner} J.~P.,  et~al., 2006, \mn@doi [\ssr] {10.1007/s11214-006-8315-7},
  \href {https://ui.adsabs.harvard.edu/abs/2006SSRv..123..485G} {123, 485}

\bibitem[\protect\citeauthoryear{{Gensior}, {Feldmann}, {Mayer}, {Wetzel},
  {Hopkins}  \& {Faucher-Gigu{\`e}re}}{{Gensior} et~al.}{2023}]{Gensior2023}
{Gensior} J.,  {Feldmann} R.,  {Mayer} L.,  {Wetzel} A.,  {Hopkins} P.~F.,
  {Faucher-Gigu{\`e}re} C.-A.,  2023, \mn@doi [\mnras]
  {10.1093/mnrasl/slac138}, \href
  {https://ui.adsabs.harvard.edu/abs/2023MNRAS.518L..63G} {518, L63}

\bibitem[\protect\citeauthoryear{{Genzel} et~al.,}{{Genzel}
  et~al.}{2010}]{Genzel2010}
{Genzel} R.,  et~al., 2010, \mn@doi [\mnras]
  {10.1111/j.1365-2966.2010.16969.x}, \href
  {https://ui.adsabs.harvard.edu/abs/2010MNRAS.407.2091G} {407, 2091}

\bibitem[\protect\citeauthoryear{{Gill}, {Knebe}  \& {Gibson}}{{Gill}
  et~al.}{2004}]{Gill2004}
{Gill} S. P.~D.,  {Knebe} A.,   {Gibson} B.~K.,  2004, \mn@doi [\mnras]
  {10.1111/j.1365-2966.2004.07786.x}, \href
  {https://ui.adsabs.harvard.edu/abs/2004MNRAS.351..399G} {351, 399}

\bibitem[\protect\citeauthoryear{{Goto}}{{Goto}}{2005}]{Goto2005}
{Goto} T.,  2005, \mn@doi [\mnras] {10.1111/j.1365-2966.2005.08701.x}, \href
  {https://ui.adsabs.harvard.edu/abs/2005MNRAS.357..937G} {357, 937}

\bibitem[\protect\citeauthoryear{{Haggar}, {Pearce}, {Gray}, {Knebe}  \&
  {Yepes}}{{Haggar} et~al.}{2021}]{Haggar2021}
{Haggar} R.,  {Pearce} F.~R.,  {Gray} M.~E.,  {Knebe} A.,   {Yepes} G.,  2021,
  \mn@doi [\mnras] {10.1093/mnras/stab064}, \href
  {https://ui.adsabs.harvard.edu/abs/2021MNRAS.502.1191H} {502, 1191}

\bibitem[\protect\citeauthoryear{Hahn \& Abel}{Hahn \& Abel}{2011}]{Hahn2011}
Hahn O.,  Abel T.,  2011, \mn@doi [Monthly Notices of the Royal Astronomical
  Society] {10.1111/j.1365-2966.2011.18820.x}, 415, 2101

\bibitem[\protect\citeauthoryear{{Hayes} et~al.,}{{Hayes}
  et~al.}{2014}]{Hayes2014}
{Hayes} M.,  et~al., 2014, \mn@doi [\apj] {10.1088/0004-637X/782/1/6}, \href
  {https://ui.adsabs.harvard.edu/abs/2014ApJ...782....6H} {782, 6}

\bibitem[\protect\citeauthoryear{{He}, {Bottrell}, {Wilson}, {Moreno},
  {Burkhart}, {Hayward}, {Hernquist}  \& {Twum}}{{He} et~al.}{2023}]{He2023}
{He} H.,  {Bottrell} C.,  {Wilson} C.,  {Moreno} J.,  {Burkhart} B.,  {Hayward}
  C.~C.,  {Hernquist} L.,   {Twum} A.,  2023, \mn@doi [\apj]
  {10.3847/1538-4357/acca76}, \href
  {https://ui.adsabs.harvard.edu/abs/2023ApJ...950...56H} {950, 56}

\bibitem[\protect\citeauthoryear{Ho}{Ho}{1995}]{Ho1995}
Ho T.~K.,  1995, in Proceedings of 3rd International Conference on Document
  Analysis and Recognition. pp 278--282 vol.1,
  \mn@doi{10.1109/ICDAR.1995.598994}

\bibitem[\protect\citeauthoryear{Hopkins}{Hopkins}{2015}]{Hopkins2015}
Hopkins P.~F.,  2015, \mn@doi [Monthly Notices of the Royal Astronomical
  Society] {10.1093/mnras/stv195}

\bibitem[\protect\citeauthoryear{Hopkins, Quataert  \& Murray}{Hopkins
  et~al.}{2012}]{Hopkins2012a}
Hopkins P.~F.,  Quataert E.,   Murray N.,  2012, \mn@doi [Monthly Notices of
  the Royal Astronomical Society] {10.1111/J.1365-2966.2012.20578.X}, 421, 3488

\bibitem[\protect\citeauthoryear{Hopkins, Kere{\v{s}}, O{\~{n}}orbe,
  Faucher-Gigu{\`{e}}re, Quataert, Murray  \& Bullock}{Hopkins
  et~al.}{2014}]{Hopkins2014a}
Hopkins P.~F.,  Kere{\v{s}} D.,  O{\~{n}}orbe J.,  Faucher-Gigu{\`{e}}re C.~A.,
   Quataert E.,  Murray N.,   Bullock J.~S.,  2014, \mn@doi [Monthly Notices of
  the Royal Astronomical Society] {10.1093/mnras/stu1738}

\bibitem[\protect\citeauthoryear{Hopkins et~al.,}{Hopkins
  et~al.}{2018}]{Hopkins2018}
Hopkins P.~F.,  et~al., 2018, \mn@doi [Monthly Notices of the Royal
  Astronomical Society] {10.1093/mnras/sty1690}, 480, 800

\bibitem[\protect\citeauthoryear{{Hopkins} et~al.,}{{Hopkins}
  et~al.}{2023}]{Hopkins2023}
{Hopkins} P.~F.,  et~al., 2023, \mn@doi [\mnras] {10.1093/mnras/stad1902},
  \href {https://ui.adsabs.harvard.edu/abs/2023MNRAS.tmp.1847H} {}

\bibitem[\protect\citeauthoryear{{Hung} et~al.,}{{Hung}
  et~al.}{2013}]{Hung2013}
{Hung} C.-L.,  et~al., 2013, \mn@doi [\apj] {10.1088/0004-637X/778/2/129},
  \href {https://ui.adsabs.harvard.edu/abs/2013ApJ...778..129H} {778, 129}

\bibitem[\protect\citeauthoryear{Hunter}{Hunter}{2007}]{Hunter2007}
Hunter J.~D.,  2007, \mn@doi [Computing in Science and Engineering]
  {10.1109/MCSE.2007.55}, 9, 90

\bibitem[\protect\citeauthoryear{{Jackson}, {Martin}, {Kaviraj}, {Laigle},
  {Devriendt}, {Dubois}  \& {Pichon}}{{Jackson} et~al.}{2019}]{Jackson2019}
{Jackson} R.~A.,  {Martin} G.,  {Kaviraj} S.,  {Laigle} C.,  {Devriendt}
  J.~E.~G.,  {Dubois} Y.,   {Pichon} C.,  2019, \mn@doi [\mnras]
  {10.1093/mnras/stz2440}, \href
  {https://ui.adsabs.harvard.edu/abs/2019MNRAS.489.4679J} {489, 4679}

\bibitem[\protect\citeauthoryear{Kennicutt}{Kennicutt}{1998}]{KennicuttJr.1998}
Kennicutt Jr. R.~C.,  1998, \mn@doi [The Astrophysical Journal]
  {10.1086/305588}, 498, 541

\bibitem[\protect\citeauthoryear{{Kennicutt} \& {De Los Reyes}}{{Kennicutt} \&
  {De Los Reyes}}{2021}]{Kennicutt2021}
{Kennicutt} Robert~C. J.,  {De Los Reyes} M. A.~C.,  2021, \mn@doi [\apj]
  {10.3847/1538-4357/abd3a2}, \href
  {https://ui.adsabs.harvard.edu/abs/2021ApJ...908...61K} {908, 61}

\bibitem[\protect\citeauthoryear{{Kennicutt} \& {Evans}}{{Kennicutt} \&
  {Evans}}{2012}]{Kennicutt&Evans2012}
{Kennicutt} R.~C.,  {Evans} N.~J.,  2012, \mn@doi [\araa]
  {10.1146/annurev-astro-081811-125610}, \href
  {https://ui.adsabs.harvard.edu/abs/2012ARA&A..50..531K} {50, 531}

\bibitem[\protect\citeauthoryear{Knollmann \& Knebe}{Knollmann \&
  Knebe}{2009}]{Knollmann2009}
Knollmann S.~R.,  Knebe A.,  2009, \mn@doi [Astrophysical Journal, Supplement
  Series] {10.1088/0067-0049/182/2/608}, 182, 608

\bibitem[\protect\citeauthoryear{Kriek et~al.,}{Kriek et~al.}{2010}]{Kriek2010}
Kriek M.,  et~al., 2010, \mn@doi [The Astrophysical Journal Letters]
  {10.1088/2041-8205/722/1/L64}, 722, 64

\bibitem[\protect\citeauthoryear{Krumholz \& Gnedin}{Krumholz \&
  Gnedin}{2011}]{Krumholz&Gnedin2011}
Krumholz M.~R.,  Gnedin N.~Y.,  2011, \mn@doi [Astrophys. J.]
  {10.1088/0004-637X/729/1/36}, 729, 36

\bibitem[\protect\citeauthoryear{Krumholz, McKee  \& Tumlinson}{Krumholz
  et~al.}{2008}]{Krumholz2008}
Krumholz M.~R.,  McKee C.~F.,   Tumlinson J.,  2008, \mn@doi [Astrophys. J.]
  {10.1086/592490}, 689, 865

\bibitem[\protect\citeauthoryear{{Krumholz}, {Dekel}  \& {McKee}}{{Krumholz}
  et~al.}{2012}]{Krumholz2012}
{Krumholz} M.~R.,  {Dekel} A.,   {McKee} C.~F.,  2012, \mn@doi [\apj]
  {10.1088/0004-637X/745/1/69}, \href
  {https://ui.adsabs.harvard.edu/abs/2012ApJ...745...69K} {745, 69}

\bibitem[\protect\citeauthoryear{{Lapiner} et~al.,}{{Lapiner}
  et~al.}{2023}]{Lapiner2023}
{Lapiner} S.,  et~al., 2023, \mn@doi [\mnras] {10.1093/mnras/stad1263}, \href
  {https://ui.adsabs.harvard.edu/abs/2023MNRAS.522.4515L} {522, 4515}

\bibitem[\protect\citeauthoryear{{Lee} et~al.,}{{Lee} et~al.}{2015}]{Lee2015}
{Lee} N.,  et~al., 2015, \mn@doi [\apj] {10.1088/0004-637X/801/2/80}, \href
  {https://ui.adsabs.harvard.edu/abs/2015ApJ...801...80L} {801, 80}

\bibitem[\protect\citeauthoryear{{Leroy}, {Walter}, {Brinks}, {Bigiel}, {de
  Blok}, {Madore}  \& {Thornley}}{{Leroy} et~al.}{2008}]{Leroy2008}
{Leroy} A.~K.,  {Walter} F.,  {Brinks} E.,  {Bigiel} F.,  {de Blok} W.~J.~G.,
  {Madore} B.,   {Thornley} M.~D.,  2008, \mn@doi [\aj]
  {10.1088/0004-6256/136/6/2782}, \href
  {https://ui.adsabs.harvard.edu/abs/2008AJ....136.2782L} {136, 2782}

\bibitem[\protect\citeauthoryear{{Leroy} et~al.,}{{Leroy}
  et~al.}{2021}]{Leroy2021}
{Leroy} A.~K.,  et~al., 2021, \mn@doi [\apjs] {10.3847/1538-4365/ac17f3}, \href
  {https://ui.adsabs.harvard.edu/abs/2021ApJS..257...43L} {257, 43}

\bibitem[\protect\citeauthoryear{{Leslie} et~al.,}{{Leslie}
  et~al.}{2020}]{Leslie2020}
{Leslie} S.~K.,  et~al., 2020, \mn@doi [\apj] {10.3847/1538-4357/aba044}, \href
  {https://ui.adsabs.harvard.edu/abs/2020ApJ...899...58L} {899, 58}

\bibitem[\protect\citeauthoryear{Lewis, Challinor  \& Lasenby}{Lewis
  et~al.}{2000}]{Lewis2000}
Lewis A.,  Challinor A.,   Lasenby A.,  2000, \mn@doi [The Astrophysical
  Journal] {10.1086/309179}, 538, 473

\bibitem[\protect\citeauthoryear{Lewis, Challinor  \& Hanson}{Lewis
  et~al.}{2011}]{Lewis2011}
Lewis A.,  Challinor A.,   Hanson D.,  2011, \mn@doi [Journal of Cosmology and
  Astroparticle Physics] {10.1088/1475-7516/2011/03/018}, 2011

\bibitem[\protect\citeauthoryear{{Li}, {Ho}  \& {Shangguan}}{{Li}
  et~al.}{2023}]{Li2023}
{Li} Y.~A.,  {Ho} L.~C.,   {Shangguan} J.,  2023, \mn@doi [arXiv e-prints]
  {10.48550/arXiv.2307.13462}, \href
  {https://ui.adsabs.harvard.edu/abs/2023arXiv230713462L} {p. arXiv:2307.13462}

\bibitem[\protect\citeauthoryear{{Lotz}, {Jonsson}, {Cox}  \& {Primack}}{{Lotz}
  et~al.}{2008}]{Lotz2008}
{Lotz} J.~M.,  {Jonsson} P.,  {Cox} T.~J.,   {Primack} J.~R.,  2008, \mn@doi
  [\mnras] {10.1111/j.1365-2966.2008.14004.x}, \href
  {https://ui.adsabs.harvard.edu/abs/2008MNRAS.391.1137L} {391, 1137}

\bibitem[\protect\citeauthoryear{Ma et~al.,}{Ma et~al.}{2018a}]{Ma2018a}
Ma X.,  et~al., 2018a, \mn@doi [Monthly Notices of the Royal Astronomical
  Society] {10.1093/mnras/sty684}, 477, 219

\bibitem[\protect\citeauthoryear{Ma et~al.,}{Ma et~al.}{2018b}]{Ma2018}
Ma X.,  et~al., 2018b, \mn@doi [Monthly Notices of the Royal Astronomical
  Society] {10.1093/mnras/sty1024}, 478, 1694

\bibitem[\protect\citeauthoryear{McKee \& Krumholz}{McKee \&
  Krumholz}{2010}]{McKee&Krumholz2010}
McKee C.~F.,  Krumholz M.~R.,  2010, \mn@doi [Astrophys. J.]
  {10.1088/0004-637X/709/1/308}, 709, 308

\bibitem[\protect\citeauthoryear{{Michiyama} et~al.,}{{Michiyama}
  et~al.}{2016}]{Michiyama2016}
{Michiyama} T.,  et~al., 2016, \mn@doi [\pasj] {10.1093/pasj/psw087}, \href
  {https://ui.adsabs.harvard.edu/abs/2016PASJ...68...96M} {68, 96}

\bibitem[\protect\citeauthoryear{{Moreno}}{{Moreno}}{2012}]{Moreno2012}
{Moreno} J.,  2012, \mn@doi [\mnras] {10.1111/j.1365-2966.2011.19706.x}, \href
  {https://ui.adsabs.harvard.edu/abs/2012MNRAS.419..411M} {419, 411}

\bibitem[\protect\citeauthoryear{{Moreno}, {Bluck}, {Ellison}, {Patton},
  {Torrey}  \& {Moster}}{{Moreno} et~al.}{2013}]{Moreno2013}
{Moreno} J.,  {Bluck} A. F.~L.,  {Ellison} S.~L.,  {Patton} D.~R.,  {Torrey}
  P.,   {Moster} B.~P.,  2013, \mn@doi [\mnras] {10.1093/mnras/stt1694}, \href
  {https://ui.adsabs.harvard.edu/abs/2013MNRAS.436.1765M} {436, 1765}

\bibitem[\protect\citeauthoryear{{Moreno}, {Torrey}, {Ellison}, {Patton},
  {Bluck}, {Bansal}  \& {Hernquist}}{{Moreno} et~al.}{2015}]{Moreno2015}
{Moreno} J.,  {Torrey} P.,  {Ellison} S.~L.,  {Patton} D.~R.,  {Bluck} A.
  F.~L.,  {Bansal} G.,   {Hernquist} L.,  2015, \mn@doi [\mnras]
  {10.1093/mnras/stv094}, \href
  {https://ui.adsabs.harvard.edu/abs/2015MNRAS.448.1107M} {448, 1107}

\bibitem[\protect\citeauthoryear{{Moreno} et~al.,}{{Moreno}
  et~al.}{2019}]{Moreno2019}
{Moreno} J.,  et~al., 2019, \mn@doi [\mnras] {10.1093/mnras/stz417}, \href
  {https://ui.adsabs.harvard.edu/abs/2019MNRAS.485.1320M} {485, 1320}

\bibitem[\protect\citeauthoryear{{Moreno} et~al.,}{{Moreno}
  et~al.}{2021}]{Moreno2021}
{Moreno} J.,  et~al., 2021, \mn@doi [\mnras] {10.1093/mnras/staa2952}, \href
  {https://ui.adsabs.harvard.edu/abs/2021MNRAS.503.3113M} {503, 3113}

\bibitem[\protect\citeauthoryear{{Muratov}, {Kere{\v{s}}},
  {Faucher-Gigu{\`e}re}, {Hopkins}, {Quataert}  \& {Murray}}{{Muratov}
  et~al.}{2015}]{Muratov2015}
{Muratov} A.~L.,  {Kere{\v{s}}} D.,  {Faucher-Gigu{\`e}re} C.-A.,  {Hopkins}
  P.~F.,  {Quataert} E.,   {Murray} N.,  2015, \mn@doi [\mnras]
  {10.1093/mnras/stv2126}, \href
  {https://ui.adsabs.harvard.edu/abs/2015MNRAS.454.2691M} {454, 2691}

\bibitem[\protect\citeauthoryear{{Noeske} et~al.,}{{Noeske}
  et~al.}{2007}]{Noeske2007}
{Noeske} K.~G.,  et~al., 2007, \mn@doi [\apjl] {10.1086/517926}, \href
  {https://ui.adsabs.harvard.edu/abs/2007ApJ...660L..43N} {660, L43}

\bibitem[\protect\citeauthoryear{Orr et~al.,}{Orr et~al.}{2018}]{Orr2018}
Orr M.~E.,  et~al., 2018, \mn@doi [Monthly Notices of the Royal Astronomical
  Society] {10.1093/MNRAS/STY1241}, 478, 3653

\bibitem[\protect\citeauthoryear{{Orr} et~al.,}{{Orr} et~al.}{2021}]{Orr2021}
{Orr} M.~E.,  et~al., 2021, \mn@doi [\apjl] {10.3847/2041-8213/abdebd}, \href
  {https://ui.adsabs.harvard.edu/abs/2021ApJ...908L..31O} {908, L31}

\bibitem[\protect\citeauthoryear{{{\"O}stlin} et~al.,}{{{\"O}stlin}
  et~al.}{2014}]{Ostlin2014}
{{\"O}stlin} G.,  et~al., 2014, \mn@doi [\apj] {10.1088/0004-637X/797/1/11},
  \href {https://ui.adsabs.harvard.edu/abs/2014ApJ...797...11O} {797, 11}

\bibitem[\protect\citeauthoryear{{Pan} et~al.,}{{Pan} et~al.}{2018}]{Pan2018}
{Pan} H.-A.,  et~al., 2018, \mn@doi [\apj] {10.3847/1538-4357/aaeb92}, \href
  {https://ui.adsabs.harvard.edu/abs/2018ApJ...868..132P} {868, 132}

\bibitem[\protect\citeauthoryear{{Pandya} et~al.,}{{Pandya}
  et~al.}{2021}]{Pandya2021}
{Pandya} V.,  et~al., 2021, \mn@doi [\mnras] {10.1093/mnras/stab2714}, \href
  {https://ui.adsabs.harvard.edu/abs/2021MNRAS.508.2979P} {508, 2979}

\bibitem[\protect\citeauthoryear{{Patton}, {Torrey}, {Ellison}, {Mendel}  \&
  {Scudder}}{{Patton} et~al.}{2013}]{Patton2013}
{Patton} D.~R.,  {Torrey} P.,  {Ellison} S.~L.,  {Mendel} J.~T.,   {Scudder}
  J.~M.,  2013, \mn@doi [\mnras] {10.1093/mnrasl/slt058}, \href
  {https://ui.adsabs.harvard.edu/abs/2013MNRAS.433L..59P} {433, L59}

\bibitem[\protect\citeauthoryear{{Patton}, {Qamar}, {Ellison}, {Bluck},
  {Simard}, {Mendel}, {Moreno}  \& {Torrey}}{{Patton}
  et~al.}{2016}]{Patton2016}
{Patton} D.~R.,  {Qamar} F.~D.,  {Ellison} S.~L.,  {Bluck} A. F.~L.,  {Simard}
  L.,  {Mendel} J.~T.,  {Moreno} J.,   {Torrey} P.,  2016, \mn@doi [\mnras]
  {10.1093/mnras/stw1494}, \href
  {https://ui.adsabs.harvard.edu/abs/2016MNRAS.461.2589P} {461, 2589}

\bibitem[\protect\citeauthoryear{{Patton} et~al.,}{{Patton}
  et~al.}{2020}]{Patton2020}
{Patton} D.~R.,  et~al., 2020, \mn@doi [\mnras] {10.1093/mnras/staa913}, \href
  {https://ui.adsabs.harvard.edu/abs/2020MNRAS.494.4969P} {494, 4969}

\bibitem[\protect\citeauthoryear{{Pawlik} et~al.,}{{Pawlik}
  et~al.}{2018}]{Pawlik2018}
{Pawlik} M.~M.,  et~al., 2018, \mn@doi [\mnras] {10.1093/mnras/sty589}, \href
  {https://ui.adsabs.harvard.edu/abs/2018MNRAS.477.1708P} {477, 1708}

\bibitem[\protect\citeauthoryear{{Pearson} et~al.,}{{Pearson}
  et~al.}{2019}]{Pearson2019}
{Pearson} W.~J.,  et~al., 2019, \mn@doi [\aap] {10.1051/0004-6361/201936337},
  \href {https://ui.adsabs.harvard.edu/abs/2019A&A...631A..51P} {631, A51}

\bibitem[\protect\citeauthoryear{{Pereira-Santaella}
  et~al.,}{{Pereira-Santaella} et~al.}{2021}]{Pereira-Santaella2021}
{Pereira-Santaella} M.,  et~al., 2021, \mn@doi [\aap]
  {10.1051/0004-6361/202140955}, \href
  {https://ui.adsabs.harvard.edu/abs/2021A&A...651A..42P} {651, A42}

\bibitem[\protect\citeauthoryear{Price, Kriek, Feldmann, Quataert, Hopkins,
  Faucher-Gigu{\`{e}}re, Kere{\v{s}}  \& Barro}{Price et~al.}{2017}]{Price2017}
Price S.~H.,  Kriek M.,  Feldmann R.,  Quataert E.,  Hopkins P.~F.,
  Faucher-Gigu{\`{e}}re C.-A.,  Kere{\v{s}} D.,   Barro G.,  2017, \mn@doi
  [ApJ] {10.3847/2041-8213/aa7d4b}, 844, L6

\bibitem[\protect\citeauthoryear{{Puschnig} et~al.,}{{Puschnig}
  et~al.}{2023}]{Puschnig2023}
{Puschnig} J.,  et~al., 2023, \mn@doi [\mnras] {10.1093/mnras/stad1820}, \href
  {https://ui.adsabs.harvard.edu/abs/2023MNRAS.524.3913P} {524, 3913}

\bibitem[\protect\citeauthoryear{{Renaud}, {Bournaud}, {Kraljic}  \&
  {Duc}}{{Renaud} et~al.}{2014}]{Renaud2014}
{Renaud} F.,  {Bournaud} F.,  {Kraljic} K.,   {Duc} P.~A.,  2014, \mn@doi
  [\mnras] {10.1093/mnrasl/slu050}, \href
  {https://ui.adsabs.harvard.edu/abs/2014MNRAS.442L..33R} {442, L33}

\bibitem[\protect\citeauthoryear{{Renaud}, {Bournaud}, {Agertz}, {Kraljic},
  {Schinnerer}, {Bolatto}, {Daddi}  \& {Hughes}}{{Renaud}
  et~al.}{2019}]{Renaud2019}
{Renaud} F.,  {Bournaud} F.,  {Agertz} O.,  {Kraljic} K.,  {Schinnerer} E.,
  {Bolatto} A.,  {Daddi} E.,   {Hughes} A.,  2019, \mn@doi [\aap]
  {10.1051/0004-6361/201935222}, \href
  {https://ui.adsabs.harvard.edu/abs/2019A&A...625A..65R} {625, A65}

\bibitem[\protect\citeauthoryear{{Rinaldi}, {Caputi}, {van Mierlo}, {Ashby},
  {Caminha}  \& {Iani}}{{Rinaldi} et~al.}{2022}]{Rinaldi2022}
{Rinaldi} P.,  {Caputi} K.~I.,  {van Mierlo} S.~E.,  {Ashby} M. L.~N.,
  {Caminha} G.~B.,   {Iani} E.,  2022, \mn@doi [\apj]
  {10.3847/1538-4357/ac5d39}, \href
  {https://ui.adsabs.harvard.edu/abs/2022ApJ...930..128R} {930, 128}

\bibitem[\protect\citeauthoryear{{Rodighiero} et~al.,}{{Rodighiero}
  et~al.}{2011}]{Rodighiero2011}
{Rodighiero} G.,  et~al., 2011, \mn@doi [\apjl] {10.1088/2041-8205/739/2/L40},
  \href {https://ui.adsabs.harvard.edu/abs/2011ApJ...739L..40R} {739, L40}

\bibitem[\protect\citeauthoryear{{Rodr{\'\i}guez Montero}, {Dav{\'e}}, {Wild},
  {Angl{\'e}s-Alc{\'a}zar}  \& {Narayanan}}{{Rodr{\'\i}guez Montero}
  et~al.}{2019}]{Rodriguez-Montero2019}
{Rodr{\'\i}guez Montero} F.,  {Dav{\'e}} R.,  {Wild} V.,
  {Angl{\'e}s-Alc{\'a}zar} D.,   {Narayanan} D.,  2019, \mn@doi [\mnras]
  {10.1093/mnras/stz2580}, \href
  {https://ui.adsabs.harvard.edu/abs/2019MNRAS.490.2139R} {490, 2139}

\bibitem[\protect\citeauthoryear{{Rohr} et~al.,}{{Rohr}
  et~al.}{2022}]{Rohr2022}
{Rohr} E.,  et~al., 2022, \mn@doi [\mnras] {10.1093/mnras/stab3625}, \href
  {https://ui.adsabs.harvard.edu/abs/2022MNRAS.510.3967R} {510, 3967}

\bibitem[\protect\citeauthoryear{{Safronov}}{{Safronov}}{1960}]{Safronov1960}
{Safronov} V.~S.,  1960, Annales d'Astrophysique, \href
  {https://ui.adsabs.harvard.edu/abs/1960AnAp...23..979S} {23, 979}

\bibitem[\protect\citeauthoryear{{Saintonge} et~al.,}{{Saintonge}
  et~al.}{2011a}]{Saintonge2011a}
{Saintonge} A.,  et~al., 2011a, \mn@doi [\mnras]
  {10.1111/j.1365-2966.2011.18677.x}, \href
  {https://ui.adsabs.harvard.edu/abs/2011MNRAS.415...32S} {415, 32}

\bibitem[\protect\citeauthoryear{{Saintonge} et~al.,}{{Saintonge}
  et~al.}{2011b}]{Saintonge2011b}
{Saintonge} A.,  et~al., 2011b, \mn@doi [\mnras]
  {10.1111/j.1365-2966.2011.18823.x}, \href
  {https://ui.adsabs.harvard.edu/abs/2011MNRAS.415...61S} {415, 61}

\bibitem[\protect\citeauthoryear{{Salim} et~al.,}{{Salim}
  et~al.}{2007}]{Salim2007}
{Salim} S.,  et~al., 2007, \mn@doi [\apjs] {10.1086/519218}, \href
  {https://ui.adsabs.harvard.edu/abs/2007ApJS..173..267S} {173, 267}

\bibitem[\protect\citeauthoryear{{Sanders} \& {Mirabel}}{{Sanders} \&
  {Mirabel}}{1996}]{Sanders&Mirabel1996}
{Sanders} D.~B.,  {Mirabel} I.~F.,  1996, \mn@doi [\araa]
  {10.1146/annurev.astro.34.1.749}, \href
  {https://ui.adsabs.harvard.edu/abs/1996ARA&A..34..749S} {34, 749}

\bibitem[\protect\citeauthoryear{Sargent, B{\'{e}}thermin, Daddi  \&
  Elbaz}{Sargent et~al.}{2012}]{Sargent2012}
Sargent M.~T.,  B{\'{e}}thermin M.,  Daddi E.,   Elbaz D.,  2012, \mn@doi [The
  Astrophysical Journal Letters] {10.1088/2041-8205/747/2/L31}, 747, 31

\bibitem[\protect\citeauthoryear{Sargent et~al.,}{Sargent
  et~al.}{2014}]{Sargent2014}
Sargent M.~T.,  et~al., 2014, \mn@doi [Astrophysical Journal]
  {10.1088/0004-637X/793/1/19}, 793, 19

\bibitem[\protect\citeauthoryear{Schmidt}{Schmidt}{1959}]{Schmidt1959}
Schmidt M.,  1959, \mn@doi [The Astrophysical Journal] {10.1086/146614}, 129,
  243

\bibitem[\protect\citeauthoryear{{Schreiber} et~al.,}{{Schreiber}
  et~al.}{2015}]{Schreiber2015}
{Schreiber} C.,  et~al., 2015, \mn@doi [\aap] {10.1051/0004-6361/201425017},
  \href {https://ui.adsabs.harvard.edu/abs/2015A&A...575A..74S} {575, A74}

\bibitem[\protect\citeauthoryear{Scoville et~al.,}{Scoville
  et~al.}{2016}]{Scoville2016}
Scoville N.,  et~al., 2016, \mn@doi [The Astrophysical Journal]
  {10.3847/0004-637x/820/2/83}, 820, 83

\bibitem[\protect\citeauthoryear{Scoville et~al.,}{Scoville
  et~al.}{2017}]{Scoville2017}
Scoville N.,  et~al., 2017, \mn@doi [The Astrophysical Journal]
  {10.3847/1538-4357/aa61a0}, 837, 150

\bibitem[\protect\citeauthoryear{{Segovia Otero}, {Renaud}  \&
  {Agertz}}{{Segovia Otero} et~al.}{2022}]{SegoviaOtero2022}
{Segovia Otero} {\'A}.,  {Renaud} F.,   {Agertz} O.,  2022, arXiv e-prints,
  \href {https://ui.adsabs.harvard.edu/abs/2022arXiv220608379S} {p.
  arXiv:2206.08379}

\bibitem[\protect\citeauthoryear{{Shah} et~al.,}{{Shah}
  et~al.}{2020}]{Shah2020}
{Shah} E.~A.,  et~al., 2020, \mn@doi [\apj] {10.3847/1538-4357/abbf59}, \href
  {https://ui.adsabs.harvard.edu/abs/2020ApJ...904..107S} {904, 107}

\bibitem[\protect\citeauthoryear{Silverman et~al.,}{Silverman
  et~al.}{2015}]{Silverman2015}
Silverman J.~D.,  et~al., 2015, \mn@doi [Astrophysical Journal Letters]
  {10.1088/2041-8205/812/2/L23}, 812

\bibitem[\protect\citeauthoryear{Silverman et~al.,}{Silverman
  et~al.}{2018}]{Silverman2018}
Silverman J.~D.,  et~al., 2018, \mn@doi [The Astrophysical Journal]
  {10.3847/1538-4357/aae25e}, 867, 92

\bibitem[\protect\citeauthoryear{{Smercina} et~al.,}{{Smercina}
  et~al.}{2022}]{Smercina2022}
{Smercina} A.,  et~al., 2022, \mn@doi [\apj] {10.3847/1538-4357/ac5d5f}, \href
  {https://ui.adsabs.harvard.edu/abs/2022ApJ...929..154S} {929, 154}

\bibitem[\protect\citeauthoryear{{Sofue}, {Wakamatsu}, {Taniguchi}  \&
  {Nakai}}{{Sofue} et~al.}{1993}]{Sofue1993}
{Sofue} Y.,  {Wakamatsu} K.-I.,  {Taniguchi} Y.,   {Nakai} N.,  1993, \pasj,
  \href {https://ui.adsabs.harvard.edu/abs/1993PASJ...45...43S} {45, 43}

\bibitem[\protect\citeauthoryear{{Solomon} \& {Sage}}{{Solomon} \&
  {Sage}}{1988}]{Solomon_&_Sage1998}
{Solomon} P.~M.,  {Sage} L.~J.,  1988, \mn@doi [\apj] {10.1086/166865}, \href
  {https://ui.adsabs.harvard.edu/abs/1988ApJ...334..613S} {334, 613}

\bibitem[\protect\citeauthoryear{{Sparre} \& {Springel}}{{Sparre} \&
  {Springel}}{2016}]{Sparre&Springel2016}
{Sparre} M.,  {Springel} V.,  2016, \mn@doi [\mnras] {10.1093/mnras/stw1793},
  \href {https://ui.adsabs.harvard.edu/abs/2016MNRAS.462.2418S} {462, 2418}

\bibitem[\protect\citeauthoryear{{Sparre}, {Hayward}, {Feldmann},
  {Faucher-Gigu{\`e}re}, {Muratov}, {Kere{\v{s}}}  \& {Hopkins}}{{Sparre}
  et~al.}{2017}]{Sparre2017}
{Sparre} M.,  {Hayward} C.~C.,  {Feldmann} R.,  {Faucher-Gigu{\`e}re} C.-A.,
  {Muratov} A.~L.,  {Kere{\v{s}}} D.,   {Hopkins} P.~F.,  2017, \mn@doi
  [\mnras] {10.1093/mnras/stw3011}, \href
  {https://ui.adsabs.harvard.edu/abs/2017MNRAS.466...88S} {466, 88}

\bibitem[\protect\citeauthoryear{{Speagle}, {Steinhardt}, {Capak}  \&
  {Silverman}}{{Speagle} et~al.}{2014}]{Speagle2014}
{Speagle} J.~S.,  {Steinhardt} C.~L.,  {Capak} P.~L.,   {Silverman} J.~D.,
  2014, \mn@doi [\apjs] {10.1088/0067-0049/214/2/15}, \href
  {https://ui.adsabs.harvard.edu/abs/2014ApJS..214...15S} {214, 15}

\bibitem[\protect\citeauthoryear{Springel}{Springel}{2005}]{Springel2005}
Springel V.,  2005, {The cosmological simulation code GADGET-2},
  \mn@doi{10.1111/j.1365-2966.2005.09655.x}

\bibitem[\protect\citeauthoryear{{Stinson}, {Dalcanton}, {Quinn}, {Kaufmann}
  \& {Wadsley}}{{Stinson} et~al.}{2007}]{Stinson2007}
{Stinson} G.~S.,  {Dalcanton} J.~J.,  {Quinn} T.,  {Kaufmann} T.,   {Wadsley}
  J.,  2007, \mn@doi [\apj] {10.1086/520504}, \href
  {https://ui.adsabs.harvard.edu/abs/2007ApJ...667..170S} {667, 170}

\bibitem[\protect\citeauthoryear{Suess, Kriek, Price  \& Barro}{Suess
  et~al.}{2019}]{Suess2019}
Suess K.~A.,  Kriek M.,  Price S.~H.,   Barro G.,  2019, \mn@doi [The
  Astrophysical Journal] {10.3847/1538-4357/ab1bda}, 877, 103

\bibitem[\protect\citeauthoryear{{Tacchella}, {Dekel}, {Carollo}, {Ceverino},
  {DeGraf}, {Lapiner}, {Mandelker}  \& {Primack Joel}}{{Tacchella}
  et~al.}{2016}]{Tacchella2016a}
{Tacchella} S.,  {Dekel} A.,  {Carollo} C.~M.,  {Ceverino} D.,  {DeGraf} C.,
  {Lapiner} S.,  {Mandelker} N.,   {Primack Joel} R.,  2016, \mn@doi [\mnras]
  {10.1093/mnras/stw131}, \href
  {https://ui.adsabs.harvard.edu/abs/2016MNRAS.457.2790T} {457, 2790}

\bibitem[\protect\citeauthoryear{Tacconi et~al.,}{Tacconi
  et~al.}{2018}]{Tacconi2018}
Tacconi L.~J.,  et~al., 2018, \mn@doi [The Astrophysical Journal]
  {10.3847/1538-4357/aaa4b4}, 853, 179

\bibitem[\protect\citeauthoryear{{Tacconi}, {Genzel}  \& {Sternberg}}{{Tacconi}
  et~al.}{2020}]{Tacconi2020}
{Tacconi} L.~J.,  {Genzel} R.,   {Sternberg} A.,  2020, \mn@doi [\araa]
  {10.1146/annurev-astro-082812-141034}, \href
  {https://ui.adsabs.harvard.edu/abs/2020ARA&A..58..157T} {58, 157}

\bibitem[\protect\citeauthoryear{{Toomre}}{{Toomre}}{1964}]{Toomre1964}
{Toomre} A.,  1964, \mn@doi [\apj] {10.1086/147861}, \href
  {https://ui.adsabs.harvard.edu/abs/1964ApJ...139.1217T} {139, 1217}

\bibitem[\protect\citeauthoryear{{Violino}, {Ellison}, {Sargent}, {Coppin},
  {Scudder}, {Mendel}  \& {Saintonge}}{{Violino} et~al.}{2018}]{Violino2018}
{Violino} G.,  {Ellison} S.~L.,  {Sargent} M.,  {Coppin} K. E.~K.,  {Scudder}
  J.~M.,  {Mendel} T.~J.,   {Saintonge} A.,  2018, \mn@doi [\mnras]
  {10.1093/mnras/sty345}, \href
  {https://ui.adsabs.harvard.edu/abs/2018MNRAS.476.2591V} {476, 2591}

\bibitem[\protect\citeauthoryear{Vogelsberger et~al.,}{Vogelsberger
  et~al.}{2014}]{Vogelsberger2014}
Vogelsberger M.,  et~al., 2014, \mn@doi [Monthly Notices of the Royal
  Astronomical Society] {10.1093/mnras/stu1536}

\bibitem[\protect\citeauthoryear{Wetzel, Hopkins, Kim, Faucher-Giguere, Keres
  \& Quataert}{Wetzel et~al.}{2016}]{Wetzel2016}
Wetzel A.~R.,  Hopkins P.~F.,  Kim J.-h.,  Faucher-Giguere C.-A.,  Keres D.,
  Quataert E.,  2016, \mn@doi [The Astrophysical Journal]
  {10.3847/2041-8205/827/2/L23}

\bibitem[\protect\citeauthoryear{{Whitaker}, {van Dokkum}, {Brammer}  \&
  {Franx}}{{Whitaker} et~al.}{2012}]{Whitaker2012}
{Whitaker} K.~E.,  {van Dokkum} P.~G.,  {Brammer} G.,   {Franx} M.,  2012,
  \mn@doi [\apjl] {10.1088/2041-8205/754/2/L29}, \href
  {https://ui.adsabs.harvard.edu/abs/2012ApJ...754L..29W} {754, L29}

\bibitem[\protect\citeauthoryear{{Wilkinson}, {Pimbblet}, {Stott}, {Few}  \&
  {Gibson}}{{Wilkinson} et~al.}{2018}]{Wilkinson2018}
{Wilkinson} C.~L.,  {Pimbblet} K.~A.,  {Stott} J.~P.,  {Few} C.~G.,   {Gibson}
  B.~K.,  2018, \mn@doi [\mnras] {10.1093/mnras/sty1493}, \href
  {https://ui.adsabs.harvard.edu/abs/2018MNRAS.479..758W} {479, 758}

\bibitem[\protect\citeauthoryear{{Wilkinson}, {Ellison}, {Bottrell}, {Bickley},
  {Gwyn}, {Cuillandre}  \& {Wild}}{{Wilkinson} et~al.}{2022}]{Wilkinson2022}
{Wilkinson} S.,  {Ellison} S.~L.,  {Bottrell} C.,  {Bickley} R.~W.,  {Gwyn} S.,
   {Cuillandre} J.-C.,   {Wild} V.,  2022, \mn@doi [\mnras]
  {10.1093/mnras/stac1962}, \href
  {https://ui.adsabs.harvard.edu/abs/2022MNRAS.516.4354W} {516, 4354}

\bibitem[\protect\citeauthoryear{{Zabludoff}, {Zaritsky}, {Lin}, {Tucker},
  {Hashimoto}, {Shectman}, {Oemler}  \& {Kirshner}}{{Zabludoff}
  et~al.}{1996}]{Zabludoff1996}
{Zabludoff} A.~I.,  {Zaritsky} D.,  {Lin} H.,  {Tucker} D.,  {Hashimoto} Y.,
  {Shectman} S.~A.,  {Oemler} A.,   {Kirshner} R.~P.,  1996, \mn@doi [\apj]
  {10.1086/177495}, \href
  {https://ui.adsabs.harvard.edu/abs/1996ApJ...466..104Z} {466, 104}

\bibitem[\protect\citeauthoryear{Zolotov et~al.,}{Zolotov
  et~al.}{2015}]{Zolotov2015}
Zolotov A.,  et~al., 2015, \mn@doi [MNRAS] {10.1093/MNRAS/STV740}, 450, 2327

\makeatother
\end{thebibliography}


\appendix
\section{Difference in the evolution of interacting and non-interacting SB galaxies}\label{app:INT_split}
\begin{figure}
    \centering
	\includegraphics[width=\columnwidth]{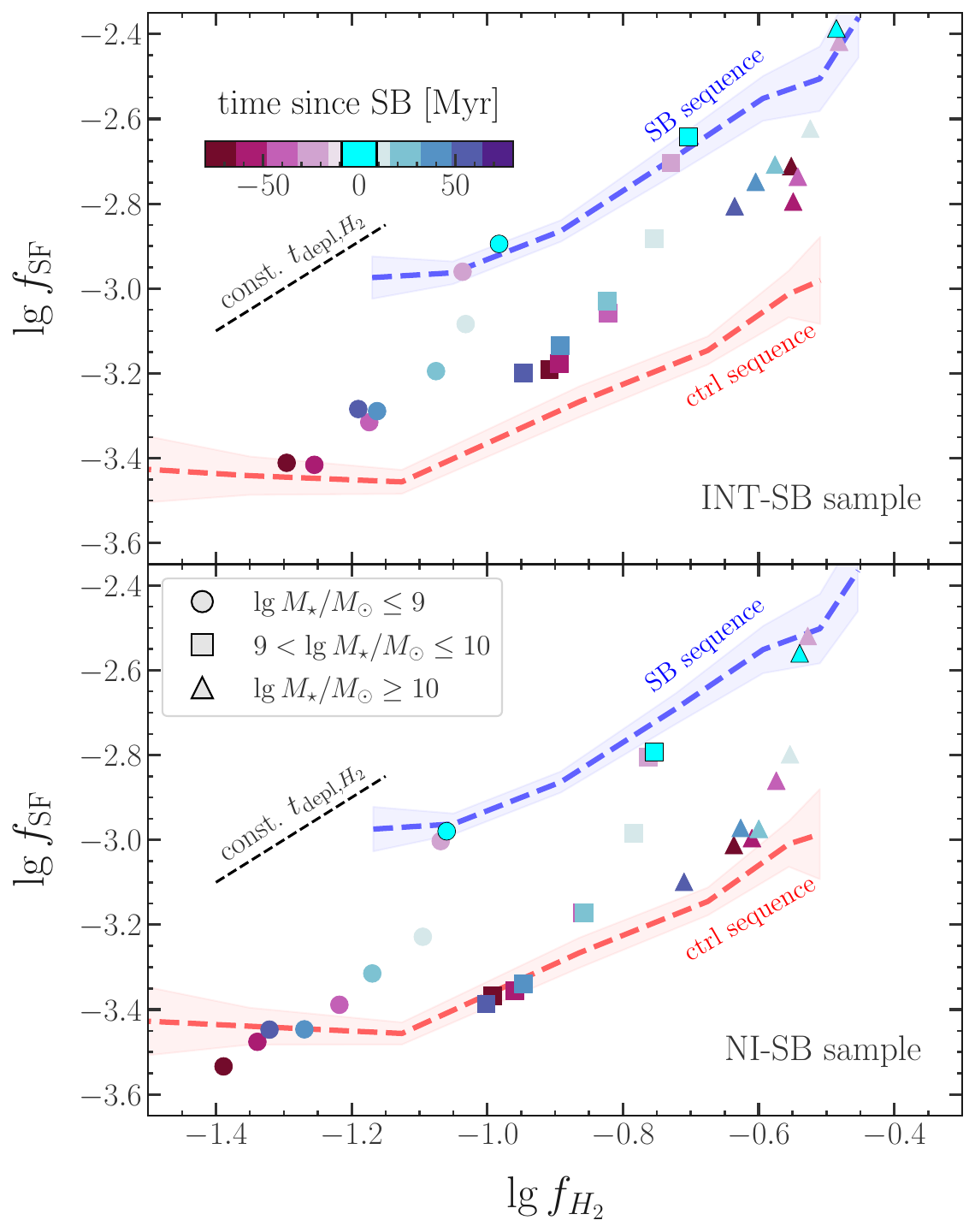}
    \caption{Median evolution of the mass fraction of molecular gas ($\fmol$) and the fraction of gas mass eligible for star formation in our model ($f_\SF$) in  of interacting (top panel; INT-SB sample) and non-interacting (bottom panel; NI-SB sample) SB galaxies, in a period of time $\sim 150\Myr$ centred around the time when the galaxies are identified as a SB, colour-coded by the time difference with respect to the beginning of the SB event. The blue and red dashed curves represent the median trends for SB and control galaxies, respectively, parameterised by their $M_\star$ (i.e., connecting the median values in this plane for SB and control galaxies in different stellar mass bins), with $2\sigma$ bootstrapped error (shaded area). The different marker styles refer to different stellar mass bins. Galaxies move towards the SB sequence by increasing $\fmol$ and, disproportionately $f_\SF$, i.e., resulting in a decrease in their $\tdepl$. Following the SB event, galaxies galaxies move towards the control sequence along the same tracks, reducing both $\fmol$ and $f_\SF$. On average, at $\sim 60\Myr$ prior to and after the SB, INT-SB galaxies have a larger $f_\SF$ by $\sim 0.2$ dex larger compared to both control and NI-SB galaxies. Furthermore, at $\sim 60\Myr$ prior to and after the SB, NI-SB galaxies have comparable $f_\SF$, on average, to control galaxies.}
    \label{fig:fSF_vs_fH2_INTsplit}
\end{figure}

In this section, we investigate the different behaviour of interacting and non-interacting SB galaxies before and after the beginning of SB event, expanding what explored in Section~\ref{sec:results_interactions} on the role of interactions in driving SBs. Figure~\ref{fig:fSF_vs_fH2_INTsplit} shows the median evolution of both the fraction of molecular gas ($\fmol$) and fraction of high-density gas that is eligible for star formation in our model ($f_\SF$) in interacting (top panel; INT-SB sample) and non-interacting (bottom panel; NI-SB sample) SB galaxies, for a time period of $\sim 150\Myr$ centred around the beginning of the SB. The blue and red dashed curves represent the median trends for all (i.e., irrespective of whether they are interacting or non-interacting) SB and control galaxies, respectively, parameterised by their $M_\star$, i.e., connecting the median values for SB and control galaxies in this plane in different mass bins. In general, by moving towards the SB sequence, SB galaxies increase both their $f_\SF$ and $\fmol$, as previously noted in Section~\ref{sec:results_fSF}. NI-SB galaxies evolve from the control sequence to the SB sequence in the $\sim 70\Myr$ prior to the beginning of the SB event. Conversely, INT-SB galaxies at $\sim 70\Myr$ prior to the SB are offset from the control sequence by $\sim 0.1-0.3$ dex, for increasing stellar mass, exhibiting higher median $f_\SF$ and $\fmol$. The same behaviour is also visible in the $\sim 70\Myr$ after the SB. Hence, interactions in $M_\star\gtrsim 10^{9}\Msun$ galaxies can drive SBs only if these are already in excess of high-density and molecular gas, compared to control galaxies. This result combined with the increasing fraction of INT-SB galaxies with stellar mass (see Figure~\ref{fig:INT_fraction_1}) results in the trends observed in Figure~\ref{fig:fSF_vs_fH2}, where massive ($M_\star\gtrsim 10^{10}\Msun$) and hence likely interacting SB galaxies, exhibit large median values of $f_\SF$ and $\fmol$ at $\sim 70\Myr$ prior to and after the SB, compared to control galaxies. 

\begin{figure}
    \centering
    \includegraphics[width=\columnwidth]{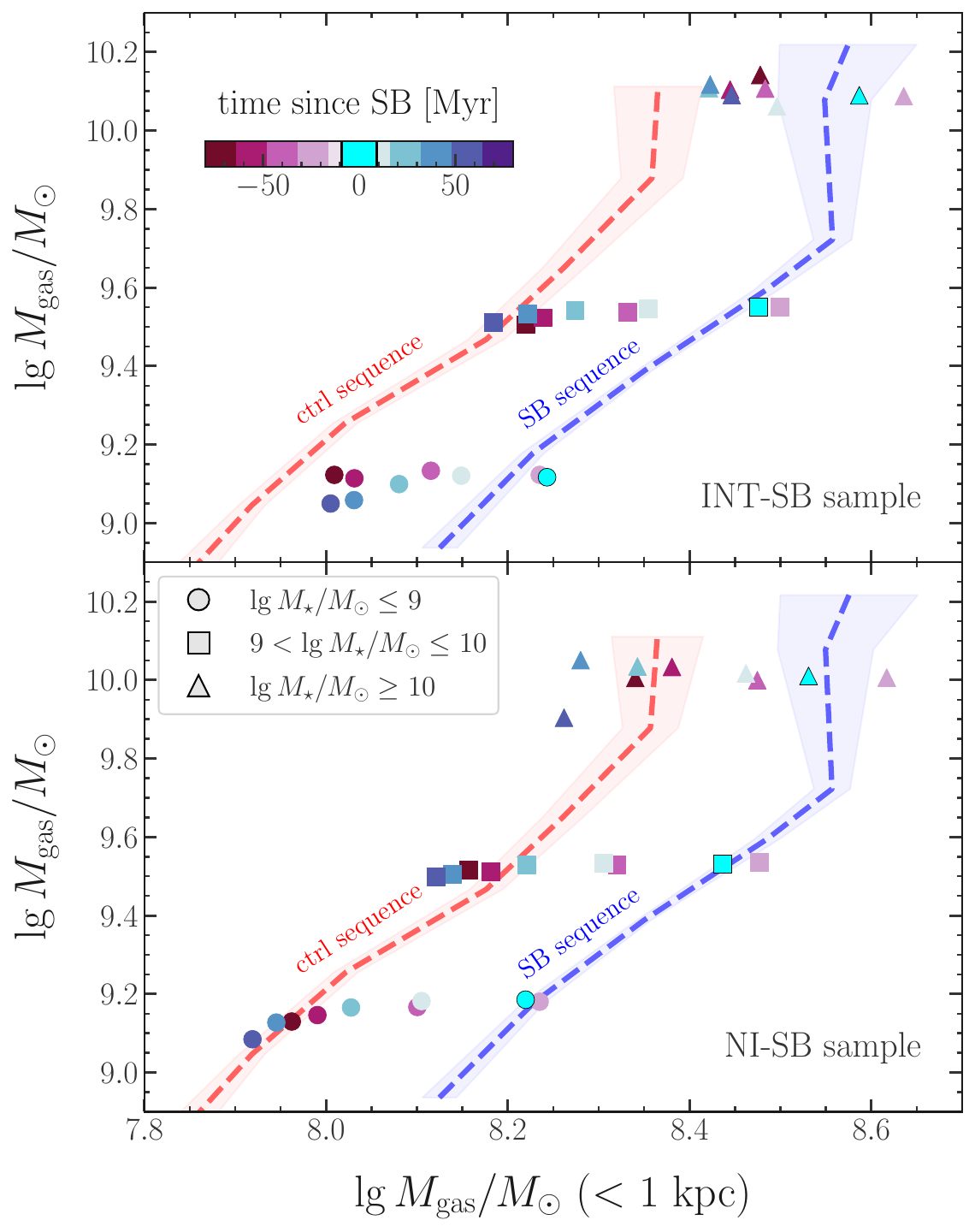}
    \caption{Median evolution of the total gas mass ($M_{\rm gas}$) and gas mass within the central kpc ($M_{\rm gas}\left(<1\kpc\right)$) of interacting (top panel; INT-SB sample) and non-interacting (bottom panel; NI-SB sample) SB galaxies, in a period of time $\sim 150\Myr$ centred around the time when the galaxies are identified as a SB, colour-coded by the time difference with respect to the beginning of the SB event. The blue and red dashed curves represent the median trends for SB and control galaxies, respectively, parameterised by their $M_\star$ (i.e., connecting the median values in this plane for SB and control galaxies in different stellar mass bins), with $2\sigma$ bootstrapped error (shaded area). On average, at $\sim 60\Myr$ prior to and after the SB, INT-SB galaxies have a larger $M_{\rm gas}\left(<1\kpc\right)$ by $\sim 0.5$ dex compared to both control and NI-SB galaxies. Furthermore, at $\sim 60\Myr$ prior to and after the SB, NI-SB galaxies have comparable $M_{\rm gas}\left(<1\kpc\right)$, on average, to control galaxies.}
    \label{fig:Mgas_vs_Mgas_1kpc_INTsplit}
\end{figure}

Figure~\ref{fig:Mgas_vs_Mgas_1kpc_INTsplit},shows the median evolution of the total (all phases) gas mass within the central kpc ($M_{\rm gas}\left(<1\kpc\right)$) and within the galaxy radius (i.e., within $0.1\,R_{\rm vir}$; $M_{\rm gas}$) in interacting (top panel; INT-SB sample) and non-interacting (bottom panel; NI-SB sample) SB galaxies, for a time period of $\sim 150\Myr$ centred around the beginning of the SB. As in Figure~\ref{fig:fSF_vs_fH2_INTsplit}, the blue and red dashed curves represent the median trends for all SB and control galaxies, respectively, parameterised by their $M_\star$. In general, INT-SB galaxies have a median $M_{\rm gas}\left(<1\kpc\right)$ that is larger by $\sim 0.1$ dex than the median values for NI-SB galaxies. Interactions are therefore slightly more effective in driving SBs in galaxies with an already relatively large amount of gas concentrated in their central regions, whereas NI-SB galaxies can experience larger fluctuations in $M_{\rm gas}\left(<1\kpc\right)$. Combining this result with the dependency of the fraction of INT-SB galaxies on stellar mass we can recover the behaviour in Figure~\ref{fig:Mgas_vs_Mgas_1kpc}.

\bsp	
\label{lastpage}
\end{document}